%% file: elliptic_draft_brief.tex
\documentclass[12pt]{article}
 
\input{epsf}
\usepackage{epsfig}
\usepackage{amssymb}
\usepackage{amsfonts}
\usepackage{amsbsy}
\usepackage[cmtip,all]{xy}
\usepackage{amsmath}
\usepackage{esint}
\usepackage{collectbox}
\usepackage{amscd}
\usepackage{mathrsfs}
\usepackage{amsmath,amsthm}
\usepackage{enumitem}
\usepackage{dsfont}
\usepackage{soul}
\usepackage{comment}
\usepackage{cite} 
\usepackage[most]{tcolorbox}
\usepackage{stmaryrd}

\usepackage{xcolor}
\definecolor{burgundy}{rgb}{0.5, 0.0, 0.13}
\definecolor{olive}{rgb}{0.50, 0.50, 0.0}
\usepackage[linktocpage=true,colorlinks=true,linkcolor=burgundy,citecolor=black!20!blue,urlcolor=violet]{hyperref}

\usepackage{tikz}
\usepackage{tikz-cd}
\usetikzlibrary{arrows}
\usetikzlibrary{arrows.meta}
\usetikzlibrary{positioning}
\usetikzlibrary{shapes,snakes}
\usetikzlibrary{fit}
\usetikzlibrary{decorations.pathmorphing,decorations.pathreplacing,decorations.markings}

\DeclareMathAlphabet{\mathpzc}{OT1}{pzc}{m}{it}

\def\be{ \begin{equation} }
\def\ee{ \end{equation}}

\def\exp{{\rm exp}}

\def\I{{\rm i}}

\def\log{{\rm log}}

\def\Tr{{\rm Tr}}

\def\p{\partial}

\def\CA{{\cal A}}

\def\CC {{\cal C}}
\def\CD {{\cal D}}

\def\CG {{\cal G}}
\def\CH {{\cal H}}
\def\CI {{\cal I}}
\def\CJ {{\cal J}}

\def\CM {{\cal M}}
\def\CN {{\cal N}}
\def\CO {{\cal O}}

\def\CO {{\cal O}}

\def\CG {{\cal G}}
\def\CH {{\cal H}}
\def\CI {{{\cal I}}}

\def\CQ {{\cal Q}}

\def\IC{\mathbb{C}}

\def\IN{\mathbb{N}}

\def\IP{\mathbb{P}}
\def\IR{{\mathbb{R}}}

\def\IZ{{\mathbb{Z}}}

\def\fg{\mathfrak{g}}

\def\fl{\mathfrak{l}}
\def\fm{\mathfrak{m}}
\def\fn{\mathfrak{n}}

\def\fp{\mathfrak{p}}
\def\fq{\mathfrak{q}}

\def\fs{\mathfrak{s}}
\def\ft{\mathfrak{t}}

\def\fp{\mathfrak{p}}
\def\fq{\mathfrak{q}}

\def\fs{\mathfrak{s}}
\def\ft{\mathfrak{t}}

\def\fR{\mathfrak{R}}

\def\bPsi{{\boldsymbol{\Psi}}}

\def\bvphi{{\boldsymbol{\varphi}}}

\def\Dslash{\,{\raise.15ex\hbox{/}\mkern-12mu \CD}}

\def\lm{\limits}

\numberwithin{equation}{section}

\newcommand\Kappa{\mathrm{K}}
\def\myE{\mathsf{E}}
\def\myT{\mathsf{T}}
\def\myY{\mathsf{Y}}
\def\myA{\mathsf{A}}
\def\myequiv{:= }

\def\myh{\mathsf{h}}

\def\mye{e}
\def\myf{f}
\def\myk{\psi}

\def\lbr{\llbracket}
\def\rbr{\rrbracket}
\def\mys{\texttt{s}}

\def\myphi{\bvphi}
\def\myp{{\fp}}
\def\extra{{\ft}}
\def\myblue{white!40!blue}
\newcommand{\longsquiggly}{\xymatrix{{}\ar@{~>}[r]&{}}}

\tcbset{boxrule=0.6pt,colback=white!97!blue}

\newcommand\sqbox[1]{{
	\setbox0=\hbox{\mbox{$\Box$}}
	\setbox1=\hbox{\mbox{\raisebox{0.35ex}{\tiny #1}}}
	\mbox{\raisebox{-0.2ex}{\rlap{\hbox to \wd0{\hss{\box1}\hss}}\box0}}
}}

\def\bckcolor{white}

\begin{document}
	
\pagenumbering{Alph} 
\begin{titlepage} 
		
\vskip 1.5in
\begin{center}
		
{\bf\Large{Toroidal and Elliptic \\ \vskip 5mm Quiver BPS Algebras and Beyond}}
\vskip 1cm 
\renewcommand{\thefootnote}{\fnsymbol{footnote}}
{Dmitry Galakhov$^{1,2,}$\footnote[2]{e-mail: dmitrii.galakhov@ipmu.jp; galakhov@itep.ru},  Wei Li$^{3}$\footnote[3]{e-mail: weili@mail.itp.ac.cn} and Masahito Yamazaki$^{1,}$\footnote[4]{e-mail: masahito.yamazaki@ipmu.jp}} 
\vskip 0.2in 
\renewcommand{\thefootnote}{\roman{footnote}}
{\small{ 
                \textit{$^1$Kavli Institute for the Physics and Mathematics of the Universe (WPI), }\vskip -.4cm
                \textit{University of Tokyo, Kashiwa, Chiba 277-8583, Japan}
                \vskip 0 cm 
                \textit{$^2$Institute for Information Transmission Problems,}\vskip -.4cm
                \textit{ Moscow, 127994, Russia}
                 \vskip 0 cm 
                \textit{$^3$Institute for Theoretical Physics, Chinese Academy of Sciences}\vskip -.4cm
                \textit{ Beijing, 100190, China}
}}
\end{center}

\vskip 0.5in
\baselineskip 16pt

\begin{abstract}

The quiver Yangian, an infinite-dimensional algebra introduced recently in \cite{Li:2020rij}, is the algebra underlying BPS state counting problems for toric Calabi-Yau three-folds. We introduce  trigonometric and elliptic analogues of quiver Yangians, which we call  toroidal quiver algebras and elliptic  quiver algebras, respectively. We construct the representations of the shifted toroidal and elliptic algebras in terms of the statistical model of crystal melting. We also derive the algebras and their representations from equivariant localization of three-dimensional $\mathcal{N}=2$ supersymmetric quiver gauge theories, and their dimensionally-reduced counterparts. The analysis of supersymmetric gauge theories suggests that there exist even richer classes of algebras associated with higher-genus Riemann surfaces and generalized cohomology theories.

\end{abstract}

\date{August, 2021}
\end{titlepage}
\pagenumbering{arabic} 
	
\newpage
\tableofcontents
\newpage
	
	
\section{Introduction and summary}

For centuries physics and mathematics are tied together and their mutual influence 
goes in both directions. While physicists have often used known mathematical results to derive new results in physics,
physicists can also use their insights to create {\it new} mathematics, which can be of independent interest to mathematicians.

The \emph{quiver Yangian} (introduced in \cite{Li:2020rij} and studied further in \cite{Galakhov:2020vyb,Galakhov:2021xum}) in our opinion is an excellent example for such a 
new mathematical output arising from the physics of supersymmetric gauge theories and string theory. The quiver Yangian $\myY(Q, W)$ is an infinite-dimensional algebra
defined from a quiver $Q$ and a superpotential $W$, and 
can be regarded as a vast generalization of the affine Yangians for Lie superalgebras $\mathfrak{gl}_{m|n}$.
The consistency of the algebras, as well as of their representations on the statistical-mechanical model of crystal melting,
on the one hand can be checked algebraically from the definitions, and on the other hand 
can be derived physically in terms of equivariant localization of supersymmetric quantum mechanics \cite{Galakhov:2020vyb,Galakhov:2021xum}.

In the realm of integrable models it has long been known that Yangians lie at the bottom of the 
elliptic--trigonometric--rational hierarchy of algebras underlying those models (cf.\ \cite{Belavin-Drinfeld}).
It is therefore a natural question to ask if there exist trigonometric and elliptic counterparts of quiver Yangians.

In this paper we answer this question positively.
We define elliptic and trigonometric counterparts $\myE_{\tau}(Q, W)$ and $\myT_{\beta}(Q, W)$ of quiver Yangians,
thus completing the elliptic-trigonometric-rational hierarchy for quiver Yangians. (In the following we sometimes suppress the 
dependence on the superpotential $W$ and use simplified notations
$\myE_{\tau}(Q)$ and $\myT_{\beta}(Q)$.) 
As in the case of quiver Yangians, our algebras can be regarded as 
generalizations of elliptic algebras and quantum toroidal algebras for $\mathfrak{gl}_{m|n}$
discussed previously in the literature (see e.g.\ \cite{MR1324698,Ding:1996mq,Miki2007,MR2793271,Feigin:2013fga,MR2566895,Bezerra:2019dmp}).
We hope that the study of 
our general elliptic and toroidal quiver algebras and their representation theories
will grow into a new exciting research area.

In physics language, 
these algebras are realized by 3d/2d/1d quantum field theories with four supercharges,
in which the algebras are related by dimensional reductions. Moreover all these theories are effective field theories on the D-branes probing toric Calabi-Yau three-folds:

\be\label{red_diagram}
\begin{array}{c}
        \begin{tikzpicture}
                \begin{scope}[scale=0.8]
                        \begin{scope}[yscale=0.5,scale=0.8]
                                \draw[thick,fill=white!70!blue] (0,0) circle (1.5);
                        \end{scope}
                        \begin{scope}[shift={(0,0)}]
                                \begin{scope}[xscale=2,yscale=1.5]
                                        \draw[thick,fill=white] (-0.2,0) to[out=315,in=180] (0,-0.1) to[out=0,in=225] (0.2,0) to[out=135,in=0] (0,0.1) to[out=180,in=45] (-0.2,0);
                                        \draw[thick] (-0.25,0.05) -- (-0.2,0) to[out=315,in=180] (0,-0.1) to[out=0,in=225] (0.2,0) -- (0.25,0.05) (-0.2,0) to[out=45,in=180] (0,0.1) to[out=0,in=135] (0.2,0);
                                \end{scope}
                        \end{scope}
                \end{scope}
                \node at (0,0.8) {$\myE_{\tau}(Q,W)$};
                \node(A) at (0,-0.8) {\bf\tiny D8-D6-D4-D2-D0};
                \begin{scope}[shift={(3.5,0)}]
                        \draw[thick] (0,0) circle (0.3);
                        \node at (0,0.8) {$\myT_{\beta}(Q,W)$};
                        \node(B) at (0,-0.8) {\bf\tiny D7-D5-D3-D1};
                \end{scope}
                \begin{scope}[shift={(7,0)}]
                        \draw[fill=black] (0,0) circle (0.08);
                        \node at (0,0.8) {$\myY(Q,W)$};
                        \node(C) at (0,-0.8) {\bf\tiny D6-D4-D2-D0};
                \end{scope}
                \draw[-stealth] (1.3,0) -- (2.2,0) node[pos=0.5,above] {\scalebox{0.7}{$\begin{array}{c}
                                \mbox{ dim.}\\ \mbox{red.}\\
                        \end{array}$}};
                \draw[-stealth] (4.8,0) -- (5.7,0) node[pos=0.5,above] {\scalebox{0.7}{$\begin{array}{c}
                                        \mbox{ dim.}\\ \mbox{red.}\\
                                \end{array}$}};
                %
                \node at (-1.3,0.3) {$\left[\rule{0cm}{1cm}\right.$}; 
                \node at (8.0,0.3) {$\left.\rule{0cm}{1cm}\right]$}; 
                \node[right] at (8.0,0) {$\times \genfrac{(}{)}{0pt}{0}{\textrm{toric Calabi-Yau}}{\textrm{3-fold}}$};
                \node[below] at (A) {$\begin{array}{c}
                                \Theta_q(z)\\
                                3{\rm d}\;\CN=2\\
                        \end{array}$};
                \node[below] at (B) {$\begin{array}{c}
                                2 \sinh(\beta z/2)\\
                                2{\rm d}\;\CN=(2,2)\\
                        \end{array}$};
                \node[below] at (C) {$\begin{array}{c}
                                z\\
                                1{\rm d}\;\CN=4\\
                        \end{array}$};
        \end{tikzpicture}
\end{array}
\! .
\ee
Here the elliptic--trigonometric-rational hierarchy is represented by three odd functions,
namely the theta function $\zeta(z)=\Theta_q(z)$ (where as we will see $q=\exp(2\pi \I \tau)$ with $\tau$ being the modulus of the torus) for the elliptic case,
the hyperbolic sine function $\zeta(z)=2\sinh (\beta z/2)$ (where $\beta$ determines the periodicity along the circle direction of the cylinder) for the trigonometric case,
and $\zeta(z)=z$ for the rational case. We will see that the physics of supersymmetric gauge theories gives  not only these algebras and their crystal-melting representations,
but also their Hopf-algebra and shuffle-algebra structures.

Somewhat surprisingly, the physics derivation from quiver gauge theories actually points to 
even richer possibilities: we can consider 
more general Riemann surfaces (of genus greater than one, and also possibly with punctures):
\be
\begin{array}{c}
        \begin{tikzpicture}
                \begin{scope}[scale=0.8]
                        \begin{scope}[yscale=0.5]
                                \draw[thick,fill=white!70!blue] (0,0) circle (1.5);
                        \end{scope}
                        \begin{scope}[shift={(-0.4,-0.2)}]
                                \begin{scope}[xscale=2,yscale=1.5]
                                        \draw[thick,fill=white] (-0.2,0) to[out=315,in=180] (0,-0.1) to[out=0,in=225] (0.2,0) to[out=135,in=0] (0,0.1) to[out=180,in=45] (-0.2,0);
                                        \draw[thick] (-0.25,0.05) -- (-0.2,0) to[out=315,in=180] (0,-0.1) to[out=0,in=225] (0.2,0) -- (0.25,0.05) (-0.2,0) to[out=45,in=180] (0,0.1) to[out=0,in=135] (0.2,0);
                                \end{scope}
                        \end{scope}
                        \begin{scope}[shift={(0.4,0.2)}]
                                \begin{scope}[xscale=2,yscale=1.5]
                                        \draw[thick,fill=white] (-0.2,0) to[out=315,in=180] (0,-0.1) to[out=0,in=225] (0.2,0) to[out=135,in=0] (0,0.1) to[out=180,in=45] (-0.2,0);
                                        \draw[thick] (-0.25,0.05) -- (-0.2,0) to[out=315,in=180] (0,-0.1) to[out=0,in=225] (0.2,0) -- (0.25,0.05) (-0.2,0) to[out=45,in=180] (0,0.1) to[out=0,in=135] (0.2,0);
                                \end{scope}
                        \end{scope}
                \end{scope}
        \end{tikzpicture}
\end{array}\quad\leftrightsquigarrow
\quad \textrm{(generalized cohomology theories)} \;.
\ee
While we leave many questions for future work, we argue in this paper  that we can define quiver BPS algebras for more general cases,
associated with moduli spaces of supersymmetric field theories and/or 
generalized cohomology theories (and ``formal group laws'').
For this reason the possible family of quiver BPS algebras will go well beyond the traditional
elliptic-trigonometric-rational hierarchy of quiver BPS algebras (and hence the word `beyond' in the title of this paper).\footnote{
Related to this point, it is 
worth pointing out that 
the ``lens-elliptic'' extension of the elliptic case has been studied in the literature \cite{Kels:2018xnf}, by replacing the theta function
by the ``lens-theta'' functions introduced in \cite{Kels:2015bda,Kels:2017toi}.}

The rest of this paper is organized as follows.
In section \ref{sec:alg} we define elliptic and toroidal quiver algebras.
In section \ref{sec:rep} we construct representations of the elliptic and toroidal quiver algebras
from the statistical model of crystal melting. 
In section \ref{sec:D-brane} we discuss the derivations of the quiver BPS algebras 
from the analysis of supersymmetric quiver gauge theories arising from D-branes.
This analysis is extended further in section \ref{sec:Hopf}, where we derive Hopf-algebra structures 
and shuffle-algebra structures. Section \ref{sec:hyperelliptic} is about the generalization to 
higher genus, and connections to generalized cohomology theories and formal group laws.
We have also included appendices on more technical materials.

This paper is intended for a mixed readership of both mathematicians and physicists.
For mathematicians interested in our algebras and their representation theories, note that 
sections \ref{sec:alg} and \ref{sec:rep} do not require any knowledge of supersymmetric field theories,
and will serve as a self-contained guide to the algebras. Sections \ref{sec:D-brane} and \ref{sec:Hopf}, while written partly in the language of physics,
contain more geometrical and algebraic discussions of the algebras.
For physicists, one option is to start with the definitions of the algebras in sections \ref{sec:alg} and \ref{sec:rep},
however those wishing to get intuitions and understand the physics contexts first might prefer to start with section \ref{sec:D-brane},
where we discuss the first-principle derivations of the quiver BPS algebras from supersymmetric quiver gauge theories.

{\it Note added: } During the final stages of this project, a preprint \cite{Noshita:2021ldl} appeared in the arXiv, which discusses the toroidal version of our quiver algebras but only for the unshifted cases $\mys^{(a)}=0$ and with the focus on the non-chiral quivers;
the overlaps are within sections \ref{sec:alg} and \ref{sec:rep} of this paper.

\section{Elliptic and toroidal quiver algebras}\label{sec:alg}

In this section we define elliptic and toroidal quiver BPS algebras $\myE_{\tau}(Q,W)$ and $\myT_{\beta}(Q,W)$,
generalizing the quiver Yangian $\myY(Q,W)$ for the rational case.
The quiver Yangian can be regarded as a reduction of the toroidal quiver algebra,
which in turn is a reduction of the elliptic quiver algebra. In fact, we will see that all the three algebras can be defined in a unified manner
with essentially the same ingredients, if we keep in mind some important differences between them.

Let us here comment on the naming conventions of the algebras.
Since we have the elliptic-trigonometric-rational hierarchy, one systematic naming scheme is to
call our algebras  ``elliptic/trigonometric/rational quiver BPS algebra,'' or simply 
``elliptic/trigonometric/rational quiver algebra,'' respectively. 

While our elliptic/trigonometric algebras are in general new, 
the complication is that they (as we will see in section \ref{subsec:comparison}) coincide with 
known algebras for Lie superalgebras $\fg\fl_{n|m}$ when the associated toric Calabi-Yau manifold has no compact four-cycles;
in these cases various names are already given to the 
algebras. For the rational case the algebra is called the affine Yangian, the spherical Hecke central (SHc) algebra 
or the universal enveloping algebra for the $W_{1+\infty}$-algebra  \cite{MR3951025,MR3150250,Tsymbaliuk,Tsymbaliuk:2014fvq,Prochazka:2015deb,Gaberdiel:2017dbk}, while for the trigonometric case\footnote{For the trigonometric case
the algebra is sometimes denoted as $U_{q,p}(\ddot{\fg\fl}_{m|n})$ or $U_{q,p}(\widehat{\widehat{\fg\fl}}_{m|n})$.}
it  is called the quantum toroidal algebra\cite{MR1324698,Feigin:2013fga,Awata:2017lqa} or Ding-Iohara-Miki (DIM) algebra \cite{Ding:1996mq,Miki2007,Awata:2018svb} (for the case of $\mathfrak{gl}_1$);\footnote{Strictly speaking people mean slightly different algebras by these terminologies, e.g.\ quantum toroidal algebras (resp.\ DIM algebras) do (resp.\ do not) include the Serre relations.}  
the elliptic case of the DIM algebra \cite{MR2566895,MR3262444} is sometimes called the elliptic DIM algebra (again for $\mathfrak{gl}_1$), for example.
In order to reflect this state-of-the-art in the community, in this paper
we call the rational quiver algebra as the quiver Yangian (as in our previous papers),
the trigonometric quiver algebra as the \emph{toroidal quiver algebra},
while we will use the terminology \emph{elliptic quiver algebra} for the elliptic case.\footnote{If we do not mind longer names,
we can also use ``quiver affine Yangian'' for the rational case and ``quiver quantum toroidal algebra'' for the trigonometric case.}

\bigskip
\noindent
\begin{tabular}{c|c|c}
rational & trigonometric & elliptic \\
\hline
\hline
rational quiver algebra & trigonometric quiver algebra & \emph{elliptic quiver algebra} \\
= \emph{quiver Yangian} & = \emph{toroidal quiver algebra} & 
\end{tabular}

\subsection{Quivers and equivariant parameters}\label{subsec:parameters}

As in the case of quiver Yangians \cite{Li:2020rij}, we start with a pair $(Q,W)$ of a quiver $Q$ and 
a superpotential $W$.\footnote{
Mathematically a quiver is an oriented graph, and a superpotential is a formal linear sum (over $\mathbb{C}$) of closed paths of the quiver, where we do not specify the starting/ending point of the closed path.}
While our definition applies to general quivers and superpotentials, 
in practice our focus will be on quivers whose vacuum moduli spaces describe toric Calabi-Yau three-folds (see \cite{Kennaway:2007tq,Yamazaki:2008bt} and references therein). 
We denote the set of vertices of the quiver by $Q_0$ and that of the arrows by $Q_1$.

We introduce an equivariant parameter $h_I$ for each arrow $I\in Q_1$.
We impose two constraints on $h_I$. One is the loop constraint, which says that we have 
\begin{equation}\label{eq:loop-constraint}
\textrm{ loop constraint: } \quad \sum_{I\in \textrm{loop }L} h_I=0 
\end{equation}
for each closed loop $L$ associated with a term in the superpotential $W$.
Namely for each closed path in the superpotential the weights $h_I$'s for the arrows in the path
add up to zero. Another is the vertex constraint
\begin{equation}\label{eq-vertex-constraint-toric}
\textrm{ vertex constraint: } \quad
\sum_{I\in a} \textrm{sign}_a(I) \, h_I=0
\end{equation}
for each vertex $a$, where $\textrm{sign}_a(I)$ is $+1$ if the arrow $I$ flows towards the vertex $a$, $-1$ if the arrow $I$ flows outwards from the vertex $a$, and $0$ otherwise.
For quivers and superpotentials associated with toric Calabi-Yau three-folds,
we can show that we are left with two independent  parameters $\myh_1$ and $\myh_2$ when both loop and vertex constraints are imposed \cite{Li:2020rij}  ---  the two parameters represent equivariant torus actions on the toric Calabi-Yau three-fold.

Throughout the paper we adopt the following notations:
\be
\begin{array}{cl}
     \{a\to b\}: &\mbox{  the set of all arrows pointing  from node }a\mbox{ to node }b \;,\\
     |a\to b|: &\mbox{  the total number of arrows pointing from node }a\mbox{ to node }b \;.
\end{array}
\ee
For a pair of nodes $a,b\in Q_0$ we define their chirality parameter as:
\be\label{eq-def-chi}
\chi_{ab}\myequiv|a\to b|-|b\to a|\,,
\ee
which satisfies $\chi_{aa}=\chi_{ab}+\chi_{ba}=0$.
The quiver is called \textit{non-chiral} when $|a\rightarrow b|=|b\rightarrow a|$ for any pair of nodes $a,b\in Q_0$; 
otherwise the quiver is called \textit{chiral}. 
A quiver that corresponds to a toric Calabi-Yau three-fold with (resp.\ without) compact $4$-cycles is chiral (resp.\ non-chiral).


\subsection{Trigonometric and elliptic versions of quiver Yangians}
	
In this section we summarize the definitions of the quiver Yangians and their trigonometric and elliptic generalizations in a unified manner.
Since all the algebras in this paper incorporate the shifts, we will not distinguish between the unshifted and the shifted version, and will omit the adjective ``shifted" from the names.

From now on we use the convention that variables written in the upper case (such as $Z$, $W$, $H$, $C$) are exponentiated versions of variables  written in the lower case (such as $z$, $w$, $h$, $c$): 
	\begin{equation}
	    Z=e^{\beta z}\,,\qquad W=e^{\beta w}\,,\qquad H=e^{\beta h}\,,\qquad 
	    C=e^{\beta c}\,,\qquad
	    \textrm{etc.} \;.
	\end{equation}
	The choice of $\beta$ will not play a significant role in our consideration and can be re-absorbed by a proper re-scaling of variables; we keep it to make a comparison to known algebras easier.
Throughout this paper, we use the following modified hyperbolic sine function and $q$-theta function:\footnote{
Recall that the ordinary $q$-theta function is defined as $\theta_q(z)\myequiv  (Z;q)_{\infty}  (qZ^{-1};q)_{\infty}$, where the infinite $q$-Pochhammer symbol $(X;q)_{\infty}\myequiv \prod^{\infty}_{n=0}(1-X q^n)$ for $|q|\leq 1$.}
	\be\label{notations}
	\begin{split}
	\textrm{Sin}_{\beta}(z)&\myequiv 2\sinh{\frac{\beta z}{2}}= Z^{\frac{1}{2}}-Z^{-\frac{1}{2}} \,,\\
	\Theta_q(z)&\myequiv- \frac{\theta_q(z)}{Z^{1/2}}= (Z^{\frac{1}{2}}-Z^{-\frac{1}{2}})\prod_{n=1}^{\infty} \left(1-Z^{-1} q^n\right) \left(1-Zq^{n}\right) \,.
	\end{split}
	\ee
Apart from the normalization factors  that do not play a role in the later analysis, we divide the canonical $q$-theta function by the factor of $Z^{1/2}$ in order to convert it into an \textit{odd} function of $z$ --- this property will be important when we derive the quiver BPS algebras.
	
The three types of algebras: the quiver Yangians, the trigonometric quiver Yangians, and the elliptic quiver Yangians, are generated by triplets of element $(\mye^{(a)}(z),\myf^{(a)}(z),\myk_\pm^{(a)}(z))$ and can then all be presented in a unified manner as
	\begin{tcolorbox}[ams equation]
		\begin{aligned}\label{eq-summary}
			\myk^{(a)}_{\epsilon}(z)\, \myk^{(b)}_{\epsilon}(w) &\simeq C^{\epsilon\,\extra\,\chi_{ab}}\,\myk^{(b)}_{\epsilon}(w)\, \myk^{(a)}_{\epsilon}(z) \qquad \epsilon = \pm \,, \\
			\myk^{(a)}_{+}(z)\, \myk^{(b)}_{-}(w) &\simeq \frac{ \myphi^{a\Leftarrow b}\left(z+\frac{c}{2},w-\frac{c}{2}\right) }{\myphi^{a\Leftarrow b}\left(z-\frac{c}{2},w+\frac{c}{2}\right)}\, \myk^{(b)}_{-}(w)\, \myk^{(a)}_{+}(z)\,,\\
			\myk^{(a)}_{\pm}(z)\mye^{(b)}(w) &\simeq \myphi^{a\Leftarrow b}\left(z\pm \frac{c}{2},w\right) \mye^{(b)}(w) \myk^{(a)}_{\pm}(z)\,,\\
			\myk^{(a)}_{\pm}(z)\myf^{(b)}(w) &\simeq \myphi^{a\Leftarrow b}\left(z\mp \frac{c}{2},w\right)^{-1} \myf^{(b)}(w) \myk^{(a)}_{\pm}(z)\,,\\
			\mye^{(a)}(z)\mye^{(b)}(w) &\simeq (-1)^{|a||b|}\,\myphi^{a\Leftarrow b}(z,w) \mye^{(b)}(w) \mye^{(a)}(z)\,,\\
			\myf^{(a)}(z)\myf^{(b)}(w) &\simeq (-1)^{|a||b|}\, \myphi^{a\Leftarrow b}(z,w)^{-1} \myf^{(b)}(w) \myf^{(a)}(z)\,,\\
			\left[\mye^{(a)}(z)\,,\myf^{(b)}(w)\right\}&\simeq -\delta_{a,b} \left(\myp(\Delta-c) \myk^{(a)}_{+}\left(z-\frac{c}{2}\right)- \myp(\Delta+c)  \myk^{(a)}_{-}\left(w-\frac{c}{2}\right) \right)\,,
		\end{aligned}
	\end{tcolorbox}
	\noindent where for the rational case, there is one set of $\psi$, namely $\psi_{+}=\psi_{-}$. These relations have a lot of additional ingredients, let us explain an application of  them all one by one.

Let us first explain the common features shared by the three classes of, namely elliptic/trigonometric/rational,  algebras in the definition \eqref{eq-summary}. 	
The $\mathbb{Z}_2$ grading (i.e.\ the Bose/Fermi statistics) of the $\mye^{(a)}$ and $\myf^{(a)}$ generators is denoted by $|a|$, and is defined as 
	\begin{equation}
		|a| =\left(|a\to a| + 1\right) \;{\rm mod}\; 2 \;,
	\end{equation}
	while the $\myk^{(a)}_{\pm}(z)$ generators are always even.
Namely, the $\mye^{(a)}$ and $\myf^{(a)}$ generators are bosonic (resp.\ fermionic) when $|a|=0$ (resp.\ $|a|=1$) while the $\myk^{(a)}_{\pm}(z)$ generators are always bosonic.
The notation $[\star,\star\}$ in \eqref{eq-summary} denotes the super-commutator, namely it is an anti-commutator for two fermions and a commutator otherwise.
	And throughout the paper, we define
	\begin{equation}
		\Delta\myequiv z-w\,.
	\end{equation}

The properties of the quiver BPS algebras are encoded in the so-called bond factor --- a function constructed from the quiver $Q$ according to the following rule
    \be
    \label{eq-charge-atob}
		\varphi^{a\Leftarrow b} (u)\myequiv (-1)^{|b\rightarrow a| \chi_{ab}} \frac{\prod_{I\in \{a\rightarrow b\}} \zeta (u+h_{I}) }{\prod_{J\in \{b\rightarrow a\}} \zeta(u-h_{J}) } \;,
    \ee
    \noindent where the form of the weight factor $\zeta(u)$ from each arrow  depends on whether we are considering the rational, trigonometric, or elliptic case.
    Expression \eqref{eq-charge-atob} is a direct generalization of the bond factor function (see, say, \cite[equation (2.9)]{Galakhov:2021xum}) from the rational case to the case of a generic odd function $\zeta(u)$.
    Also we have introduced a supplementary sign factor, so that the bond factor satisfies a reciprocity condition for a quiver of arbitrary chirality:
        \begin{equation}\label{eq-reciprocity-bare}
   \varphi^{a\Leftarrow b} (u) \, \varphi^{b\Leftarrow a} (-u)
   =1\,.
    \end{equation}
Clearly, we have a freedom in the definition \eqref{eq-charge-atob} to multiply the r.h.s.\ expression by a sign function $(-1)^{f_{ab}}$, subject to $f_{ab}=f_{ba}\,({\rm mod}\,2)$.
For non-chiral quivers associated with Calabi-Yau 3-folds without compact 4-cycles, the choice of the function $f_{ab}$ can be fixed by a comparison with a canonical definition of quantum rational/trigonometric/elliptic algebras associated with $\widehat{\fg\fl}_{m|n}$.
For algebras associated with chiral quivers, there is no other a priori definition, hence there is no preferred choice of the sign function $(-1)^{f_{ab}}$. We propose \eqref{eq-charge-atob} as a canonical definition of the algebras while keeping in mind that, in principle, we can simultaneously define the whole family of algebras that differ by a sign choice $(-1)^{f_{ab}}$ for each arrow between quiver nodes.

Unlike the case of quiver Yangians, the bond factor \eqref{eq-charge-atob} actually has to be slightly modified before we can use it to define the algebras and study their representations, especially for the general case of the chiral quivers. 
Therefore, we will call bond factor \eqref{eq-charge-atob} \emph{bare}.
As we will show later,   to deal with chiral quivers and to study general representations (for both non-chiral and chiral quivers),\footnote{As shown in \cite{Galakhov:2021xum}, 
the representations of a quiver algebra are related to the framing of the quivers, which in general results in the shift of the algebra. 
The framed quiver is in general chiral, even when the corresponding unframed quiver is non-chiral.
The balanced bond factor of the framed quiver enters the description of the representations of the algebra, see \eqref{eq-GS-FQ}.}
we need to introduce the \emph{balanced} bond factor:
    \begin{tcolorbox}[ams equation]\label{eq-def-balanced}
    \myphi^{a\Leftarrow b}(z,w)\myequiv \left(Z W\right)^{\frac{\extra}{2}\chi_{ab}}\varphi^{a\Leftarrow b}(z-w)\;,
    \end{tcolorbox}
    \noindent  where $\extra$ is defined as
    		\begin{equation}\label{eq-def-extra}
			\extra\myequiv\begin{cases}
				\begin{aligned}
					&0&\qquad &\textrm{rational}\\
					&       1&\qquad &\textrm{trig./elliptic}\,.
				\end{aligned}
			\end{cases}
		\end{equation}
    The prefactor $\left(Z W\right)^{\frac{\extra}{2}\chi_{ab}}$ in \eqref{eq-def-balanced} is called the ``balancing factor", its key role is to balance a Laurent expansion of \eqref{eq-charge-atob} since for chiral quivers it may go over half-integer powers of $Z$, and we would prefer to work with Laurent series in $Z$.
    For the rational case (i.e.\ the quiver Yangian) or when the quiver is non-chiral, the balancing factor equals $1$ and the balanced bond factor reduces to the bare one.
    The balanced bond factor \eqref{eq-def-balanced} satisfies the reciprocity condition 
    \begin{equation}\label{eq-reciprocity-final}
   \myphi^{a\Leftarrow b} (z,w)\,\myphi^{b\Leftarrow a} (w,z)=1\,,
    \end{equation}
where we have used the reciprocity for the bare bond factor \eqref{eq-reciprocity-bare} and that $\chi_{ab}+\chi_{ba}=0$.
	
The difference among the three classes of algebras are in  (1) the weight factor $\zeta(u)$ in the definition of the balanced bond factor $\myphi^{a\Leftarrow b}(z,w)$, (2) the value of the central term $c$, (3) the mode expansions of the generators $(\mye^{(a)}(z), \myk^{(a)}_{\pm}(z),\myf^{(a)}(z))$, (4) the meaning of $\simeq$ in \eqref{eq-summary} and (5) the definition of the propagator $\myp(z)$.
	Let us now explain them in turn. 
	\begin{enumerate}[leftmargin=5mm]
		\item
		In the bare bond factor \eqref{eq-charge-atob} and hence in the balanced bond factor \eqref{eq-def-balanced}, the weight factor $\zeta(u)$ for each arrow is 
		\begin{equation}\label{eq-def-zeta}
			\zeta(u)\myequiv\begin{cases}
				\begin{aligned}
					&u &\qquad &\textrm{rational}\\
					&\textrm{Sin}_{\beta}{(u)} &\qquad &\textrm{trig.}\\
					&\Theta_q(u) &\qquad &\textrm{elliptic}\,,
				\end{aligned}
			\end{cases} 
		\end{equation}
where the functions $\textrm{Sin}_{\beta}(u)$ and $\Theta_q(u)$ are defined in \eqref{notations}.		
		
		\item The central term reads:
		\begin{equation}
			c=\begin{cases}
				\begin{aligned}
					&0&\qquad &\textrm{rational}\\
					&c&\qquad &\textrm{trig./elliptic}\,.
				\end{aligned}
			\end{cases}
		\end{equation}
		\item
		The three cases differ in the mode expansions for their generators. 
	For the $\mye^{(a)}(z)$ and $ \myf^{(a)}(z)$ generators, for both the non-chiral and chiral quivers, we have
	\begin{equation}\label{modes}
			\mye^{(a)}(z)=\begin{cases}
				\begin{aligned}
					&\sum_{n\in \mathbb{N}} \frac{\mye^{(a)}_n}{z^{n}}\\
					&\sum_{n\in \mathbb{Z}} \frac{\mye^{(a)}_n}{Z^{n}}\\
					&\sum_{n\in \mathbb{Z}}\sum_{\alpha\in \mathbb{Z}_{\geq 0}} \frac{\mye^{(a)}_{n,\alpha}}{Z^{n}}q^{\alpha}
				\end{aligned}
			\end{cases}\textrm{and}\quad \myf^{(a)}(z)=\begin{cases}
				\begin{aligned}
					&\sum_{n\in \mathbb{N}} \frac{\myf^{(a)}_n}{z^{n}} & &\textrm{rational}\\
					&\sum_{n\in \mathbb{Z}} \frac{\myf^{(a)}_n}{Z^{n}} & &\textrm{trig.}\\
					&\sum_{n\in \mathbb{Z}}\sum_{\alpha\in \mathbb{Z}_{\geq 0}} \frac{\myf^{(a)}_{n,\alpha}}{Z^{n}}q^{\alpha}& &\textrm{elliptic}
				\end{aligned}
			\end{cases}
		\end{equation}
		For the $\myk_{\pm}$ generators in the case of non-chiral quivers, 
		\begin{equation}
			\myk^{(a)}_{\pm}(z)=\begin{cases}
				\begin{aligned}
					&\sum_{n\in \mathbb{Z}_{\geq 0}} \myk^{(a)}_nz^{-\left(n+\mys^{(a)}\right)} &\qquad &\textrm{rational}\\
					&\sum_{n\in \mathbb{Z}_{\geq 0}} \myk^{(a)}_{\pm,n}Z^{\mp \left(n+\mys^{(a)}\right)} &\qquad &\textrm{trig.}\\
					&\sum_{n\in \mathbb{Z}_{\geq 0}} \myk^{(a)}_{
							\pm, n, 0}Z^{\mp \left(n+\mys^{(a)}\right)}+\sum\lm_{n\in\IZ}\sum\lm_{\alpha\in\IN}\myk^{(a)}_{
							\pm, n,\alpha}Z^{\mp n}q^{\alpha}&\qquad &\textrm{elliptic}
				\end{aligned}
			\end{cases}
		\end{equation}
	where we have used the fact that in the rational case,   $\myk=\myk_{\pm}$.
Note that compared with the known algebras for some of the non-chiral quivers in the literature, we have introduces the additional integer $\mys^{(a)}$ (called the ``shift") since it will be important when studying the generic, in particular non-vacuum, representations of the algebras, see \cite{Galakhov:2021xum} for the case of shifted quiver Yangians.  For the $\myk_{\pm}$ generators in the case of chiral quivers, 
		\begin{equation}
			\myk^{(a)}_{\pm}(z)=\begin{cases}
				\begin{aligned}
					&\sum_{n\in \mathbb{Z}} \myk^{(a)}_nz^{-\left(n+\mys^{(a)}\right)} &\qquad &\textrm{rational}\\
					&\sum_{n\in \mathbb{Z}} \myk^{(a)}_{\pm,n}Z^{\mp \left(n+\mys^{(a)}\right)} &\qquad &\textrm{trig.}\\
					&\sum_{n\in \mathbb{Z}} \myk^{(a)}_{
							\pm, n, 0}Z^{\mp \left(n+\mys^{(a)}\right)}+\sum\lm_{n\in\IZ}\sum\lm_{\alpha\in\IN}\myk^{(a)}_{
							\pm, n,\alpha}Z^{\mp n}q^{\alpha}&\qquad &\textrm{elliptic}
				\end{aligned}
			\end{cases}
		\end{equation}
Note that for the chiral quivers, when we consider infinite-dimensional representations, the information of the shift $\{\mys^{(a)}\}$ would be lost in the expansion of the $\psi^{(a)}_{\pm}(z)$ generators and only visible in the representation; whereas for finite-dimensional representations, the shift $\{\mys^{(a)}\}$ would be visible both in the representations and in the mode expansions of $\psi^{(a)}_{\pm}(z)$.

    \item The meaning of the ``$\simeq$" signs are also different in the three classes.     
    In the case of the rational algebra relations are among Taylor series expansions around $z=\infty$, therefore the expansions of the l.h.s.\ and the r.h.s.\ are equivalent up to terms $z^nw^{m\geq 0}$ and $z^{n\geq 0}w^m$. In the trigonometric and elliptic cases we compare Laurent series in $Z$ and $W$, namely, for a relation  $L_1\simeq (P/Q)L_2$, where $L_i$ are two Laurent series and $P/Q$ is a  ratio of two hyperbolic sines or theta-functions, the ``$\simeq$" means  that the Laurent expansions of $QL_1$ and $PL_2$ are equivalent.
	
	\item Finally, the propagator $\myp(z)$ is  a formal delta function operator, defined differently for the three cases:
		\begin{equation}\label{eq-def-p}
			\myp(z)\myequiv\begin{cases}
				\begin{aligned}
					&\frac{1}{z} &\qquad &\textrm{rational}\\
		        	&\delta(Z)\myequiv\sum\lm_{n\in\IZ}Z^{n}&\qquad &\textrm{trig.}/\textrm{elliptic}
				\end{aligned}
			\end{cases}
		\end{equation}
where the $\delta$-function is a ``formal" delta function. For details on the properties of the formal delta function $\delta(Z)$  see Appendix~\ref{app:Cyl}.

    \end{enumerate}
    
As explained in the introduction, the form of the algebras in \eqref{eq-summary} comes from the natural trigonometric and elliptic generalizations of the quiver Yangians of \cite{Li:2020rij}, i.e.\ as the uplift $u\rightarrow \textrm{Sin}_{\beta}(u)\rightarrow \Theta_{q}(u)$, which will be confirmed later from the gauge theory computation, see section \ref{sec:D-brane}.
However, the trigonometric and elliptic versions have important features that are absent in the rational case. In the algebra \eqref{eq-summary} these features manifest as the appearance of $c$ and $\chi_{ab}$, as well as the fact that it is the balanced bond factor \eqref{eq-def-balanced} instead of the bare one \eqref{eq-charge-atob} that enters the definition of the algebra \eqref{eq-summary}.
Note that some of the new ingredients, such as the factor $C^{\epsilon \,\extra\, \chi_{ab}}$ in the first equation of the \eqref{eq-summary}, have been missing in most of the existing literature; this is because for the trigonometric and elliptic versions it is often the case that only some of the non-chiral quivers have been considered in the literature. 

We perform four types of consistency checks for the algebra defined in \eqref{eq-summary}.
The most straightforward one is to explicitly check the mutual consistency of various equations of \eqref{eq-summary}, in the spirit of Jacobi identities. 
For example, if we assume the last three equations, i.e.\ the $e$-$e$, $f$-$f$, and $e$-$f$ relations,  in \eqref{eq-summary}, we can derive all the remaining ones.
(We leave this computation to Appendix~\ref{app:consistency}.)

The second consistency check is that when $c=0$, we can construct crystal-melting representations, see section \ref{sec:rep}.

Thirdly, as for the three equations that we assume for mutual consistency checks, note that since the $e$-$e$ and  $f$-$f$ relations are both independent of $c$ --- and the consistency check for the $c=0$ case is taken care of by the construction of the crystal representations --- essentially we have only one important relation, namely the $e$-$f$ relation, to justify, which we do by comparing with the known algebras, see section \ref{subsec:comparison}.
    
Lastly, one can try to realize the algebra \eqref{eq-summary} explicitly using free fields.
In applications the so-called free field representation of the discussed algebras might become useful.
As far as we are aware, a complete free field realization is known only for algebras associated with the root system of $\widehat{\fg\fl}_n$, whereas for the subalgebra generated by $\myk_\pm^{(a)}(z)$ it can be constructed for general quivers.
We discuss a free field realization for the subalgebras generated by $\myk_\pm^{(a)}(z)$ of our algebras in Appendix~\ref{app:free_field}.

\subsection{Comparison to known algebras}\label{subsec:comparison}

While our elliptic/trigonometric algebras are in general new,
we can compare our results against existing definitions in the literature,
when the toric Calabi-Yau three-fold has no compact four-cycles.
Let us verify that this is indeed the case, by extending the similar analysis for the quiver Yangians in \cite{Li:2020rij}.

Consider $\widehat{\fg\fl}_{m|n}$. In order to specify the algebra we need to choose a 
Dynkin diagram. This is determined by a signature (cf.\ \cite{Nagao:2009rq,2019arXiv191208729B}):
\be
\Sigma_{m,n}:\quad \{1,2,\ldots,m+n \}\longrightarrow \{+1,-1\},\;\mbox{so that}\;\#(+1)=m,\;\#(-1)=n \;,
\ee
where in the following we consider indices modulo $m+n$. 

\definecolor{palette1}{rgb}{0.603922, 0.466667, 0.811765}
\definecolor{palette2}{rgb}{0.329412, 0.219608, 0.517647}
\definecolor{palette3}{rgb}{0.0156863, 0.282353, 0.333333}
\definecolor{palette4}{rgb}{0.631373, 0.211765, 0.439216}
\definecolor{palette5}{rgb}{0.92549, 0.254902, 0.462745}
\definecolor{palette6}{rgb}{1., 0.643137, 0.368627}
\definecolor{palette7}{rgb}{0.313725, 0.45098, 0.85098}

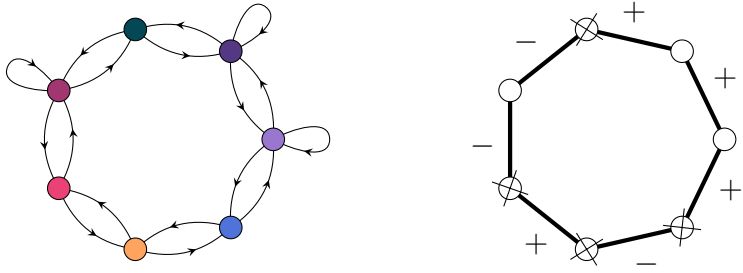
\begin{figure}[!ht]
\begin{center}
\begin{tikzpicture}[scale=1.5]
    \begin{scope}[shift={(0,0)}]
    \foreach \r in {0., 51.4286, 102.857, 154.286, 205.714, 257.143, 308.571}
    {
        \begin{scope}[rotate = \r]
            \draw[postaction={decorate},decoration={markings, mark= at position 0.6 with {\arrow{stealth}}}] (1,0) to[out=85.714, in=325.714] (0.62349, 0.781831);
            \draw[postaction={decorate},decoration={markings, mark= at position 0.4 with {\arrowreversed{stealth}}}] (1,0) to[out=145.714, in=265.714] (0.62349, 0.781831);
        \end{scope}
    }
    \foreach \r in {0., 51.4286, 154.286}
    {
        \begin{scope}[rotate = \r]
            \draw[postaction={decorate},decoration={markings, mark= at position 0.75 with {\arrow{stealth}}}] (1,0) to[out=30,in=90] (1.5,0) to[out=270,in=330] (1,0);
        \end{scope}
    }
    \foreach \r/\s in {0./palette1, 51.4286/palette2, 102.857/palette3, 154.286/palette4, 205.714/palette5, 257.143/palette6, 308.571/palette7}
    {
        \begin{scope}[rotate = \r]
            \draw[fill=\s] (1,0) circle (0.1);
        \end{scope}
    }
    \end{scope}
    \begin{scope}[shift={(4,0)}]
    \foreach \r/\s in {0./$+$, 51.4286/$+$, 102.857/$-$, 154.286/$-$, 205.714/$+$, 257.143/$-$, 308.571/$+$}
    {
        \begin{scope}[rotate = \r]
            \draw[ultra thick] (1,0) -- (0.62349, 0.781831); 
            \node at (1.0072, 0.546788) {\s};
        \end{scope}
    }
    \foreach \r in {0., 51.4286, 102.857, 154.286, 205.714, 257.143, 308.571}
    {
        \begin{scope}[rotate = \r]
            \draw[fill=white] (1,0) circle (0.1);
        \end{scope}
    }
    \foreach \r/\s in {102.857, 205.714, 257.143, 308.571}
    {
        \begin{scope}[rotate = \r]
            \begin{scope}[shift={(1,0)}]
                \draw (-0.12,-0.12) -- (0.12,0.12) (-0.12,0.12) -- (0.12,-0.12);
            \end{scope}
        \end{scope}
    }
    \end{scope}
\end{tikzpicture}
\caption{An example of a quiver and a Dynkin diagram associated with signature $\Sigma=\{+1,+1,+1,-1,-1,+1,-1\}$.}\label{fig:quiver}
\end{center}
\end{figure}

Then the quiver and the superpotential are given by (see e.g. Figure \ref{fig:quiver})
\be\renewcommand{\arraystretch}{1.3}
\begin{array}{c|c|c}
        \mbox{spin arrangement} & \sigma_i\sigma_{i+1}=1& \sigma_i\sigma_{i+1}=-1\\
        \hline
        \IZ_2\mbox{-parity} &\mbox{Even} & \mbox{Odd} \\
        \hline
        \mbox{Dynkin node} & \begin{array}{c}
                \begin{tikzpicture}
                        \draw[ultra thick] (-0.5,0) -- (0.5,0);
                        \draw[fill=white] (0,0) circle (0.15);
                \end{tikzpicture}
        \end{array}& \begin{array}{c}
                \begin{tikzpicture}
                        \draw[ultra thick] (-0.5,0) -- (0.5,0);
                        \draw[fill=white] (0,0) circle (0.15);
                        \draw (-0.2,-0.2) -- (0.2,0.2) (0.2,-0.2) -- (-0.2,0.2);
                \end{tikzpicture}
        \end{array}\\
        \hline
        \mbox{quiver node} & \begin{array}{c}
                \begin{tikzpicture}
                        \draw (0,0) circle (0.15);
                        \draw[-stealth] (-1,0.05) -- (-0.141421,0.05) node[pos=0.2,above] {$A_i$};
                        \draw[stealth-] (-1,-0.05) -- (-0.141421,-0.05) node[pos=0.2,below] {$B_i$};
                        \draw[stealth-] (1,0.05) -- (0.141421,0.05) node[pos=0.2,above] {$A_{i+1}$};
                        \draw[-stealth] (1,-0.05) -- (0.141421,-0.05) node[pos=0.2,below] {$B_{i+1}$};
                        \draw[postaction={decorate},decoration={markings, 
                                mark= at position 0.85 with {\arrow{stealth}}}] (-0.05,0.141421) to[out=120,in=180] node[pos=0.7,above left] {$C_i$} (0,0.7) to[out=0,in=60] (0.05,0.141421);
                        \draw[white] (-0.1,1.1) -- (0.1,1.1);
                \end{tikzpicture}
        \end{array}& \begin{array}{c}
                \begin{tikzpicture}
                        \draw (0,0) circle (0.15);
                        \draw[-stealth] (-1,0.05) -- (-0.141421,0.05) node[pos=0.2,above] {$A_i$};
                        \draw[stealth-] (-1,-0.05) -- (-0.141421,-0.05) node[pos=0.2,below] {$B_i$};
                        \draw[stealth-] (1,0.05) -- (0.141421,0.05) node[pos=0.2,above] {$A_{i+1}$};
                        \draw[-stealth] (1,-0.05) -- (0.141421,-0.05) node[pos=0.2,below] {$B_{i+1}$};
                        \draw[white] (-0.1,1.1) -- (0.1,1.1);
                \end{tikzpicture}
        \end{array}\\
        \hline
        \mbox{superpotential }\delta W_i & \sigma_i\,\Tr\, C_i\left(B_{i+1}A_{i+1}-A_{i}B_{i}\right)& -\sigma_i\,\Tr\,B_{i+1}A_{i+1}A_iB_i\\
\end{array}
\ee
Let us define the Cartan matrix and the auxiliary matrix following \cite{2019arXiv191208729B}:
\be
\begin{split}
        A_{i,j}^{\Sigma}&=\left(\sigma_i+\sigma_{i+1}\right)\delta_{i,j}-\sigma_i\delta_{i,j+1}-\sigma_j\delta_{i+1,j}\;,\\
        M_{i,j}^{\Sigma}&=\sigma_i\delta_{i,j+1}-\sigma_j\delta_{i+1,j}\;.
\end{split}
\ee
We can choose the following equivariant weights for the arrows:
\be
\begin{split}
        h(A_i)=\sigma_i\left(-\myh_1-\myh_2\right)\;,\quad h(B_i)=\sigma_i\left(-\myh_1+\myh_2\right)\;,\quad h(C_i)=2\sigma_i\myh_1\;.
\end{split}
\ee
The resulting relations for $\myT_{\beta}(Q_{m|n})$ can then be transformed into:
\be\label{toroidal_gl_m_n}
\scalebox{0.9}{$\bgroup\everymath{\displaystyle}\renewcommand{\arraystretch}{1.5}\setlength{\tabcolsep}{0pt}
\begin{array}{rcl}
\multicolumn{3}{c}{\myk^{(i)}_{\epsilon}(z)\, \myk^{(j)}_{\epsilon}(w) =\myk^{(j)}_{\epsilon}(w)\, \myk^{(i)}_{\epsilon}(z) \qquad \epsilon = \pm, }\\
\frac{d^{M_{i,j}^{\Sigma}}p^{A_{i,j}^{\Sigma}}Z-CW}{d^{M_{i,j}^{\Sigma}}Z-p^{A_{i,j}^{\Sigma}}CW}\myk^{(i)}_{\epsilon}(z)\, \myk^{(j)}_{\epsilon}(w) &=&\frac{d^{M_{i,j}^{\Sigma}}p^{A_{i,j}^{\Sigma}}CZ- W}{d^{M_{i,j}^{\Sigma}}C Z-p^{A_{i,j}^{\Sigma}}W}\myk^{(j)}_{-}(w)\, \myk^{(i)}_{+}(z), \\
\left(d^{M_{i,j}^{\Sigma}}C^{\pm\frac{1}{2}}Z-p^{A_{i,j}^{\Sigma}}W\right)\, \myk_{\pm}^{(i)}(z)\mye^{(j)}(w)&=&\left(d^{M_{i,j}^{\Sigma}}p^{A_{i,j}^{\Sigma}}C^{\pm\frac{1}{2}}Z-W\right)\, \mye^{(j)}(w)\myk_{\pm}^{(i)}(z),\\
\left(d^{M_{i,j}^{\Sigma}}Z-p^{A_{i,j}^{\Sigma}}W\right)\, \mye^{(i)}(z)\mye^{(j)}(w)&=&(-1)^{|i||j|}\left(d^{M_{i,j}^{\Sigma}}p^{A_{i,j}^{\Sigma}}Z-W\right) \, \mye^{(j)}(w)\mye^{(i)}(z),\\
\left(d^{M_{i,j}^{\Sigma}}C^{\mp\frac{1}{2}}Z-p^{-A_{i,j}^{\Sigma}}W\right)\,  \myk_\pm^{(i)}(z)\myf^{(j)}(w)&=&\left(d^{M_{i,j}^{\Sigma}}p^{-A_{i,j}^{\Sigma}}C^{\mp\frac{1}{2}}Z-W\right) \, \myf^{(j)}(w)\myk_{\pm}^{(i)}(z),\\
\left(d^{M_{i,j}^{\Sigma}}Z-p^{-A_{i,j}^{\Sigma}}W\right)\, \myf^{(i)}(z)\,\myf^{(j)}(w)&=&(-1)^{|i||j|}\left(d^{M_{i,j}^{\Sigma}}p^{-A_{i,j}^{\Sigma}}Z-W\right) \, \myf^{(j)}(w)\,\myf^{(i)}(z)\;,\\
\multicolumn{3}{c}{\left[\mye^{(a)}(z)\,,\myf^{(b)}(w)\right\}= -\delta_{i,j} \left(\myp(\Delta-c)\, \myk^{(a)}_{+}\left(z-\dfrac{c}{2}\right)- \myp(\Delta+c) \, \myk^{(a)}_{-}\left(w-\dfrac{c}{2}\right) \right)\,.}
\end{array}
\egroup$}
\ee
where
\be
p=e^{\beta \myh_1},\quad d=e^{\beta \myh_2}.
\ee
On the other hand, the resulting relations for $\myE_{\tau}(Q_{m|n})$ can be transformed into:
\be\label{elliptic_gl_m_n}
\scalebox{0.8}{$\bgroup\everymath{\displaystyle}\renewcommand{\arraystretch}{1.7}\setlength{\tabcolsep}{0pt}
\begin{array}{rcl}
\multicolumn{3}{c}{\myk^{(i)}_{\epsilon}(z)\, \myk^{(j)}_{\epsilon}(w) =\myk^{(j)}_{\epsilon}(w)\, \myk^{(i)}_{\epsilon}(z) \qquad \epsilon = \pm, }\\
\frac{\Theta_q\left(z-w+c-A_{i,j}^{\Sigma}\mathsf{h}_1+M_{i,j}^{\Sigma}\mathsf{h}_2\right)}{\Theta_q\left(z-w-c-A_{i,j}^{\Sigma}\mathsf{h}_1+M_{i,j}^{\Sigma}\mathsf{h}_2\right)}\psi_+^{(i)}(z)\psi_-^{(j)}(w)&=&\frac{\Theta_q\left(z-w+c+A_{i,j}^{\Sigma}\mathsf{h}_1+M_{i,j}^{\Sigma}\mathsf{h}_2\right)}{\Theta_q\left(z-w-c+A_{i,j}^{\Sigma}\mathsf{h}_1+M_{i,j}^{\Sigma}\mathsf{h}_2\right)}\psi_-^{(j)}(w)\psi_+^{(i)}(z)\;,\\
d^{M_{i,j}^{\Sigma}}C^{\pm\frac{1}{2}}Z\frac{\left(p^{A_{i,j}^{\Sigma}}d^{-M_{i,j}^{\Sigma}}C^{\mp\frac{1}{2}}W/Z|q\right)_{\infty}}{\left(qp^{-A_{i,j}^{\Sigma}}d^{-M_{i,j}^{\Sigma}}C^{\mp\frac{1}{2}}W/Z|q\right)_{\infty}}\myk_{\pm}^{(i)}(z)\,\mye^{(j)}(w)&=&-W\frac{\left(p^{A_{i,j}^{\Sigma}}d^{M_{i,j}^{\Sigma}}C^{\pm\frac{1}{2}}Z/W|q\right)_{\infty}}{\left(qp^{-A_{i,j}^{\Sigma}}d^{M_{i,j}^{\Sigma}}C^{\pm\frac{1}{2}}Z/W|q\right)_{\infty}} \mye^{(j)}(w)\,\myk_{\pm}^{(i)}(z)\;,\\
d^{M_{i,j}^{\Sigma}}Z\frac{\left(p^{A_{i,j}^{\Sigma}}d^{-M_{i,j}^{\Sigma}}W/Z|q\right)_{\infty}}{\left(qp^{-A_{i,j}^{\Sigma}}d^{-M_{i,j}^{\Sigma}}W/Z|q\right)_{\infty}}\mye^{(i)}(z)\,\mye^{(j)}(w)&=&-(-1)^{|i||j|}W\frac{\left(p^{A_{i,j}^{\Sigma}}d^{M_{i,j}^{\Sigma}}Z/W|q\right)_{\infty}}{\left(qp^{-A_{i,j}^{\Sigma}}d^{M_{i,j}^{\Sigma}}Z/W|q\right)_{\infty}}  \mye^{(j)}(w)\,\mye^{(i)}(z)\;,\\
d^{M_{i,j}^{\Sigma}}C^{\mp\frac{1}{2}}Z\frac{\left(p^{-A_{i,j}^{\Sigma}}d^{-M_{i,j}^{\Sigma}}C^{\pm\frac{1}{2}}W/Z|q\right)_{\infty}}{\left(qp^{A_{i,j}^{\Sigma}}d^{-M_{i,j}^{\Sigma}}C^{\pm\frac{1}{2}}W/Z|q\right)_{\infty}} \myk_\pm^{(i)}(z)\,\myf^{(j)}(w)&=&-W\frac{\left(p^{-A_{i,j}^{\Sigma}}d^{M_{i,j}^{\Sigma}}C^{\mp\frac{1}{2}}Z/W|q\right)_{\infty}}{\left(qp^{A_{i,j}^{\Sigma}}d^{M_{i,j}^{\Sigma}}C^{\mp\frac{1}{2}}Z/W|q\right)_{\infty}}  \myf^{(j)}(w)\,\myk_{\pm}^{(i)}(z)\;,\\
d^{M_{i,j}^{\Sigma}}Z\frac{\left(p^{-A_{i,j}^{\Sigma}}d^{-M_{i,j}^{\Sigma}}W/Z|q\right)_{\infty}}{\left(qp^{A_{i,j}^{\Sigma}}d^{-M_{i,j}^{\Sigma}}W/Z|q\right)_{\infty}}\myf^{(i)}(z)\,\myf^{(j)}(w)&=&-(-1)^{|i||j|}W\frac{\left(p^{-A_{i,j}^{\Sigma}}d^{M_{i,j}^{\Sigma}}Z/W|q\right)_{\infty}}{\left(qp^{A_{i,j}^{\Sigma}}d^{M_{i,j}^{\Sigma}}Z/W|q\right)_{\infty}}  \myf^{(j)}(w)\,\myf^{(i)}(z)\;,\\
\multicolumn{3}{c}{\left[\mye^{(a)}(z)\,,\myf^{(b)}(w)\right\}= -\delta_{i,j} \left(\myp(\Delta-c)\, \myk^{(a)}_{+}\left(z-\dfrac{c}{2}\right)- \myp(\Delta+c) \, \myk^{(a)}_{-}\left(w-\dfrac{c}{2}\right) \right)\,.}
\end{array}
\egroup$}
\ee
These relations can now be compared with (3.55) -- (3.62) in \cite{Konno:2016fmh}.

\subsection{Serre relations}\label{s:Serre}

The relations given in \eqref{eq-summary} can in general be supplemented by further relations, often called the Serre relations,
without spoiling the consistency of the algebras. When (a maximal set of) such relations are imposed we obtain the 
\emph{reduced} elliptic quiver algebra $\underline{\myE}_{\tau}(Q,W)$:
\begin{align}
\underline{\myE}_{\tau}(Q,W) = \myE_{\tau}(Q,W) / \{\textrm{(Serre relations)} \} \;.
\end{align}
We can call the original algebra $\myE_{\tau}(Q, W)$ the \emph{unreduced} elliptic quiver algebra
when we want to emphasize that the Serre relations are not imposed.
We can similarly define the reduced quantum-toroidal quiver algebra $\underline{\myT}_{\beta}(Q, W)$
by imposing Serre relations on $\myT_{\beta}(Q, W)$.

Serre relations are established naturally for the quivers corresponding to $\widehat{\fg\fl}_{m|n}$ discussed in section \ref{subsec:comparison},
see \cite{2019arXiv191208729B} for the quantum-toroidal cases,
and \cite{Konno:2016fmh} for the elliptic case with $\widehat{\mathfrak{gl}}_n$.
For algebras associated with chiral quivers Serre relations are unknown in general since there is no apparent bridge to a Lie superalgebra counterpart.
Nevertheless we present our proposal for relations that could play a role of the Serre relations in the cases of Calabi-Yau three-folds with compact four-cycles in Appendix~\ref{app:Serre}.
To describe this, let us 
define the root pairing by
\be
\langle \alpha_i|\alpha_j\rangle=A_{i,j}^{\Sigma}\;,
\ee
and extend it over the weights by linearity.
We define the weight operators following \cite{2019arXiv191208729B}:
\be
t_it_j=t_jt_i,\quad t_i\, \mye^{(j)}(z) \,t_i^{-1}=q^{\langle\alpha_i|\alpha_j\rangle} \mye^{(j)}(z)\;,
\ee
so that we can define a weight $w(X)$ of the operator $X$ through a relation:
\be
t_i\,X\,t_i^{-1}=q^{\langle \alpha_i|w(X)\rangle}X\;.
\ee
Finally we introduce a graded $q$-commutator:
\be
\lbr X,Y\rbr=XY-(-1)^{|X||Y|}q^{-\langle X|Y\rangle}Y X\;.
\ee

\begin{subequations}
In terms of these we can write down the following Serre relations for $\myT_{\beta}(Q_{m|n})$:
        \be\label{Serre_a}
        {\rm Sym}_{z_1,z_2}\left\lbr \mye^{(i)}(z_1),\left\lbr \mye^{(i)}(z_2),\mye^{(i\pm 1)} (w)\right\rbr\right\rbr=0,\quad A_{i,i}^{\Sigma}\neq 0\;.
        \ee
        If $mn\neq 2$, $m+n>2$:
        \be
        {\rm Sym}_{z_1,z_2}\left\lbr \mye^{(i)}(z_1),\left\lbr \mye^{(i+1)}(w_1),\left\lbr \mye^{(i)}(z_2),\mye^{(i-1)}(w_2)\right\rbr\right\rbr\right\rbr=0,\quad A_{i,i}^{\Sigma}=0\;.
        \ee
        If $mn=2$:
        \be
        \begin{split}
              {\rm Sym}_{z_1,z_2}{\rm Sym}_{w_1,w_2}&\left\lbr\mye^{(i-1)}(z_1),\left\lbr\mye^{(i+1)}(w_1),\left\lbr\mye^{(i-1)}(z_2),\left\lbr\mye^{(i+1)}(w_2),\mye^{(i)}(y)\right\rbr\right\rbr\right\rbr\right\rbr=\\
              &\qquad \qquad  =\left\{z_1\leftrightarrow w_1, z_2\leftrightarrow w_2\right\},\quad A_{i,i}^{\Sigma}\neq 0\;.
        \end{split}
        \ee
        If $m|n=1|1$:
        \be
        {\rm Sym}_{z_1,z_2}{\rm Sym}_{w_1,w_2}\left\lbr \mye^{(1)}(z_1),\left\lbr \mye^{(2)}(w_1),\left\lbr \mye^{(1)}(z_2),\mye^{(2)}(w_2)\right\rbr\right\rbr\right\rbr=0\;.
        \ee
\end{subequations}
In these expressions ${\rm Sym}_{z_1,z_2}$ denotes the symmetrization of the expression with respect to the exchange of $z_1$ and $z_2$. 

Here we should mention that our derivation of these relations works for an analytic continuation of \eqref{eq-summary} when we treat the l.h.s.\ and the r.h.s.\ as functions rather than Laurent series, and we use the bond factors simply as ratios of functions $\zeta(u)$ that are defined in \eqref{eq-def-zeta}.
In practice this implies that in addition to this derivation one needs to check Serre relations for the first few generators in the mode expansion \eqref{modes}.

To derive the Serre relations for $\myE_{\tau}(Q_{m|n})$ let us introduce free field operators $\chi_i(z)$ commuting with $\myE_{\tau}(Q_{m|n})$ and having the following correlators:
\be
e^{\langle\chi_i(z_1)\chi_j(z_2)\rangle}=\frac{\left(q p^{A_{i,j}^{\Sigma}}d^{-M_{i,j}^{\Sigma}}Z_2/Z_1|q\right)_{\infty}}{\left(q p^{-A_{i,j}^{\Sigma}}d^{-M_{i,j}^{\Sigma}}Z_2/Z_1|q\right)_{\infty}} \;.
\ee
Using the vertex operators of $\chi_i(z)$ we can define new algebra generators analogous to ``dressed" generators in \cite{konno2009elliptic}: 
\be
E^{(i)}(z)\myequiv e^{\chi_i(z)}\mye^{(i)}(z)  \;.
\ee
Then if the correlators are applied to the free fields, the new generators behave as those of $\myT_{\beta}(Q_{m|n})$:
\be
\begin{split}
        &\left(d^{M_{i,j}^{\Sigma}}Z-p^{A_{i,j}^{\Sigma}}W\right)\left\langle E^{(i)}(z_1)E^{(j)}(z_2)\right\rangle=\\
        &=(-1)^{|i||j|}\left(d^{M_{i,j}^{\Sigma}}p^{A_{i,j}^{\Sigma}}Z-W\right) \left\langle E^{(j)}(z_2)E^{(i)}(z_1)\right\rangle  \;.
\end{split}
\ee

We can derive the corresponding Serre relations for the elliptic case by simply starting with a relation for $\myT_{\beta}(Q_{m|n})$, substituting the generators $\mye^{(i)}(z)$ with $E^{(i)}(z)$ and taking correlators $\langle *\rangle$ of the whole expressions.
The correlators will produce the necessary Pochhammer factors in the expressions.
For instance, let us consider the simplest relation \eqref{Serre_a}.
After following the procedure described above, one obtains the following Serre relation in $\myE_{\tau}(Q_{m|n})$:
\be
\begin{split}
        \frac{\left(q p^2 Z_2/Z_1|q\right)_{\infty}}{\left(q p^{-2} Z_2/Z_1|q\right)_{\infty}}\Bigg(\frac{\left(q p^{-1} Z_1/W|q\right)_{\infty}}{\left(q p Z_1/W|q\right)_{\infty}}\frac{\left(q p^{-1} Z_2/W|q\right)_{\infty}}{\left(q p Z_2/W|q\right)_{\infty}}\mye^{(i\pm 1)}(w)\mye^{(i)}(z_1)\mye^{(i)}(z_2)-\\
        -\left(p+p^{-1}\right)\frac{\left(q p^{-1} W/Z_1|q\right)_{\infty}}{\left(q p W/Z_1|q\right)_{\infty}}\frac{\left(q p^{-1} Z_2/W|q\right)_{\infty}}{\left(q p Z_2/W|q\right)_{\infty}}\mye^{(i)}(z_1)\mye^{(i\pm 1)}(w)\mye^{(i)}(z_2)+\\
        +\frac{\left(q p^{-1} W/Z_1|q\right)_{\infty}}{\left(q p W/Z_1|q\right)_{\infty}}\frac{\left(q p^{-1} W/Z_2|q\right)_{\infty}}{\left(q p W/Z_2|q\right)_{\infty}}\mye^{(i)}(z_1)\mye^{(i)}(z_2)\mye^{(i\pm 1)}(w)
        \Bigg)+\{z_1\leftrightarrow z_2\}=0  \;.
\end{split}
\ee
We can compare this relation to (3.24) in \cite{Konno:2016fmh} when $n=0$ (or $m=0$).

\section{Representations from crystals} \label{sec:rep}

In this section we construct representations of the quiver BPS algebras in terms of the statistical-mechanical model of crystal melting~\cite{Okounkov:2003sp,Iqbal:2003ds,Szendroi,MR2836398,MR2592501,Ooguri:2009ijd,Yamazaki:2010fz}, generalizing the 
discussion for quiver Yangians in \cite{Li:2020rij}.

For the quiver Yangians, the $\myk^{(a)}(z)$ are Cartan generators, and the crystal configurations are eigenstates of $\myk^{(a)}(z)$.
Based on a certain ansatz, one can bootstrap the entire algebra using these crystal representations, see \cite{Li:2020rij}  for the unshifted quiver Yangian and \cite{Galakhov:2021xum} for the shifted quiver Yangian.

By contrast, for the trigonometric and elliptic quiver Yangians, the $\myk^{(a)}_{+}$ and $\myk^{(b)}_{-}$ generators do not commute for general $c$. 
As a consequence, the crystals are {\it not} representations of the trigonometric and elliptic quiver Yangians, for general $c$.
However, when $c=0$, there are enough degeneration in the algebraic relations \eqref{eq-summary} so that the  $\myk^{(a)}_{+}$ and $\myk^{(b)}_{-}$ generators mutually commute, and the form of the algebraic relations are almost identical to those for the rational quiver Yangian; as a result, the molten crystals that are representations of the quiver Yangians are also representations of the trigonometric and elliptic quiver Yangians when $c=0$.
In this section, we will first review the crystals (see \cite{Ooguri:2009ijd,Yamazaki:2010fz}), then show that they are representations of trigonometric and elliptic quiver BPS algebras when $c=0$.

\subsection{Crystals from quivers}

As we have explained, a quiver with a superpotential can encode the information of a toric Calabi-Yau three-fold geometry.
Such a quiver-superpotential pair actually has two equivalent descriptions: as an abstract quiver graph with a superpotential $(Q,W)$ or a periodic quiver on a torus $\tilde Q=(Q_0,Q_1,Q_2)$, where $Q_2$ is the set of loops in the fundamental domain of $\tilde{Q}$.
(The equivalence between these two descriptions can be explained by brane constructions from which the quiver description arises, see \cite{Yamazaki:2008bt}.)
For the purpose of studying the crystal representations, we will need the description of the quiver-superpotential pair as a periodic quiver $\tilde Q=(Q_0,Q_1,Q_2)$.

To a pair $(Q,W)$ one associates a geometric object known as the \emph{quiver path algebra}.
This construction goes as follows.
Consider an algebra $\IC Q$ of formal variables associated to paths in a quiver. We impose a simple multiplication rule dictating that the result of multiplication of two paths $p_1$ and $p_2$ is the joint path $p_1p_2$ if $p_1$ and $p_2$ can be concatenated, and zero otherwise.
In this language the superpotential $W$ is an element in $\IC Q$ that is a formal combination of loops in the quiver $Q$.
It is natural to define a formal derivation of a loop path $\ell$ with respect to any other path $p$ as the sum of all possible complements of $p$ in loop $\ell$.
Denote $\CI_W$ the two-sided ideal in $\IC Q$ generated by $\p_I W$ for all $I\in Q_1$.
The quiver path algebra is then defined as:
\be
\wp(Q,W)\myequiv\IC Q/\CI_W  \;.
\ee

Each quiver arrow $I\in Q_1$ has two physical characteristics: the equivariant weight (complex mass)  and the R-charge.
The former characteristic is a complex parameter $h_I$ satisfying loop and vertex constraints discussed in section \ref{subsec:parameters}. The latter characteristic is assigned to $I$ in such a way that the total R-charges of all loop paths contributing to $W$ is $+2$ (instead of $0$ for equivariant weights).

Consider all monomials in $\wp(Q,W)$ corresponding to paths starting with one quiver vertex $v\in Q_0$. All these monomials $m$ can be uniquely characterized by their equivariant weight $h(m)$ and R-charge $R(m)$.
The set of possible values 
\begin{equation}
    (h(m),R(m))\in \IC\times \IR 
\end{equation}
forms a lattice $\CC_0$ in the 3d space.
The $R$-charges can be chosen in such a way that $R(I)\geq 0$, so that $\CC_0$ is a part of a lattice filling only a convex region of the 3d space.
By forgetting the R-charge we obtain a 2d projection of $\CC_0$ to the $h$-plane, and this projection can be identified with a periodic quiver $\tilde Q$.
Therefore an alternative definition of $\CC_0$ is a lift of paths in $\tilde Q$ to 3d where the third coordinate is given by twice the number of loops around faces the path is encircling.

The crystal $\CC_0$ defines the vacuum representation of the quiver algebra, as studied in \cite{Li:2020rij,Galakhov:2020vyb} for the quiver Yangians, and is named the \emph{canonical crystal} in \cite{Galakhov:2021xum}.
As shown in \cite{Galakhov:2021xum}, any subcrystal ${}^{\sharp}\CC\subset \CC_0$ can be used to define a representation (although not every subcrystal can give rise to an irreducible representations).
Different subcrystals ${}^{\sharp}\CC\in \CC_0$ correspond to different \emph{framings} of the same quiver.\footnote{Actually in the construction of the canonical crystal $\CC_0$, there is already a choice for the color of the leading atom, which corresponds to the choice of the framed vertex $\mathfrak{f}\in Q_0$ in the  \emph{canonical framing}. }
A triplet of quiver, superpotential and framing defines a subcrystal ${}^{\sharp}\CC\in \CC_0$ and thus defines a representation of a shifted quiver Yangian \cite{Galakhov:2021xum}.
Here we would like to use a similar identification between crystals and representations of $\myT_{\beta}(Q)$ and $\myE_{\tau}(Q)$.

A crystal ${}^{\sharp}\CC$ is a configuration of points in 3d. We will call these points ``atoms" and denote them by $\Box$.
The atoms are ``colored" by quiver vertices $v\in Q_0$.
The coloring function works as follows: when we project an atom $\Box$ to a vertex in $\tilde Q$, this vertex is a lift of some vertex $v\in Q_0$ to the periodic quiver $\tilde Q$ covering $Q$, as a result we claim the color of the atom $\Box$ is $v$ and denote this as $\sqbox{$v$}$.
The edges of the crystal correspond to multiplication of corresponding monomials in $\wp(Q,W)$ by $I\in Q_1$.
We use the notation $I\cdot\Box$ to denote  a new atom that corresponds to a monomial in $\wp(Q,W)$ that is given by multiplying  the monomial associated with $\Box$ by $I$ on the right.

For a crystal ${}^{\sharp}\CC$ we define a set of molten crystals $\Kappa$ as a finite subset of ${}^{\sharp}\CC$ satisfying the \emph{melting rule}:
\be\label{melting}
\begin{array}{c}
\mbox{If  }I\cdot\Box\mbox{ is in }\Kappa \mbox{ for an atom } \Box\mbox{ and for an edge }I\in Q_1 \;,\\
\mbox{ then }\Box\mbox{ itself is in }\Kappa  \;.
\end{array}
\ee

Let us define the set ${\rm Add}(\Kappa)$ (resp.  ${\rm Rem}(\Kappa)$) as the set of atoms that can be added (resp. subtracted) to (resp. from) $\Kappa$ so that the resulting set is again a molten crystal.

An extensive selection of  such  crystal constructions can be found in \cite{Yamazaki:2010fz,Li:2020rij,Galakhov:2021xum}.
Let us mention here the simplest example of crystals associated with $\IC^3$.
The corresponding unframed quiver is a trefoil quiver with a canonical superpotential, that corresponds to $\tilde Q$ represented by a triangular lattice:
\be
\begin{array}{c}
\begin{tikzpicture}
        \tikzset{arr/.style={ 
                        postaction={decorate},
                        decoration={markings, mark= at position 0.6 with {\arrow{stealth}}},
                        thick}}
        \draw[thick, red, postaction={decorate},decoration={markings, 
                mark= at position 0.7 with {\arrow{stealth}}}] (0,0) to[out=60,in=0] (0,1) to[out=180,in=120] (0,0);
        \node[above] at (0,1) {$B_1$};
        \begin{scope}[rotate=120]
                \draw[thick, blue, postaction={decorate},decoration={markings, 
                        mark= at position 0.7 with {\arrow{stealth}}}] (0,0) to[out=60,in=0] (0,1) to[out=180,in=120] (0,0);
                \node[below left] at (0,1) {$B_2$};
        \end{scope}
        \begin{scope}[rotate=240]
                \draw[thick, orange, postaction={decorate},decoration={markings, 
                        mark= at position 0.7 with {\arrow{stealth}}}] (0,0) to[out=60,in=0] (0,1) to[out=180,in=120] (0,0);
                \node[below right] at (0,1) {$B_3$};
        \end{scope}
        \draw[fill=white] (0,0) circle (0.08);
        \node[below] at (0,-1) {$W=\Tr \,B_1\left[B_2,B_3\right]$};
        \begin{scope}[shift={(5,0)}]
                \draw[arr,red] (0,0) -- (0.866025, 0.5);
                \draw[arr,orange] (0,1) -- (0,0);
                \draw[arr,blue] (0,0) -- (-0.866025, 0.5);
                \draw[arr,red] (-0.866025, -0.5) -- (0,0);
                \draw[arr,orange] (0,0) -- (0,-1);
                \draw[arr,blue] (0.866025, -0.5) -- (0,0);
                \draw[arr,blue] (0.866025, 0.5) -- (0,1);
                \draw[arr,blue] (0,-1) -- (-0.866025, -0.5);
                \draw[arr,orange] (0.866025, 0.5) -- (0.866025, -0.5);
                \draw[arr,orange] (-0.866025, 0.5) -- (-0.866025, -0.5);
                \draw[arr,red] (0,-1) -- (0.866025, -0.5);
                \draw[arr,red] (-0.866025, 0.5) -- (0,1);
                \begin{scope}[shift={(1.73205,0)}]
                        \draw[arr,red] (0,0) -- (0.866025, 0.5);
                        \draw[arr,orange] (0,1) -- (0,0);
                        \draw[arr,blue] (0,0) -- (-0.866025, 0.5);
                        \draw[arr,red] (-0.866025, -0.5) -- (0,0);
                        \draw[arr,orange] (0,0) -- (0,-1);
                        \draw[arr,blue] (0.866025, -0.5) -- (0,0);
                        \draw[arr,blue] (0.866025, 0.5) -- (0,1);
                        \draw[arr,blue] (0,-1) -- (-0.866025, -0.5);
                        \draw[arr,orange] (0.866025, 0.5) -- (0.866025, -0.5);
                        \draw[arr,red] (0,-1) -- (0.866025, -0.5);
                        \draw[arr,red] (-0.866025, 0.5) -- (0,1);
                \end{scope}
                \draw[fill=white] (0,0) circle (0.08) (0.866025, 0.5) circle (0.08) (0.866025, -0.5) circle (0.08) (-0.866025, 0.5) circle (0.08) (-0.866025, -0.5) circle (0.08) (0,1) circle (0.08) (0,-1) circle (0.08);
                \begin{scope}[shift={(1.73205,0)}]
                        \draw[fill=white] (0,0) circle (0.08) (0.866025, 0.5) circle (0.08) (0.866025, -0.5) circle (0.08) (0,1) circle (0.08) (0,-1) circle (0.08);
                \end{scope}
        \end{scope}
        \draw[<->] (1.5,0) -- (3.5,0);
\end{tikzpicture}
\end{array}.
\ee
A canonical crystal fills the first octant in a cubic lattice $\IZ^3$.
The molten crystals are represented by plane partitions where atoms are depicted as boxes piling up in a corner of a 3d room (see Figure \ref{fig:pile}).
The corresponding set ${\rm Add}(\Kappa)$ (resp. ${\rm Rem}(\Kappa)$) includes boxes that can be added (resp.\ removed) to (resp.\ from) the pile such that the resulting construction remains stable.

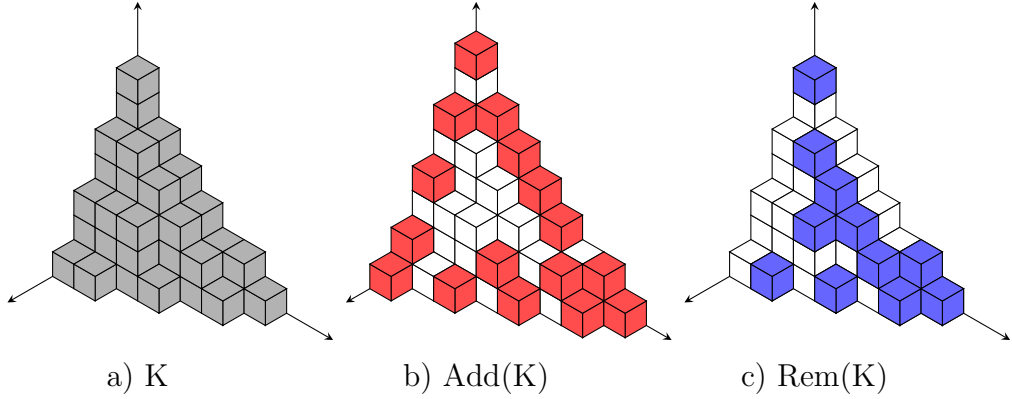
\begin{figure}
\begin{center}
		\begin{tikzpicture}
			\begin{scope}
				\draw[-stealth] (0,0) -- (0,3);
				\draw[-stealth] (0,0) -- (2.59808, -1.5);
				\draw[-stealth] (0,0) -- (-1.73205, -1.);
				\begin{scope}[scale=0.4]
					\tikzset{style/.style={fill=white!40!gray}}
					\input{figs/fig_partition.tex}
				\end{scope}
				\node at (0,-2) {a) $\Kappa$}; 
			\end{scope}
			\begin{scope}[shift={(4.5,0)}]
				\draw[-stealth] (0,0) -- (0,3);
				\draw[-stealth] (0,0) -- (2.59808, -1.5);
				\draw[-stealth] (0,0) -- (-1.73205, -1.);
				\begin{scope}[scale=0.4]
					\tikzset{style1/.style={fill=\bckcolor}}
					\tikzset{style2/.style={fill=white!30!red}}
					\input{figs/fig_partition_1.tex}
				\end{scope}
				\node at (0,-2) {b) ${\rm Add}(\Kappa)$}; 
			\end{scope}
			\begin{scope}[shift={(9,0)}]
				\draw[-stealth] (0,0) -- (0,3);
				\draw[-stealth] (0,0) -- (2.59808, -1.5);
				\draw[-stealth] (0,0) -- (-1.73205, -1.);
				\begin{scope}[scale=0.4]
					\tikzset{style1/.style={fill=\bckcolor}}
					\tikzset{style2/.style={fill=\myblue}}
					\input{figs/fig_partition_2.tex}
				\end{scope}
				\node at (0,-2) {c) ${\rm Rem}(\Kappa)$}; 
			\end{scope}
		\end{tikzpicture}
        \caption{Plane partition $\Kappa=\{\{7, 6, 4, 2\}, \{6, 5, 3\}, \{4, 4, 3\}, \{3, 3\}, \{2, 2\}, \{1\}, \{1\}\}$ and sets ${\rm Add}(\Kappa)$ (red boxes) and ${\rm Rem}(\Kappa)$ (blue boxes).}\label{fig:pile}
\end{center}
\end{figure}

\subsection{Representations for \texorpdfstring{$c=0$}{c=0}}

In this subsection, we determine crystal representations of the algebra \eqref{eq-summary} when the central charge is $c=0$.

A molten crystal representation ${}^{\sharp}\textrm{Rep}$  is defined by a crystal ${}^{\sharp}\CC$, which is a subcrystal of the canonical crystal $\CC_0$.
(For the special case of ${}^{\sharp}\CC=\CC_0$, the representation ${}^{\sharp}\textrm{Rep}$ is  the vacuum representation.) 
The module ${}^{\sharp}\textrm{Rep}$ is spanned by vectors labeled  by molten crystal states $|\Kappa\rangle$ from ${}^{\sharp}\CC$.

Let us explain the ansatz for how the generators $(\mye, \myk_{\pm},\myf)$ should act on the set of crystal states $\{|\Kappa\rangle\}$. 
The reasoning is almost identical to the one for the rational case of the quiver Yangian, see \cite[section 6.1]{Li:2020rij}.

\begin{enumerate}[leftmargin=5mm]
\item 
For $c=0$, all the $\psi^{\pm}(z)$ generators are Cartan generators. 
\begin{equation}
\begin{aligned}
\myk^{\epsilon}(z)\myk^{\epsilon'}(w) &= \myk^{\epsilon'}(w)\myk^{\epsilon}(z) \;, \qquad \epsilon, \epsilon' = \pm \;. \\
\end{aligned}
\end{equation}
Therefore, they can have simultaneous eigenstates. 
Moreover, for $c=0$, $\myk_{+}(z)$ and $\myk_{-}(z)$ are actually indistinguishable, therefore the two have the same action:
\begin{equation}\label{eq-reps-ansatz-psi}
\myk^{(a)}_{\pm}(u)|\Kappa\rangle= \left[\Psi^{(a)}_{\Kappa}(u)\right]_\pm|\Kappa\rangle\,,
\end{equation}
\noindent where $\left[\star\right]_{\pm}$ are two expansions in the Laurent series defined in Appendix~\ref{app:Cyl} for the cases of trigonometric and elliptic algebras.
In the case of a rational algebra there is no need to consider a Laurent expansions and both $\left[\star\right]_{\pm}$ coincide with the very argument $\star$.

\item The generator $e^{(a)}(u)$ adds an atom to $|\Kappa \rangle$  and $f^{(a)}(u)$ removes an atom from $|\Kappa\rangle$:
\begin{equation}\label{eq-reps-ansatz-ef}
\begin{aligned}
\mye^{(a)}(u)|\Kappa\rangle &= \sum_{\sqbox{$a$} \in \textrm{Add}(\Lambda)} \myp\left(u-h_{\sqbox{$a$}}\right)\, \textrm{E}(\Kappa\rightarrow \Kappa+\sqbox{$a$})\, |\Kappa+\sqbox{$a$} \rangle\,,\\
\myf^{(a)}(u)|\Kappa\rangle &= \sum_{\sqbox{$a$} \in \textrm{
Rem}(\Kappa)}\myp\left(u-h_{\sqbox{$a$}}\right)\, \textrm{F}(\Kappa\rightarrow \Kappa-\sqbox{$a$})\,  |\Kappa-\sqbox{$a$} \rangle\,,
\end{aligned}
\end{equation}
where $\myp(u)$ captures the dependence on the spectral parameter and $\textrm{E}(\Kappa\rightarrow \Kappa+\sqbox{$a$})$ and $
\textrm{F}(\Kappa\rightarrow \Kappa+\sqbox{$a$})$ describe the coefficients for different final states.
We further separate the statistical factor $\pm$ from the modulus of the coefficients by defining
\begin{equation}
    \begin{aligned}
    \textrm{E}(\Kappa\rightarrow \Kappa+\sqbox{$a$})&= \epsilon(K\rightarrow K+\sqbox{$a$})\, \lbr\Kappa\rightarrow \Kappa + \sqbox{$a$}\rbr\,,\\
    \textrm{F}(\Kappa\rightarrow \Kappa+\sqbox{$a$})&= \epsilon(K\rightarrow K-\sqbox{$a$})\, \lbr\Kappa\rightarrow \Kappa - \sqbox{$a$}\rbr\,,
    \end{aligned}
\end{equation}
with
\begin{equation}
  \epsilon(K\rightarrow K+\sqbox{$a$})=\pm \qquad \textrm{and} \qquad  \epsilon(K\rightarrow K-\sqbox{$a$})=\pm\,.  
\end{equation}
(This ansatz takes exactly the same form as the one for the quiver Yangian, see \cite[eqs.\ (6.2) and (6.3)]{Li:2020rij}.)
\end{enumerate}
The goal now is to fix $\Psi^{(a)}$,  $\lbr\Kappa\rightarrow \Kappa \pm \sqbox{$a$}\rbr$, and $\epsilon(K\rightarrow K\pm\sqbox{$a$})$.
This takes five steps.
\begin{enumerate}[leftmargin=5mm]
\item From the $\myk-\mye$ and $\myk-\myf$ relations, we have
\begin{equation}
\Psi^{(a)}_{\Kappa+\sqbox{$b$}}(z)\, \myp\left(w-h_{\sqbox{$b$}}\right)\sim \myp\left(w-h_{\sqbox{$b$}}\right)\,\myphi^{a\Leftarrow b}(z, w)\, \Psi^{(a)}_{\Kappa}(z)\,.
\end{equation}
Given the form of the function $\myp(u)$, taking $\sim$ picks up the $w\rightarrow h_{\sqbox{$a$}}$ value, and we have
\begin{equation}
\Psi^{(a)}_{\Kappa+\sqbox{$b$}}(z)=\myphi^{a\Leftarrow b}(z, h_{\sqbox{$b$}})\, \Psi^{(a)}_{\Kappa}(z)\,,
\end{equation}
which gives
\begin{equation}
\Psi^{(a)}_{\Kappa}(u)={}^{\sharp}\psi^{(a)}_0(u)\cdot \prod_{b\in Q_0} \prod_{\sqbox{$b$}\in\Kappa} \myphi^{a\Leftarrow b} \left(u, h_{\sqbox{$b$}}\right) \,.
\end{equation}

The prefactor  ${}^{\sharp}\psi^{(a)}_0(u)$ describes the contribution from the ground state of the representation ${}^{\sharp}\textrm{Rep}$, and is determined by the shape of the crystal ${}^{\sharp}\CC$.
For the rational case, it is given by \cite[eq.\ (3.7)]{Galakhov:2021xum}, and uplifting which gives its expression for the general algebra \eqref{eq-summary}.
It is easiest to express $ {}^{\sharp}\psi^{(a)}_0(u)$
in the language of the framed quiver,\footnote{
For how to translate the shape of the crystal ${}^{\sharp}\CC$ to the framed quiver, see \cite[section 3]{Galakhov:2021xum}.} as
\begin{equation}\label{eq-GS-FQ}
   {}^{\sharp}\psi^{(a)}_0(u)=\myphi^{a\Leftarrow \mathfrak{f}}(u, 0) \,,
\end{equation}
where $\mathfrak{f}$ denotes the framing vertex. 
\item Let us define the notation
\begin{equation}\label{eq-def-Ktwosteps}
\begin{aligned}
\lbr\Kappa\rightarrow \Kappa+\sqbox{$a$} \rightarrow \Kappa+\sqbox{$a$}+\sqbox{$b$}\rbr &\myequiv \lbr\Kappa\rightarrow \Kappa+\sqbox{$a$} \rbr \cdot \lbr\Kappa+\sqbox{$a$} \rightarrow \Kappa+\sqbox{$a$}+\sqbox{$b$}\rbr \,,\\
\lbr\Kappa\rightarrow \Kappa-\sqbox{$a$} \rightarrow \Kappa-\sqbox{$a$}-\sqbox{$b$}\rbr &\myequiv \lbr\Kappa\rightarrow \Kappa-\sqbox{$a$} \rbr \cdot \lbr\Kappa-\sqbox{$a$} \rightarrow \Kappa-\sqbox{$a$}-\sqbox{$b$}\rbr \,,
\end{aligned}
\end{equation}
and 
\begin{equation}
\begin{aligned}
\epsilon(\Kappa\rightarrow \Kappa+\sqbox{$a$} \rightarrow \Kappa+\sqbox{$a$}+\sqbox{$b$}) &\myequiv \epsilon(\Kappa\rightarrow \Kappa+\sqbox{$a$} ) \cdot \epsilon(\Kappa+\sqbox{$a$} \rightarrow \Kappa+\sqbox{$a$}+\sqbox{$b$}) \,,\\
\epsilon(\Kappa\rightarrow \Kappa-\sqbox{$a$} \rightarrow \Kappa-\sqbox{$a$}-\sqbox{$b$}) &\myequiv \epsilon(\Kappa\rightarrow \Kappa-\sqbox{$a$} ) \cdot \epsilon(\Kappa-\sqbox{$a$} \rightarrow \Kappa-\sqbox{$a$}-\sqbox{$b$}) \,.
\end{aligned}
\end{equation}
Then from the $\mye-\mye$ and $\myf-\myf$ relations, we have the constraints on the coefficients $\lbr\Kappa\rightarrow \Kappa\pm\sqbox{$a$}\rbr$ as
\begin{equation}\label{eq-bootstrap-eeff}
\begin{aligned}
\frac{ \lbr\Kappa\rightarrow \Kappa+\sqbox{$b$} \rightarrow \Kappa+\sqbox{$b$}+\sqbox{$a$}\rbr   
}{  \lbr\Kappa\rightarrow \Kappa+\sqbox{$a$} \rightarrow \Kappa+\sqbox{$a$}+\sqbox{$b$}\rbr
} &= \myphi^{a\Leftarrow b} \left(h_{\sqbox{$a$}}, h_{\sqbox{$b$}}\right) \;,\\
\frac{ \lbr\Kappa\rightarrow \Kappa-\sqbox{$b$} \rightarrow \Kappa-\sqbox{$b$}-\sqbox{$a$}\rbr   
 }{  \lbr\Kappa\rightarrow \Kappa-\sqbox{$a$} \rightarrow \Kappa-\sqbox{$a$}-\sqbox{$b$}\rbr 
} &= \myphi^{a\Leftarrow b} \left(h_{\sqbox{$a$}}, h_{\sqbox{$b$}}\right)^{-1} \;,\\
\end{aligned}
\end{equation}
and 
\begin{equation}
\begin{aligned}\label{epsilon_equation}
\frac{ \epsilon(\Kappa\rightarrow \Kappa+\sqbox{$b$} \rightarrow \Kappa+\sqbox{$b$}+\sqbox{$a$})   
}{  \epsilon(\Kappa\rightarrow \Kappa+\sqbox{$a$} \rightarrow \Kappa+\sqbox{$a$}+\sqbox{$b$})
} &=
\frac{ \epsilon(\Kappa\rightarrow \Kappa-\sqbox{$b$} \rightarrow \Kappa-\sqbox{$b$}-\sqbox{$a$})
 }{  \epsilon(\Kappa\rightarrow \Kappa-\sqbox{$a$} \rightarrow \Kappa-\sqbox{$a$}-\sqbox{$b$})
} =(-1)^{|a||b|} \;.
\end{aligned}
\end{equation}
\item From the $\mye-\myf$ relations, we have 
\begin{equation}\label{eq-bootstrap-ef}
(-1)^{|a|}\lbr\Kappa \rightarrow \Kappa -\sqbox{$a$}\rightarrow \Kappa \rbr =\lbr\Kappa \rightarrow \Kappa +\sqbox{$a$}\rightarrow \Kappa \rbr =\lim_{u\rightarrow h_{\sqbox{$a$}}}\, \zeta(u-h_{\sqbox{$a$}})\,\Psi^{(a)}_{\Kappa}(u) \;.
\end{equation}
\item
From the definition  \eqref{eq-def-Ktwosteps} and the constraints \eqref{eq-bootstrap-eeff} and \eqref{eq-bootstrap-ef}, one can determine the solution for $\lbr\Kappa\rightarrow \Kappa \pm \sqbox{$a$}\rbr$ to be
\begin{equation}
\begin{aligned}
\lbr\Kappa\rightarrow \Kappa +\sqbox{$a$}\rbr&=\sqrt{\lim_{u\rightarrow h_{\sqbox{$a$}}}\, \zeta(u-h_{\sqbox{$a$}})\,\Psi^{(a)}_{\Kappa}(u) } \;,\\
\lbr\Kappa\rightarrow \Kappa -\sqbox{$a$}\rbr&=\sqrt{(-1)^{|a|}\lim_{u\rightarrow h_{\sqbox{$a$}}}\,\zeta(u-h_{\sqbox{$a$}})\,\Psi^{(a)}_{\Kappa}(u) } \;.\\
\end{aligned}
\end{equation}
For the reason behind the factor $(-1)^{|a|}$, see \cite[section 6.5.3.3]{Li:2020rij}. 
\item
For the prescription on how to fix the statistical factors $\epsilon(\Kappa\rightarrow\Kappa\pm \sqbox{$a$})$, see \cite[section 6]{Li:2020rij}. 
The explicit result for a specific choice of signs $f_{ab}$ (see the discussion below \eqref{eq-reciprocity-final}) is given in \cite[equation (2.62)]{Galakhov:2020vyb}, and here we generalize it for the generic sign choice $f_{ab}$  in Appendix~\ref{app:sign}:
\begin{equation}
    \epsilon(\Kappa\to\Kappa\pm\sqbox{$a$})=\mathscr{S}_{\Box}\left[|a||b|\bigg|\Kappa\right].
\end{equation}
\end{enumerate}
In summary, the crystal representations for the rational, trigonometric, and elliptic quiver BPS algebras \eqref{eq-summary} with $c=0$ are given by 
\begin{tcolorbox}[ams equation]\label{eq-summary-reps}
\begin{aligned}
\myk^{(a)}_{\pm}(z)|\Kappa\rangle&= \left[\Psi^{(a)}_{\Kappa}(z)\right]_{\pm}|\Kappa\rangle \;,\\
\Psi^{(a)}_{\Kappa}(u)&={}^{\sharp}\psi^{(a)}_0(u)\cdot \prod_{b\in Q_0} \prod_{\sqbox{$b$}\in\Kappa} \myphi^{a\Leftarrow b} \left(u, h_{\sqbox{$b$}}\right) \;, \\
\mye^{(a)}(u)|\Kappa\rangle &= \sum_{\sqbox{$a$} \in \textrm{Add}(\Kappa)} \epsilon(\Kappa\rightarrow\Kappa+ \sqbox{$a$})\,\myp\left(u-h_{\sqbox{$a$}}\right) \\\
&\qquad\qquad \qquad \qquad  \cdot \sqrt{\lim_{u\rightarrow 0}\, \zeta\left(u\right)\,\Psi^{(a)}_{\Kappa}\left(u+h_{\sqbox{$a$}}\right)}\,|\Kappa+\sqbox{$a$}\rangle \;,\\
\myf^{(a)}(u)|\Kappa\rangle &= \sum_{\sqbox{$a$} \in \textrm{
Rem}(\Kappa)}\epsilon(\Kappa\rightarrow\Kappa- \sqbox{$a$})\,\myp\left(u-h_{\sqbox{$a$}}\right) \\
&\qquad \qquad \qquad \qquad  \cdot \sqrt{(-1)^{|a|}\lim_{u\rightarrow 0}\,\zeta(u)\,\Psi^{(a)}_{\Kappa}\left(u+h_{\sqbox{$a$}}\right)}\,|\Kappa-\sqbox{$a$}\rangle \;.
\end{aligned}
\end{tcolorbox}
\noindent One can explicitly check that this representation is consistent with the algebraic relations \eqref{eq-summary}, namely that \eqref{eq-summary-reps} indeed gives a legitimate representation of the algebra \eqref{eq-summary} when $c=0$. 
Further, one can check that by choosing the $\zeta(u)$ and $\myp(u)$ to be the rational ones in \eqref{eq-def-zeta} and \eqref{eq-def-p}, respectively, the representation \eqref{eq-summary-reps} reduces to the one for the rational quiver Yangians of   
\cite{Li:2020rij,Galakhov:2021xum}.

We just saw that the $\lbr\Kappa\rightarrow \Kappa \pm\sqbox{$a$}\rbr$ are solved by demanding that the ansatz \eqref{eq-reps-ansatz-psi} and \eqref{eq-reps-ansatz-ef} indeed give a representation of the algebra \eqref{eq-summary} when $c=0$.
In the next section, we will compute these coefficients directly from gauge theories. (For the rational case, see \cite[section 2]{Galakhov:2020vyb}) for the unshifted quiver Yangians and \cite[section 4]{Galakhov:2021xum} for the shifted quiver Yangians.)
Since they acquire an independent definition in the gauge theory,  
see \eqref{BPS_rep}, we will use
$\left[\Kappa\rightarrow \Kappa \pm\sqbox{$a$}\right]$ to denote the coefficients that are computed directly from the gauge theory.

The result from the gauge theory computation is that for all the rational, toroidal and quantum elliptic quiver BPS algebras in the case $c=0$, the matrix elements satisfy: 
\be\label{matrix_coeff_relations}
\begin{split}
&[\Kappa+\sqbox{$a$}_1\to\Kappa+\sqbox{$a$}_1+\sqbox{$b$}_2\to\Kappa+\sqbox{$b$}_2]=[\Kappa+\sqbox{$a$}_1\to\Kappa\to\Kappa+\sqbox{$b$}_2] \;,\\
&\frac{[\Kappa\to\Kappa+\sqbox{$b$}_2\to\Kappa+\sqbox{$a$}_1+\sqbox{$b$}_2]}{[\Kappa\to\Kappa+\sqbox{$a$}_1\to\Kappa+\sqbox{$a$}_1+\sqbox{$b$}_2]}=\tilde\varphi^{a\Leftarrow b}\left(h_{\sqbox{$a$}_1}-h_{\sqbox{$b$}_2}\right),\\
&\frac{[\Kappa+\sqbox{$a$}_1+\sqbox{$b$}_2\to\Kappa+\sqbox{$b$}_2\to\Kappa]}{[\Kappa+\sqbox{$a$}_1+\sqbox{$b$}_2\to\Kappa+\sqbox{$a$}_1\to\Kappa]}=\tilde\varphi^{a\Leftarrow b}\left(h_{\sqbox{$a$}_1}-h_{\sqbox{$b$}_2}\right),\\
&[\Kappa\to\Kappa+\sqbox{$a$}\to\Kappa]=\lim\lm_{t\to 0}\;\zeta(t)\;\tilde\Psi_{\Kappa}^{(a)}\left(t+h_{\sqbox{$a$}}\right),
\end{split}
\ee
where we actually encounter a slightly modified bond factor (which satisfy the reciprocity condition \eqref{eq-reciprocity-final}):
\be\label{tilde_phi}
\tilde\varphi^{a\Leftarrow b}(z)=(-1)^{\delta_{a,b}}\frac{\prod\lm_{I\in\{a\to b\}}\zeta\left(-z-h_I\right)}{\prod\lm_{J\in\{b\to a\}}\zeta\left(z-h_J\right)} \;,
\ee
and correspondingly a modified charge function
\be\label{tilde_psi}
\tilde{\Psi}_{\Kappa}^{(a)}(z)=\left(\prod\lm_{I\in\{a\to a\}}\frac{1}{-\zeta(h_I)}\right)\times\prod\lm_{\sqbox{$b$}\in\Kappa}\tilde\varphi^{a\Leftarrow b}\left(z-h_{\sqbox{$b$}}\right) \;.
\ee
From \eqref{matrix_coeff_relations} we can solve  the  coefficients $\left[\Kappa\rightarrow \Kappa \pm\sqbox{$a$}\right]$ to be
\be\label{bootstrap}
\left[\Kappa\to\Kappa+\sqbox{$a$}\right]=\left[\Kappa+\sqbox{$a$}\to\Kappa\right]=\sqrt{\lim\lm_{t\to 0}\zeta(t)\tilde{\Psi}_{\Kappa}^{(a)}\left(t+h_{\sqbox{$a$}}\right)} \;.
\ee
Using this set of coefficients,
the matrix elements of generators are defined in the crystal representation in the following way:
\be\label{quiver_representation}
\begin{aligned}
\mye^{(a)}(z)|\Kappa\rangle&=\sum\lm_{k\in\IZ}\sum\lm_{\sqbox{$a$}\in{\rm Add}(\Kappa)}\sigma_+(\sqbox{$a$},\Kappa)\,\myp\left(z-h_{\sqbox{$a$}}\right)\\
	&\qquad\qquad\qquad\times\left[\Kappa\to\Kappa+\sqbox{$a$}\right]\prod\lm_{\substack{b\in Q_0\\ \sqbox{$b$}\in\Kappa}} \left(H_{\sqbox{$a$}}H_{\sqbox{$b$}}\right)^{\frac{\extra\chi_{a,b}}{4}}|\Kappa+\sqbox{$a$}\rangle \;,\\
\myf^{(a)}(z)|\Kappa\rangle&=\sum\lm_{k\in\IZ}\sum\lm_{\sqbox{$a$}\in{\rm Rem}(\Kappa)}\sigma_-(\sqbox{$a$},\Kappa)\,\myp\left(z-h_{\sqbox{$a$}}\right)\\
	&\qquad\qquad\qquad\times\left[\Kappa\to\Kappa-\sqbox{$a$}\right]\prod\lm_{\substack{b\in Q_0\\ \sqbox{$b$}\in\Kappa}} \left(H_{\sqbox{$a$}}H_{\sqbox{$b$}}\right)^{\frac{\extra\chi_{a,b}}{4}}|\Kappa-\sqbox{$a$}\rangle \;,\\
\myk_\pm^{(a)}(z)|\Kappa\rangle&=\left[\tilde\Psi^{(a)}_{\Kappa}(z)\cdot (-1)^{\sum\lm_{b\in Q_0}|a\to b|\left|\Kappa^{(b)}\right|+\left|\Kappa^{(a)}\right|}\cdot \prod\lm_{\substack{b\in Q_0\\ \sqbox{$b$}\in\Kappa}}\left(Z H_{\sqbox{$b$}}\right)^{\frac{\extra\chi_{a,b}}{2}}\right]_{\pm}|\Kappa\rangle\,,
\end{aligned}
\ee
\noindent where $\left|\Kappa^{(a)}\right|$ is the total number of atoms of color $a$, and sign corrections read (see Appendix~\ref{app:sign}):\footnote{Here we used a relation
$$
|a\to b|+\chi_{ab}|b\to a|=|a\to b|+|b\to a|+|a\to b||b\to a|\quad{\rm mod}\quad 2.
$$}
\begin{equation}
\begin{split}
    &\sigma_+\left(\Box,\Kappa\right)=\mathscr{S}_{\Box}\left[|a||b|+\delta_{a,b}+|a\to b|+|b\to a|+|a\to b||b\to a|\bigg|\Kappa\right],\\
    &\sigma_-\left(\sqbox{$a$},\Kappa\right)=\sigma_+\left(\sqbox{$a$},\Kappa\right)\times\prod\lm_{\sqbox{$b$}\in\Kappa}(-1)^{|a||b|}.
\end{split}
\end{equation}

For an unprepared reader the representation \eqref{quiver_representation} might seem more complicated than \eqref{eq-summary-reps}.
In fact, the difference between these two sets of formulas are concentrated in a few factors that we introduced to incorporate the difference between our definitions of the two bond factors, $\myphi$ in \eqref{eq-def-balanced} and $\tilde \varphi$ in  \eqref{tilde_phi}.
From the algebra point of view, the definition of $\myphi$ seems more natural, whereas from the physics point of view $\tilde \varphi$ emerges as the result of the supersymmetric localization.
As we have shown explicitly, they give rise to equivalent representations. 

\section{Toroidal and elliptic  quiver algebras from gauge theories}\label{sec:D-brane}

In this section we derive the elliptic quiver algebra $\myE_{\tau}(Q,W)$ from the analysis of the
three-dimensional $\mathcal{N}=2$ quiver gauge theory  associated with the quiver $Q$ and the superpotential $W$.
(See \cite{Yamazaki:2008bt}  and references therein for realizations of these theories from D-branes and NS5-branes.) 
The theory is then compactified on the elliptic curve of modulus $\tau$, 
along which the theory reduces to a two-dimensional $\mathcal{N}=(2,2)$ theory for the toroidal quiver algebra and 
a one-dimensional $\mathcal{N}=4$ theory for the quiver Yangian.

\subsection{Vortices on elliptic curves}

We consider the three-dimensional quiver gauge theory
associated with a quiver $Q$ and a superpotential $W$.
For a node $v\in Q_0$ we have  an $\mathcal{N}=2$ vector multiplet $(A, X, \lambda, \bar{\lambda})$ (with $A$ a gauge field, $X$ a scalar in the adjoint representation, and $\lambda, \bar{\lambda}$ are fermions), 
and an Fayet-Illiopolous (FI) parameter $r$: 
\be
A_{\mu=0,1,2}^{(v)},\; X^{(v)};\quad r^{(v)}.
\ee
Here and in the following we choose the $1$ and $2$-directions to be along the elliptic curve, with $0$ being the residual direction.
For an arrow $I\in Q_1$, we have an $\mathcal{N}=2$ chiral multiplet $(\phi, \psi, \bar{\psi})$ ($\phi$ is a complex scalar and $\psi, \bar{\psi}$ are fermions), with a flavor charge (equivariant parameter) $h_I$:
\be
\phi_I;\quad h_I \;.
\ee
In the rest of the paper we would like to work with the so-called \emph{cyclic} chamber (cf.\ \cite{Galakhov:2021xum}):
\be
r^{(v)}>0 \;,\quad v\in Q_0 \;.
\ee

The vacuum moduli space is described by a set of equations (see Appendix~\ref{app:3d}):
\be\label{vortices}
\begin{split}
F_{12}^{(v)}=r^{(v)}-&\sum\lm_{w\in Q_0}\left(\sum\lm_{I:\,w\to v}\phi_I\phi_I^{\dagger}-\sum\lm_{J:\,v\to w}\phi_J^{\dagger}\phi_J\right),\;v\in Q_0 \;,\\
X^{(w)}\phi_{(I:\,v\to w)}&-\phi_{(I:\,v\to w)}X^{(v)}=0,\;I\in Q_1 \;,\\
\p_z\phi_{(I:\,v\to w)}+\I A_z^{(w)}\phi_{(I:\,v\to w)}&-\I\phi_{(I:\,v\to w)}A_z^{(v)}+h_I\phi_{(I:\,v\to w)}=0,\;I\in Q_1 \;,\\
\p_{\phi_I}W(\phi)&=0,\;I\in Q_1\;.
\end{split}
\ee
The first set of equations in \eqref{vortices} describes vortices along the elliptic curve.
The second and the third equations combine into a single set of equations if we are discussing a four-dimensional $\mathcal{N}=1$ theory
instead of its dimensional reduction, the three-dimensional $\mathcal{N}=2$ theory discussed here.
The fourth set of equations are simply the F-term constraints for the chiral multiplet scalars.

The space of solutions to \eqref{vortices} modulo gauge equivalence gives the moduli space
\be
\CM_{\rm vortex}=\eqref{vortices}/\CG \;,
\ee
which we call the vortex moduli space.
The BPS Hilbert space is the equivariant $\CQ$-cohomology of the moduli space 
\be\label{Hilbert}
\mathscr{H}_{\rm BPS}=H_{\CG}^*\left(\CM_{\rm vortex},\CQ\right) \;,
\ee
where the explicit expressions for the supercharge $\CQ$ can be found in Appendix~\ref{app:3d}.

There is a BPS bound for the Hamiltonian eigenvalues:
\be
\CH\geq Z \;,
\ee
where the central charge reads:
\be\label{cc}
Z=\sum\lm_{v\in Q_0}\; r^{(v)}\!\int F_{12}^{(v)}\;d^2x \;.
\ee
The Bradlow limit (see e.g.\cite{Maldonado:2015gfa,Miyake:2011yr}) on the vortex number reads:
\be\label{Bradlow}
\sum\lm_{v\in Q_0}\int\Tr\,F_{12}^{(v)}\;d^2 x\leq \left(\sum\lm_{v\in Q_0}r^{(v)}\right)\times{\rm Area} \;.
\ee

The Hilbert space \eqref{Hilbert} is naturally graded by a lattice of central charges that is a lattice of vorticity quantum numbers.
The Bradlow limit \eqref{Bradlow} indicates that the number of graded components is finite in general.

The BPS Hilbert space can be calculated by usual localization techniques where wave functions will be represented by effective Gaussian fluctuations around the classical vacua.
This procedure can be systematized in the Wilsonian renormalization approach where the renormalization is one-loop exact.
The linearized deviation modes will be organized in the usual Kaluza-Klein (KK) towers of winding modes, whose temporal frequency is given by the following effective expression:
\be
\omega=p_x n+p_y m+{\rm flavor\; charge} \;.
\ee

As a result of the compactification along the cycles of the elliptic curve, either $p_x$ or $p_y$ tends to infinity so that the effective flavor charges for the corresponding higher winding modes become greater than the renormalization cutoff. Those modes thus do not contribute to the effective action.
During this procedure the area of the elliptic curve:
\be
{\rm Area}=\frac{4\pi^2}{p_xp_y} 
\ee
shrinks and does not allow vortex solutions due to \eqref{Bradlow}.

Therefore we assume that the elliptic and toroidal quiver BPS algebras correspond to the zero vorticity graded component of the BPS Hilbert space.

In the next subsection we will consider the zero vorticity BPS Hilbert space and construct the representations for $\myE_{\tau}(Q,W)$ and $\myT_{\beta}(Q,W)$ algebras; and we will return to the discussion on contributions of vortices later in subsection \ref{ssec:vortices}.

\subsection{Zero vorticity BPS subalgebra}\label{s:zero_BPS}

The energy functional is positive definite on a BPS configuration and contains a squared field-strength term:
\be
\mathscr{E}=\sum\lm_{v\in Q_0}\int d^2 x\;\Tr\,\left|F_{12}^{(v)}\right|^2+\ldots=Z \;.
\ee
For the zero vorticity component of the BPS Hilbert space $Z=0$, therefore the only solutions to the differential equations \eqref{vortices} are constant field configurations, and \eqref{vortices} reduces to a system of algebraic equations for the expectation values
(and hence with the $z$-derivatives dropped):
\be\label{quiver_rep}
\begin{split}
r^{(v)}-\sum\lm_{w\in Q_0}\left(\sum\lm_{I:\,w\to v}\phi_I\phi_I^{\dagger}-\sum\lm_{J:\,v\to w}\phi_J^{\dagger}\phi_J\right)=0,\;v\in Q_0 \;,\\
X^{(w)}\phi_{(I:\,v\to w)}-\phi_{(I:\,v\to w)}X^{(v)}=0,\;I\in Q_1 \;,\\
\I A_z^{(w)}\phi_{(I:\,v\to w)}-\I\phi_{(I:\,v\to w)}A_z^{(v)}+h_I\phi_{(I:\,v\to w)}=0,\;I\in Q_1 \;,\\
\p_{\phi_I}W(\phi)=0,\;I\in Q_1\;.
\end{split}
\ee
The discussion is then essentially the same as in the case of quantum mechanics in \cite{Galakhov:2020vyb},
if we disregard the periodicities along the elliptic curve. In particular the torus fixed points of the moduli space are given by three-dimensional crystals.

By invoking the  Narasimhan-Seshadri-Hitchin-Kobayashi correspondence, we can simplify the equations 
if we complexify the gauge transformations. This gives the moduli space of quiver representations:
\be
\fR\left(Q,W,{\vec d}\right)\myequiv\left\{\begin{array}{l}
V_{v\in Q_0},\; d_v={\rm dim}\,V_v\\
\phi_{(a:v\to w)\in Q_1}\in{\rm Hom}(V_v,V_w)\\
\p_{\phi_a}W=0\\
\end{array} \right\}\Big/ \prod\lm_{v\in Q_0} \mathrm{GL}(d_v,\IC)  \;,
\ee
where the FI parameters $r_v$ are now traded for the stability parameters $\theta_v$.
Note that more precisely we should rather consider isomorphism classes of quiver representations
where a homomorphism of quiver representations $\fR$ and $\fR'$ is defined as a collection of linear maps:
\be\label{homo}
\tau_v\in {\rm Hom}(V_v,V_v') \;,\quad v\in Q_0 \;,
\ee
such that the following diagram commutes:
\be
\begin{array}{c}
\begin{tikzpicture}
        \node (A) at (0,0) {$V_v$};
        \node (B) at (3,0) {$V_w$};
        \node (C) at (0,-1.5) {$V_v'$};
        \node (D) at (3,-1.5) {$V_w'$};
        \path (A) edge[-stealth] node[above] {$\phi_{(a:v\to w)}$} (B)  (C) edge[-stealth] node[above] {$\phi_{(a:v\to w)}'$} (D) (A) edge[-stealth] node[left] {$\tau_v$} (C) (B) edge[-stealth] node[right] {$\tau_w$} (D);
\end{tikzpicture}
\end{array}
\ee

As in \cite{Galakhov:2020vyb} we consider large values of the stability parameters $\theta_v$, and consider
the infrared effective theory where heavy degrees of freedom are integrated out. 
This leaves the infrared effective meson space
\be
\fR_{\rm IR}=\bigoplus\lm_{a=1}^{N_{\rm meson}} \{ \underset{\mbox{\tiny meson}}{\phi_a},\underset{\mbox{\tiny charge}}{h_a}\}  \;,
\ee
as a tangent bundle in $\fR$ to a fixed point defined by $\Kappa$ modulo linearized gauge transformations, and the effective wave-function $\Psi$ is a function on this moduli space.

Since we are interested in the BPS sectors the wave-function is annihilated by the 
effective supercharge $\bar\CQ_{\dot 2}^{({\rm eff})}$,
which reads in terms of meson fields as (see \eqref{su_cha} and \cite{Galakhov:2020vyb}):
\be\label{diff}
\bar\CQ_{\dot 2}^{({\rm eff})}=\sum\lm_{a=1}^{N_{\rm meson}}\int d^2z\;\left[ \delta\bar\phi^a\frac{\delta}{\delta\bar\phi_{a}} +\iota_{\delta/\delta\phi_a}\left(\p_z\phi_a+h_a\phi_a\right)\right] \;,
\ee
where we substituted the fermion fields by suitable forms on the cotangent bundle to the field configuration space. 
While this expression of the supercharge involves functional derivatives, we can be more explicit if we decompose the fields into an orthogonal basis of doubly-periodic functions:
\be
\phi(x,y)=\sum\lm_{m,n\in\IZ} \phi_{m,n}e^{2\pi \I (mx+ny)}  \;,
\ee
where we parameterize the elliptic curve with modulus $\tau$ in the following way:
\begin{align}
\begin{array}{c}
\begin{tikzpicture}[scale=0.5]
        \draw (0,0) -- (1,2) node[pos=0.5,above left] {$a\tau$};
        \draw (1,2) -- (5,2) -- (4,0);
        \draw (0,0) -- (4,0) node[pos=0.5,below] {$a$};
        \draw[thick,dotted] (1,0) -- (1,2) node[pos=0.55,right] {\!$a\;{\rm Im}\,\tau$};
\end{tikzpicture}
\end{array}\quad \begin{array}{c}
z=a(x+\tau y) \;,\\
\p_z=\frac{1}{2\I\; a\; {\rm Im}\,\tau}\left(\p_y-\bar\tau\p_x\right) \;.
\end{array}
\end{align}
The resulting supercharge decomposes over modes as
\be
\bar\CQ_{\dot 2}^{({\rm eff})}\sim\sum\lm_{a=1}^{N_{\rm meson}}\sum\lm_{m,n\in\IZ}\left(d\bar\phi_{a,m,n}\frac{\p}{\p\bar\phi_{a,m,n}}+\omega_a(m,n)\, \iota_{\p/\p\phi_{a,m,n}}\right) \;,
\ee
where the effective frequencies $\omega_a(m,n)$ are given by
\be
\omega_a(m,n)=\frac{\pi(n-\bar\tau m)}{a\;{\rm Im}\,\tau}+h_a \;.
\ee
Following \cite{Galakhov:2020vyb} (see also \cite{Cordes:1994fc}), we identify the effective BPS wave functions in fixed points with the Euler classes of cotangent bundles, and a Thom representative for such an Euler class is given by a character:
\be
{\rm Eul}\left(\fR_{\rm IR}\right)\sim \prod\lm_{a=1}^{N_{\rm meson}}\prod\lm_{m,n\in\IZ}\omega_a(m,n) \;.
\ee
Now the contribution from a single meson reads:
\be\label{theta}
\prod\lm_{m,n\in\IZ}\omega_a(m,n) \quad\overset{\mbox{\tiny $\zeta$-regularization}}{\begin{array}{c}\begin{tikzpicture}
                \draw[->] (0,0) -- (2.5,0);
\end{tikzpicture}\end{array}}\quad\vartheta_{11}\left(\frac{h_a\;{\rm Im}\,\tau}{\pi}\Big|\, \bar\tau\right)\sim\zeta(h_a) \;,
\ee
where
\be
\vartheta_{11}(z|\tau)\myequiv -2q^{\frac{1}{4}}\sin\pi z\prod\lm_{k=0}^{\infty}(1-q^{2k})\left(1-2q^{2k}\cos2\pi z+q^{4k}\right),\quad q=e^{\pi \I \tau} \;,
\ee
is the Jacobi theta function satisfying $\vartheta_{11}(-z|\tau)=-\vartheta_{11}(z|\tau)$.
In terms of $\Theta_q(z)$ in \eqref{notations} we have:
\be
\vartheta_{11}(z|\tau)=-\I q^{\frac{1}{4}}(q^2;q^2)_{\infty}\Theta_q(z) \;,
\ee
provided that the map between lower case variables and upper case variables is given by $Z=e^{2\pi \I z}$.

Apparently, the Euler class we are calculating is defined by a chiral Dirac determinant on the Riemann surface.
Strictly speaking this quantity is not a periodic holomorphic gauge-invariant scalar as the naive product formula \eqref{theta} suggests.
The $\zeta$-regularization breaks a part of symmetries. 
Rather a regularization is chosen in such a way that the result is a holomorphic section of a holomorphic line bundle on the torus Jacobian \cite{Alvarez-Gaume:1986rcs}, in this way $\zeta(z)$ acquires properties that will be crucial for us as we will see soon: $\zeta(z)$ is a holomorphic odd function. 

So far we have focused on 3d $\mathcal{N}=2$ quiver gauge theories on the elliptic curve.
However, it is now clear that the same discussion will go through in parallel for 
2d $\mathcal{N}=(2,2)$ theory on a circle, or for 1d $\mathcal{N}=4$ quantum mechanics;
such dimensional reductions lead to shrinking cycles of the elliptic curve that reduce the theta function to a hyperbolic sine function (for 2d) and a linear function (for 1d), when both or one of the cycles are contracted, respectively.
As a result we derive the sequence of reductions depicted in \eqref{red_diagram}.

Following this observation we propose a generic expression for a regularized Euler class for the meson space following \cite[eq. (3.29)]{Galakhov:2020vyb}:
\be\label{Eul}
{\rm Eul}_{\zeta}\left(\fR_{\rm IR}\right)\myequiv(-1)^{\left\lfloor\sum\lm_{a:\; h_a=0}\frac{1}{2}\right\rfloor}\prod\lm_{a:\;h_a\neq 0}\zeta(h_a) \;,
\ee
where the function $\zeta$ is defined for in \eqref{eq-def-zeta} and it depends on the choice of the algebra (equivalently the dimensionality of the gauge theory).

We define the generators of the BPS algebra following \cite{Galakhov:2020vyb,Galakhov:2021xum} (analogously to \cite{Nakajima_book,Braverman:2016wma}) as the Hecke modification operators for bundles carried by D-branes.
As a result of the action of Hecke modifications, the quiver representations are embedded into each other:
\be
\begin{array}{c}
\begin{tikzpicture}
\node (A) at (0,0) {$\fR\left(Q,W,{\vec d}\right)$};
\node (B) at (6,0) {$\fR\left(Q,W,{\vec d}+1_v\right)$};
\path ([shift={(0,0.07)}]A.east) edge[-stealth] node[above] {$\mye$} ([shift={(0,0.07)}]B.west) ([shift={(0,-0.07)}]A.east) edge[stealth-] node[below] {$\myf$} ([shift={(0,-0.07)}]B.west);
\end{tikzpicture}
\end{array},
\ee
where $1_v$ is a simple vector with a unit at position $v$ and zeros elsewhere.
By \emph{embedding} we imply that there is a \emph{homomorphism} defined in \eqref{homo} from the bigger space into the smaller one.

For fixed points on the representation, this homomorphism is possible only if one of the corresponding crystals is embedded into the other crystal.
In other words the corresponding crystals differ by just a single atom and we can denote them as
$$
\Kappa \quad \mbox{and} \quad \Kappa+\Box \;.
$$

We define the incidence locus in the following way:
\be
\fR_{\rm IR}(\Kappa,\Kappa+\Box)\myequiv\left\{\fR_{\rm IR}(\Kappa)\subset\fR_{\rm IR}(\Kappa+\Box) \right\}\subset \fR_{\rm IR}(\Kappa)\oplus\fR_{\rm IR}(\Kappa+\Box) \;.
\ee

Defined in this way the incidence locus is a subspace of a graded meson space and, therefore, it is a graded meson space itself.
Correspondingly, the Euler class can be defined for the incidence locus as well.

We define matrix elements in the following way:
\be\label{BPS_rep}
\begin{split}
\left[\Kappa\to\Kappa+\Box\right]\myequiv\frac{{\rm Eul}_{\zeta}\left({\fR}_{\rm IR}(\Kappa)\right)}{{\rm Eul}_{\zeta}\left({\fR}_{\rm IR}(\Kappa,\Kappa+\Box)\right)} \;,\\
\left[\Kappa\to\Kappa-\Box\right]\myequiv\frac{{\rm Eul}_{\zeta}\left({\fR}_{\rm IR}(\Kappa)\right)}{{\rm Eul}_{\zeta}\left({\fR}_{\rm IR}(\Kappa-\Box,\Kappa)\right)}\;.
\end{split}
\ee

While we do not have generic combinatorial closed formulas for these expressions, we have analyzed these expressions  for a large number of rather generic configurations with the help of  Mathematica. 
In all the examples presented in \cite{Galakhov:2020vyb,Galakhov:2021xum}, which include various Calabi-Yau three-folds (both without and with compact $4$-cycles) and the various framed quivers for them (which correspond to different representations), we find that the following conclusion holds: 
\begin{tcolorbox}[ams equation]
\eqref{BPS_rep}\;\mbox{satisfies}\;\eqref{matrix_coeff_relations} \;,
\end{tcolorbox}
\noindent
provided that ${\rm Eul}_{\zeta}$ is a generalized Euler class \eqref{Eul} for an \emph{arbitrary} odd function $\zeta$:
\be
\zeta\left(-z\right)=-\zeta(z) \;.
\ee

To complete this subsection let us note that in the vacuum a complexified Wilson loop:
\be\label{Wilson_line}
{\bf W}={\rm Pexp}\left(\int A_1 dx^1+\I\int A_2dx^2\right) 
\ee
commutes with the supercharge $\bar Q_{\dot 2}$, therefore it contributes to the BPS algebra.
It is natural to consider a generating function for these observables, and its action is diagonal in the crystal basis:
\be
\Tr\; \myp\left(z-{\bf W}\right)|\Kappa\rangle=\sum\lm_{\Box\in\Kappa}\myp\left(z-h_{\Box}\right)|\Kappa\rangle  \;.
\ee
By mixing these operators with \eqref{BPS_rep} one can re-construct the whole crystal representation \eqref{quiver_representation} along the lines of \cite[Section 4.3]{Galakhov:2021xum}.
Mathematically we treat these operators as a representation of a subalgebra generated by the elements $\myk_\pm^{(a)}(z)$ (see Appendix~\ref{app:free_field}).

\subsection{On vortices and 4d crystals}
\label{ssec:vortices}

The dynamics of vortices is rather rich \cite{Taubes:1979tm,Maldonado:2015gfa,Manton:2010mj,Miyake:2011yr,Eto:2005yh,Bullimore:2016hdc,Hanany:2003hp,Bullimore:2018jlp,Bullimore:2021auw} and deserves a separate discussion.
Apparently, an inclusion of non-zero-vorticity BPS states into the whole BPS dynamics will extend the BPS algebra.
As we have seen already, however, such topological configurations are a feature of super-Yang-Mills-Higgs theory in flat 2+1 space-time dimensions, 
and hence those states will not survive during a compactification from 3d to 1d --- the Bradlow limit \eqref{Bradlow} singles out states with zero vorticity.
Therefore the dynamics of the vortex BPS states seems to be irrelevant for the algebra hierarchy \eqref{red_diagram} and goes beyond the scope of the present paper.
In this subsection we will highlight briefly some natural perspectives for the BPS algebra modifications and extensions with vortices included, and leave a thorough approach to this problem for  discussions elsewhere.

Let us start with simple Abrikosov-Nielsen-Olesen (ANO) Abelian vortices in a super Yang-Mills-Higgs model with a single chiral scalar.
The vortex moduli space for this system reads:
\be
\CM_{\rm vortex}=\left\{\begin{array}{c}
	F={\bf D}\;\omega_{\Sigma}\\
	D_z\phi=0\\
	X_3=0
\end{array}\right\}\Big/\CG  \;,
\ee
\noindent where $\omega_{\Sigma}$ is the top volume form on the Riemann surface $\Sigma$.
It is well-known that this moduli space admits a smooth induced metric and is parameterized by the zeros of the chiral field $\phi$ on $\Sigma$ \cite{Taubes:1979tm,Eto:2005yh,Hanany:2003hp}.
The topological charge --- the vorticity number --- is defined by the first Chern character of the resulting gauge bundle and coincides with the number of zeros of $\phi$:
\be
\fn=\frac{1}{2\pi}\int\lm_{\Sigma}F=\#\mbox{ of zeros of }\phi \;.
\ee

At low energies the wave-functions of BPS states containing vortices localize to the cohomologies of the vortex moduli space:
\be
\mathscr{H}_{\rm BPS} \cong H^*\left(\CM_{\rm vortex}\right) \;.
\ee

Singular monopole/anti-monopole operators are defined in \cite{Borokhov:2002cg,Borokhov:2002ib,Intriligator:2013lca}.
The $\bar Q_{\dot 2}$ supersymmetry preserving version of this operators has the following form:
\be\label{mon_op}
M_{\pm} \sim \exp\left[\pm \left(2\pi X_3+ \I\gamma - 2\pi\I\int\lm_{\rm D.s.}dx^0\,{\bf D}\right)\right],
\ee
where $\p_{\mu}\gamma=\pi \epsilon_{\mu\nu\lambda}F^{\nu\lambda}$ is the dual photon field and the last term (namely the integration over the time-like Dirac string) is required by the supersymmetry.
The choice of the sign in \eqref{mon_op} depends whether a monopole or anti-monopole operator is chosen and which part of the original supersymmetry it preserves.
The (anti-)monopole operator raises (lowers) the vorticity number of a BPS state by 1:
$$
\langle \fm|M_{\pm}|\fn\rangle\sim \delta_{\fm,\fn\pm 1}\;.
$$
On a non-trivial surface the algebra of monopole operators becomes extended naturally by mixing with zero-mode fields in the vortex background.
In the simplest example of a single vortex on a torus surface, the BPS Hilbert space is identified simply with the cohomologies of the surface (see e.g.~\cite{Bullimore:2021auw}) and has dimension 4. We therefore expect the monopole operator to have, in fact, 4 non-trivial matrix elements pairing the vacuum with 4 vorticity-1 states.

In the IR the gauge group is broken first to its Abelian subgroup:
\be\label{G_IR}
G_{\rm IR}=U(1)^{\otimes |\Kappa|} \;,
\ee
where each $U(1)$ factor is associated with an atom of a crystal, and $|\Kappa|$ is the total number of atoms in crystal $\Kappa$.
Further this group is broken completely by the chiral field vacuum expectation values (VEVs).

The fields producing VEVs in the IR can be described as ``baryons". For each given atom $\Box \in\Kappa$ consider a path $\wp_{\Box}$ in the crystal $\Kappa$ connecting this atom to any root atom.
We can define baryon fields as products over matrix elements associated with arrow links $I$ in the path $\wp_{\Box}$:
\be\label{baryon}
B_{\Box}=\prod\lm_{I\in\wp_{\Box}}\phi_I \;.
\ee
\noindent Since the VEVs of $\phi_I$ satisfy the superpotential condition (the loop constraint), the expectation value \eqref{baryon} is independent of the concrete choice of the path $\wp_{\Box}$.
The baryon $B_{\Box}$ has an electric charge +1 with respect to the $U(1)_{\Box}$ gauge subgroup in \eqref{G_IR} associated with atom $\Box$. 
This subgroup is broken by the VEV of $B_{\Box}$. If $B_{\Box}$ has a $k$-th order zero at point $z\in\Sigma$ then this point is pinched by $k$ units of magnetic flux of $U(1)_{\Box}$.

In general, the dynamics of such vortices is rather involved, since we need to take into account the F-term constraints for the expectation values.
We can nevertheless calculate easily their quantum numbers.
Let us define a vector of vortex numbers associated with the atoms of the crystal $\Kappa$:
\be
\vec\CN_{\Kappa}=\left\{\fn_{\Box}\right\}_{\Box\in\Kappa},\quad \fn_{\Box}\in\IZ_{\geq 0} \;.
\ee
\noindent The numbers $\fn_{\Box}$ define the magnetic flux of $U(1)_{\Box}$ through $\Sigma$ and how many zeros the field $B_{\Box}$ has on $\Sigma$.

The factorization condition \eqref{baryon} imposes further restrictions on possible values of $\vec\CN_{\Kappa}$.
Indeed for two atoms $\Box_1$ and $\Box_2$ connected by an arrow link $I$, the baryons satisfy a factorization relation:
\be
B_{\Box_2}=\phi_I B_{\Box_1}\;.
\ee
An expectation value of the chiral field $\phi_I$ is also physically observable, therefore $\phi_I$ is not allowed to have poles on $\Sigma$, only zeros.
Therefore we derive a constraint for the zeros of the baryon fields or, equivalently, the fluxes of $U(1)$ components in \eqref{G_IR}:
\be\label{4d_cry}
\begin{split}
&\mbox{If for a pair of atoms }\Box_1\mbox{ and }\Box_2\mbox{ there is}\\
&\mbox{a link }\{\Box_1\to\Box_2\}\mbox{ in }\Kappa \quad \fn_{\Box_1}\leq\fn_{\Box_2} \;. 
\end{split}
\ee
The constraint \eqref{4d_cry} has a lot in common with the melting rule for 3d crystals \eqref{melting}.
If one considers the case of plane-partition molten crystals associated with $\IC^3$,
 then one assigns numbers to boxes of the plane partition following the rule \eqref{4d_cry} with inequality $\fn_{\Box_1}\leq\fn_{\Box_2}$ substituted by $\fn_{\Box_1}\geq\fn_{\Box_2}$. The result then will be a 3d image of a 4d solid partition.

To a pair $(K,\vec\CN_{\Kappa})$ satisfying \eqref{4d_cry} one can assign a 4d configuration of atoms in a space parameterized by the following axes:
$$
\Lambda^4=\left(\mathsf{h}_1,\mathsf{h}_2,R\mbox{-charge},\fn\right)\subset\IC\times\IR\times\IZ \;,
$$
where the atoms are arranged in the following way.
We start with embedding the 3d crystal lattice in $\Lambda^4$ by simply assigning to all the atoms $\Box\in\Kappa$ the fourth coordinate equal to 0, then we fill the points along the fourth axis of the lattice $\Lambda^4$ starting with 0 and going upwards up to $\fn'=N-\fn_{\Box}$ for some large common number $N$ (for example as $N$ we can accept the Bradlow limit \eqref{Bradlow}).
All such configurations represent molten 4d crystals in the lattice $\Lambda^4$ where the lattice is generated by the quiver maps $Q_1$ all having the fourth coordinate $\fn$ equal to zero and a new monopole map $M_+$ having coordinates (0,0,0,1).

We should stress that the appearance of 4d crystals is not surprising.
Vortices are quasi-particles in the super-Yang-Mills-Higgs theory associated to a system of D8/D6/D4/D2/D0 branes probing the whole geometry of ${\rm CY_3}\times T^2$, where the torus $T^2$ is a spatial slice of our system. 
An appearance of 4d melting crystals (also referred to as brane brick models) was already noticed in the literature (see e.g. \cite{Franco:2015tya,Franco:2019bmx}) in applications to enumerative geometry counting problems of toric Calabi-Yau four-folds \cite{cao2015donaldsonthomas,MR3861713,cao2020ktheoretic}.
Although ${\rm CY_3}\times T^2$ is not a toric CY$_4$ (for which a color/shape generalization of the solid partition should be relevant \cite{Nekrasov:2017cih,Nekrasov:2018xsb}, 
nevertheless we expect that these two setups may have common features, in their countings for BPS states of D-branes.

\section{Hopf algebra and shuffle algebra from gauge theories}\label{sec:Hopf}

In the previous section we have seen that 
the quiver gauge theories give first-principle derivations of the representations of our quiver BPS algebras.
In this section we see the added bonus that the quiver gauge theories 
also explain naturally the Hopf-algebra structures and shuffle-algebra structures.

\subsection{Coproduct on crystal representations}

For the discussion of the coproduct, the crucial ingredients are the 
representations of the shifted quiver Yangians in terms of non-canonical crystals
\cite{Galakhov:2021xum}. While we refer \cite{Galakhov:2021xum} for the details,
the essential points are that we change the framing of the quiver,
which in turn makes it possible to start growing crystals
at different multiple points (`starters') on the spectral curve.
For the toroidal case the spectral curve is a cylinder and this looks as follows:

\be\label{product}
\begin{array}{c}
\begin{tikzpicture}
\draw (-1,0.2) -- (4,0.2) (-1,-0.2) -- (4,-0.2);
\begin{scope}[shift={(-1,0)}]
\begin{scope}[xscale=0.3]
        \draw (0,0) circle (0.2);
\end{scope}
\end{scope}
\begin{scope}[shift={(4,0)}]
        \begin{scope}[xscale=0.3]
                \draw (0,0) circle (0.2);
        \end{scope}
\end{scope}
\foreach \i in {0, 1, 2}
{
\begin{scope}[shift={(1.5 * \i,0.8)}]
        \draw [dashed] (-0.25,0) -- (0,-0.8) -- (0.25,0);
\begin{scope}[scale=0.15]
        \draw[thin, fill=white] (0.707107,1.22474) -- (0.707107,2.04124) -- (0.,1.63299) -- (0.,0.816497) -- cycle;
        \draw[thin, fill=white] (-0.707107,-1.22474) -- (-0.707107,-0.408248) -- (-1.41421,-0.816497) -- (-1.41421,-1.63299) -- cycle;
        \draw[thin, fill=white] (0.707107,-1.22474) -- (0.707107,-0.408248) -- (0.,-0.816497) -- (0.,-1.63299) -- cycle;
        \draw[thin, fill=white] (0.707107,-0.408248) -- (0.707107,0.408248) -- (0.,0.) -- (0.,-0.816497) -- cycle;
        \draw[thin, fill=white] (2.12132,-0.408248) -- (2.12132,0.408248) -- (1.41421,0.) -- (1.41421,-0.816497) -- cycle;
        \draw[thin, fill=white] (2.82843,-1.63299) -- (2.82843,-0.816497) -- (2.12132,-1.22474) -- (2.12132,-2.04124) -- cycle;
        \draw[thin, fill=white] (-0.707107,1.22474) -- (-0.707107,2.04124) -- (0.,1.63299) -- (0.,0.816497) -- cycle;
        \draw[thin, fill=white] (-1.41421,0.) -- (-1.41421,0.816497) -- (-0.707107,0.408248) -- (-0.707107,-0.408248) -- cycle;
        \draw[thin, fill=white] (-2.12132,-1.22474) -- (-2.12132,-0.408248) -- (-1.41421,-0.816497) -- (-1.41421,-1.63299) -- cycle;
        \draw[thin, fill=white] (-0.707107,-1.22474) -- (-0.707107,-0.408248) -- (0.,-0.816497) -- (0.,-1.63299) -- cycle;
        \draw[thin, fill=white] (-0.707107,-0.408248) -- (-0.707107,0.408248) -- (0.,0.) -- (0.,-0.816497) -- cycle;
        \draw[thin, fill=white] (0.707107,-1.22474) -- (0.707107,-0.408248) -- (1.41421,-0.816497) -- (1.41421,-1.63299) -- cycle;
        \draw[thin, fill=white] (0.707107,-0.408248) -- (0.707107,0.408248) -- (1.41421,0.) -- (1.41421,-0.816497) -- cycle;
        \draw[thin, fill=white] (1.41421,-1.63299) -- (1.41421,-0.816497) -- (2.12132,-1.22474) -- (2.12132,-2.04124) -- cycle;
        \draw[thin, fill=white] (0.,2.44949) -- (-0.707107,2.04124) -- (0.,1.63299) -- (0.707107,2.04124) -- cycle;
        \draw[thin, fill=white] (-0.707107,1.22474) -- (-1.41421,0.816497) -- (-0.707107,0.408248) -- (0.,0.816497) -- cycle;
        \draw[thin, fill=white] (-1.41421,0.) -- (-2.12132,-0.408248) -- (-1.41421,-0.816497) -- (-0.707107,-0.408248) -- cycle;
        \draw[thin, fill=white] (0.707107,1.22474) -- (0.,0.816497) -- (0.707107,0.408248) -- (1.41421,0.816497) -- cycle;
        \draw[thin, fill=white] (0.,0.816497) -- (-0.707107,0.408248) -- (0.,0.) -- (0.707107,0.408248) -- cycle;
        \draw[thin, fill=white] (1.41421,0.816497) -- (0.707107,0.408248) -- (1.41421,0.) -- (2.12132,0.408248) -- cycle;
        \draw[thin, fill=white] (2.12132,-0.408248) -- (1.41421,-0.816497) -- (2.12132,-1.22474) -- (2.82843,-0.816497) -- cycle;
\end{scope}
\end{scope}
}
\draw[fill=black] (0,0) circle (0.07) (1.5,0) circle (0.07) (3,0) circle (0.07);
\node[below] at (0,-0.2) {$m_1$};
\node[below] at (1.5,-0.2) {$m_2$};
\node[below] (A) at (3,-0.2) {$m_3$};
\node[right] at (A.east) {$\ldots$};
\node[left] at (-1.2,0) {$\Kappa_1\sqcup\Kappa_2\sqcup\Kappa_3\sqcup \ldots\mapsto$};
\end{tikzpicture}
\end{array}
\ee
Unlike shifted quiver Yangian representations considered in \cite{Galakhov:2021xum}, the crystal 2d lattices are shifted with respect to each other along the spectral parameter cylinder for generic vectors that are not an integral combination of $\mathsf{h}_1$, $\mathsf{h}_2$ and the cylinder circumference, therefore the crystals are not intersecting in general. 

We treat a configuration like \eqref{product} as a disjoint union of crystals, we will denote a state corresponding to multiple crystals $\Kappa_i$ growing from different seeds in the following way:
$$
|\Kappa_1+\Kappa_2+\Kappa_3+\dots\rangle \;.
$$
We can observe naturally an analogy between these crystal configurations and partition vectors of \cite{Nekrasov:2003rj}. 
An application to a braiding problem \cite{Smirnov:2013hh} suggests to treat multiple crystal configurations as tensor products of representations associated with individual crystals analogously to \cite[Section 2.7]{Galakhov:2020vyb}.

There is a certain freedom in the definition of the tensor product and the coproduct action on it.
For example, there are different variants of the coproduct on quantum enveloping algebras, obtained  by twisting the canonical coproduct.
Moreover, for any definition there are two canonical ways to define a coproduct that are related by the R-matrix morphism.
For us there is a freedom in choosing a representation.
Although we expect that a bootstrap representation \eqref{bootstrap} and a representation following from the QFT localization \eqref{BPS_rep} are isomorphic, the state norms in these two representations are different.
We will concentrate this freedom in a function $u$ to be defined in such a way that the coproduct will acquire a certain canonical form.
Eventually the tensor product of representations is defined in the following way:
\be\label{tensor}
|\Kappa_1\rangle\otimes|\Kappa_2\rangle\myequiv u(\Kappa_1,\Kappa_2)|\Kappa_1+\Kappa_2\rangle \;.
\ee
In \cite[Section 2.7]{Galakhov:2020vyb} the function $u$ was treated as an additional statistical factor keeping track of ordering crystals in the tensor product.

For multiple tensor factors the relation has the following form:
\be
\bigotimes\lm_i|\Kappa_i\rangle=\left|\sum\lm_i\Kappa_i\right\rangle\prod\lm_{i<j}u(\Kappa_i,\Kappa_j) \;.
\ee

The right hand side of \eqref{tensor} is a vector of an ordinary crystal representation, and for the algebra action on it we can use expressions \eqref{quiver_representation}.
It is natural to treat this action of the generator $g$ on the left hand side of \eqref{tensor} as the coproduct $\Delta(g)$.
Notice that on the right hand side the generators satisfy algebra relations, and, therefore, $\Delta$ is an algebra homomorphism by construction.

Summarizing, we derive the following representation for the coproduct action on the bootstrap representation \eqref{bootstrap}:
\begin{subequations}
\begin{equation}
	\begin{split}
		\Delta\left(\mye^{(a)}(z)\right)&|\Kappa_1\rangle\otimes|\Kappa_2\rangle=\\
		&=\sum\lm_{{\sqbox{$a$}}\in{\rm Add}(\Kappa_1)}\frac{u(\Kappa_1,\Kappa_2)}{u(\Kappa_1+{\sqbox{$a$}},\Kappa_2)}\sigma_+({\sqbox{$a$}},\Kappa_1)\sigma_+({\sqbox{$a$}},\Kappa_2)\times\\ &\quad\times\sqrt{\fq_{a}\tilde\bPsi^{(a)}_{\Kappa_2}\left(h_{{\sqbox{$a$}}}\right)}\myp(z-h_{\sqbox{$a$}})\left[\Kappa_1\to\Kappa_1+{\sqbox{$a$}}\right]|\Kappa_1+{\sqbox{$a$}}\rangle\otimes|\Kappa_2\rangle+\\
		&+\sum\lm_{{\sqbox{$a$}}\in{\rm Add}(\Kappa_2)}\frac{u(\Kappa_1,\Kappa_2)}{u(\Kappa_1,\Kappa_2+{\sqbox{$a$}})}\sigma_+({\sqbox{$a$}},\Kappa_1)\sigma_+({\sqbox{$a$}},\Kappa_2)\times\\ &\quad\times\sqrt{\fq_{a}\tilde\bPsi^{(a)}_{\Kappa_1}\left(h_{{\sqbox{$a$}}}\right)}\fp\left(z-h_{{\sqbox{$a$}}}\right)\left[\Kappa_2\to\Kappa_2+{\sqbox{$a$}}\right]|\Kappa_1\rangle\otimes|\Kappa_2+{\sqbox{$a$}}\rangle \;,
		\end{split}
		\end{equation}
		\begin{equation}
		\begin{split}
		\Delta\left(\myf^{(a)}(z)\right)&|\Kappa_1\rangle\otimes|\Kappa_2\rangle=\\
		&=\sum\lm_{{\sqbox{$a$}}\in{\rm Rem}(\Kappa_1)}\frac{u(\Kappa_1,\Kappa_2)}{u(\Kappa_1-{\sqbox{$a$}},\Kappa_2)}\sigma_-({\sqbox{$a$}},\Kappa_1)\sigma_-({\sqbox{$a$}},\Kappa_2)\times\\ &\quad\times\sqrt{\fq_{a}\tilde\bPsi^{(a)}_{\Kappa_2}\left(h_{{\sqbox{$a$}}}\right)}\myp\left(z-h_{{\sqbox{$a$}}}\right)\left[\Kappa_1\to\Kappa_1-{\sqbox{$a$}}\right]|\Kappa_1-{\sqbox{$a$}}\rangle\otimes|\Kappa_2\rangle+\\
		&+\sum\lm_{k\in\IZ}\sum\lm_{{\sqbox{$a$}}\in\Kappa_2^-(a)}\frac{u(\Kappa_1,\Kappa_2)}{u(\Kappa_1,\Kappa_2-{\sqbox{$a$}})}\sigma_-({\sqbox{$a$}},\Kappa_1)\sigma_-({\sqbox{$a$}},\Kappa_2)\times\\ &\quad\times\sqrt{\fq_{a}\tilde\bPsi^{(a)}_{\Kappa_1}\left(h_{{\sqbox{$a$}}}\right)}\myp\left(z-h_{{\sqbox{$a$}}}\right)\left[\Kappa_2\to\Kappa_2-{\sqbox{$a$}}\right]|\Kappa_1\rangle\otimes|\Kappa_2-{\sqbox{$a$}}\rangle \;,
	\end{split}
	\end{equation}
\end{subequations}
where
\be
\fq_a\myequiv\prod\lm_{I\in\{a\to a\}}\left[-\zeta\left(h_I\right)\right] \;.
\ee
Also we used the following short-hand notations:
\begin{equation}
\begin{split}
    &\tilde\myphi^{a\Leftarrow b}(z,w):=(ZW)^{\frac{\extra}{2}\chi_{ab}}\tilde\varphi^{a\Leftarrow b}(z-w) \;,\\ 
    &\tilde\bPsi^{(a)}_{\Kappa}(z):=\tilde\Psi^{(a)}_{\Kappa}(z) \prod\lm_{b\in Q_0}\prod\lm_{\sqbox{$b$}\in\Kappa}\left(Z H_{\sqbox{$b$}}\right)^{\frac{\extra\chi_{a,b}}{2}} \;.
\end{split}
\end{equation}

We choose the following function $u$:
\be
u(\Kappa_1,\Kappa_2)=\prod\lm_{\sqbox{$a$}_1\in\Kappa_1}\prod\lm_{\sqbox{$b$}_2\in\Kappa_2}\nu_+\left(\sqbox{$b$}_2,\sqbox{$a$}_1\right)\left[\tilde\varphi^{a\Leftarrow b}\left(h_{\sqbox{$a$}_1}-h_{\sqbox{$b$}_2}\right)\right]^{\frac{1}{2}}\;.
\ee
For this choice, the expressions for the generator actions simplify:
\be\label{Delta_interm}
\scalebox{0.9}{
$\bgroup\everymath{\displaystyle}\renewcommand{\arraystretch}{1.5}
\begin{array}{l}
		\Delta\left(\mye^{(a)}(z)\right)|\Kappa_1\rangle\otimes|\Kappa_2\rangle=\\
		=\sum\lm_{{\sqbox{$a$}}\in{\rm Add}(\Kappa_1)}\sigma_+({\sqbox{$a$}},\Kappa_1)\myp\left(z-h_{{\sqbox{$a$}}}\right)\left[\Kappa_1\to\Kappa_1+{\sqbox{$a$}}\right]|\Kappa_1+{\sqbox{$a$}}\rangle\otimes|\Kappa_2\rangle+\\
		+\sum\lm_{{\sqbox{$a$}}\in{\rm Add}(\Kappa_2)}\sigma(\Kappa_1,\Kappa_2)\fq_{a}\Psi^{(a)}_{\Kappa_1}\left(h_{{\sqbox{$a$}}}\right)\sigma_+({\sqbox{$a$}},\Kappa_1)\myp\left(z-h_{{\sqbox{$a$}}}\right)\left[\Kappa_2\to\Kappa_2+{\sqbox{$a$}}\right]|\Kappa_1\rangle\otimes|\Kappa_2+{\sqbox{$a$}}\rangle \;,\\
		\Delta\left(\myf^{(a)}(z)\right)|\Kappa_1\rangle\otimes|\Kappa_2\rangle=\\
		=\sum\lm_{{\sqbox{$a$}}\in{\rm Rem}(\Kappa_1)}\sigma(\Kappa_1,\Kappa_2)\fq_{a}\Psi^{(a)}_{\Kappa_2}\left(h_{{\sqbox{$a$}}}\right)\sigma_-({\sqbox{$a$}},\Kappa_1)\myp\left(z-h_{{\sqbox{$a$}}}\right)\left[\Kappa_1\to\Kappa_1-{\sqbox{$a$}}\right]|\Kappa_1-{\sqbox{$a$}}\rangle\otimes|\Kappa_2\rangle+\\
		+\sum\lm_{{\sqbox{$a$}}\in{\rm Rem}(\Kappa_2)}\sigma_-({\sqbox{$a$}},\Kappa_1)\myp\left(z-h_{{\sqbox{$a$}}}\right)\left[\Kappa_2\to\Kappa_2-{\sqbox{$a$}}\right]|\Kappa_1\rangle\otimes|\Kappa_2-{\sqbox{$a$}}\rangle \;,
\end{array}
\egroup$}
	\ee
where
\be
\sigma(\Kappa_1,\Kappa_2)=\prod\lm_{\sqbox{$a$}_1\in\Kappa_1}\prod\lm_{\sqbox{$a$}_2\in\Kappa_2}(-1)^{|a||b|}
\ee
is a mutual statistics factor for a pair of crystals.

Let us define the following action of graded generators on the tensor product:
\be
\begin{array}{l}
(\mye^{(a)}(z)\otimes 1)|\Kappa_1\rangle\otimes|\Kappa_2\rangle=\left(\mye^{(a)}(z)|\Kappa_1\rangle\right)\otimes|\Kappa_2\rangle  \;,\\
(1\otimes \mye^{(a)}(z))|\Kappa_1\rangle\otimes|\Kappa_2\rangle=\sigma(\Kappa_1,\Kappa_2)|\Kappa_1\rangle\otimes\left(\mye^{(a)}(z)|\Kappa_2\rangle\right) \;,
\\
(\myf^{(a)}(z)\otimes 1)|\Kappa_1\rangle\otimes|\Kappa_2\rangle=\sigma(\Kappa_1,\Kappa_2)\left(\myf^{(a)}(z)|\Kappa_1\rangle\right)\otimes|\Kappa_2\rangle \;, \\
(1\otimes \myf^{(a)}(z))|\Kappa_1\rangle\otimes|\Kappa_2\rangle=|\Kappa_1\rangle\otimes\left(\myf^{(a)}(z)|\Kappa_2\rangle\right) \;.
\end{array}
\ee
The tensor product is defined canonically for homogeneous elements in the following way:
\be
\left(x_1\otimes y_1\right)\left(x_2\otimes y_2\right)=(-1)^{|x_2||y_1|}\left(x_1x_2\right)\otimes \left(y_1y_2\right) \;.
\ee

The function $\Psi$ in expression \eqref{Delta_interm} is a product:
\be
\Psi\left(h_{\sqbox{$a$}}\right)\sim \prod\lm_{\sqbox{$b$}'}\varphi^{a\Leftarrow b}(h_{\sqbox{$a$}}-h_{\sqbox{$b$}'}) \;,
\ee
where $\Box\in\Kappa_1$ and $\Box'\in\Kappa_2$ for the raising generators $\Delta(\mye)$, and the situation is opposite for the lowering generators $\Delta(\myf)$.
We assume that the crystals are widely separated:
\be
|m_1|\ll|m_2|\ll|m_3|\ll\dots \;.
\ee
This allows one to approximate the following ratios:
\be
\begin{array}{cc}
H_{\Box}/H_{\Box'}\sim |m_1|/|m_2|\ll 1 \;,&\mbox{for }\Box\in\Kappa_1,\,\Box'\in\Kappa_2 \;,\\
H_{\Box}/H_{\Box'}\sim |m_2|/|m_1|\gg 1 \;,&\mbox{for }\Box\in\Kappa_2,\,\Box'\in\Kappa_1\;,
\end{array}
\ee
so that the functions $\Psi(h_{\Box})$  in \eqref{Delta_interm} can be represented by the  corresponding convergent series $\left[\Psi(h_{\Box})\right]_{\pm}$.
The Taylor series can be combined with the following Laurent series of the delta function:
$$
\left[\Psi(h_{\Box})\right]_{\pm}\myp(z-h_{\Box})=\left[\Psi(z)\right]_{\pm}\myp(z-h_{\Box})\;.
$$
Using these manipulations, we derive the following closed-form expressions for the coproducts of the raising/lowering generators:
\be
\begin{split}
\Delta\left(\mye^{(a)}(z)\right)&=\mye^{(a)}(z)\otimes 1+\fq_a\myk_-^{(a)}(z)\otimes \mye^{(a)}(z) \;,\\
\Delta\left(\myf^{(a)}(z)\right)&=\fq_a\myf^{(a)}(z)\otimes \myk_+^{(a)}(z)+1\otimes \myf^{(a)}(z)\;.
\end{split}
\ee

The corresponding expressions for the coproducts for the Cartan generators can also be easily derived since the Cartan generators act diagonally in the crystal basis.

Having our quiver BPS algebra $\CA$ we can introduce a co-unit $\varepsilon: \,\CA\to\CA\otimes \CA$ and an antipode $S:\,\CA\to \CA$ satisfying \cite{MR1492989}:
\be
\begin{split}
&m\circ\left(\varepsilon\otimes{\rm id}\right)\circ \Delta=m\circ\left(\varepsilon\otimes{\rm id}\right)\circ \Delta={\rm id}\;,\\
&\left[m\circ(S\otimes{\rm id})\circ \Delta\right](a)=\left[m\circ({\rm id}\otimes S)\circ \Delta\right](a)=\varepsilon(a)\cdot 1\;,
\end{split}
\ee
where $m$ is a multiplication.
Therefore our BPS algebra is a Hopf algebra.

For future reference let us list here the whole set of relations (compare to \cite[Section 4.2]{2019arXiv191208729B}):
\be
\begin{split}
\Delta\left(\mye^{(a)}(z)\right)&=\mye^{(a)}(z)\otimes 1+\fq_a\myk_-^{(a)}(z)\otimes \mye^{(a)}(z) \;,\\
\Delta\left(\myf^{(a)}(z)\right)&=\fq_a\myf^{(a)}(z)\otimes \myk_+^{(a)}(z)+1\otimes \myf^{(a)}(z) \;,\\
\Delta\left(\myk_{\pm}^{(a)}(z)\right)&=\fq_a\myk_{\pm}^{(a)}(z)\otimes \myk_{\pm}^{(a)}(z) \;,\\
\varepsilon\left(\mye^{(a)}(z)\right)&=\varepsilon\left(\myf^{(a)}(z)\right)=0,\;\varepsilon\left(\myk^{(a)}_{\pm}(z)\right)=\fq_a^{-1} \;,\\
S\left(\mye^{(a)}(z)\right)&=-\left(\fq_a\myk_-^{(a)}(z)\right)^{-1}e^{(a)}(z) \;,\\
S\left(\myf^{(a)}(z)\right)&=-f^{(a)}(z)\left(\fq_a\myk_+^{(a)}(z)\right)^{-1} \;,\\
S\left(\myk_\pm^{(a)}(z)\right)&=\left(\fq_a^2\myk_{\pm}^{(a)}(z)\right)^{-1}\;.
\end{split}
\ee

As a result the coproduct $\Delta$ and the co-unit $\varepsilon$ are algebra homomorphisms, whereas the antipode $S$ is an algebra anti-homomorphism:
$$
S(xy)=(-1)^{|x||y|}S(y)S(x)\;.
$$

We should stress that the resulting coproduct structure is a bit naive.
It works well for the quantum toroidal algebras (see e.g.\cite{Bezerra:2019dmp}).
However the canonical coproduct structures chosen for affine Yangians \cite{Prochazka:2015deb} and quantum elliptic algebras \cite{Jimbo:1998bi,Jimbo:1999zz,Mironov:2021sfo} are more intricate.

\def\shuffle{\raisebox{\depth}{\rotatebox{270}{\scalebox{0.8}{$\exists$}}}}
\subsection{Shuffle algebra}

Consider a modification of the Borel positive part of the crystal representation \eqref{quiver_representation} where extra factors are absent:
\be
\hat\mye^{(a)}(z)|\Kappa\rangle=\sum\lm_{\sqbox{$a$}\in{\rm Add}(\Kappa)}\myp(z-h_{\sqbox{$a$}})\left[\Kappa\to\Kappa+\sqbox{$a$}\right] |\Kappa+\sqbox{$a$}\rangle \;.
\ee
These generators satisfy a different algebra:
\be
\hat\mye^{(a)}(z)\hat\mye^{(b)}(w)=\tilde \varphi^{a\Leftarrow b}(z-w)\;\hat\mye^{(b)}(w)\hat\mye^{(a)}(z) \;,
\ee
where the modified bond factor $\tilde\varphi$ is defined in \eqref{tilde_phi}.
We will call this algebra $\hat\myE_{\tau}^+(Q)$.

Following \cite{FT,1995q.alg.....9021F} we define a function:
\be
\lambda_{a,b}(z,w)\myequiv\frac{\prod\lm_{I\in\{b\to a\}}\zeta\left(z-w-h_I\right)}{\zeta\left(z-w\right)^{\delta_{a,b}}}\;.
\ee

Using this definition it is possible to recombine the generator relations as
\be\label{E_plus}
\lambda_{a,b}(z,w)\;\hat\mye^{(a)}(z)\hat\mye^{(b)}(w)=\lambda_{b,a}(w,z)\;\hat\mye^{(b)}(w)\hat\mye^{(a)}(z)\;.
\ee

In what follows we will construct a \emph{shuffle algebra} ${\bf Sh}(Q)$ associated to a quiver $Q$ following \cite{Kontsevich:2010px}.
It will be natural to identify this algebra with an algebra of the Coulomb branch, similar to the one constructed in \cite{Galakhov:2018lta}, whereas the BPS algebra constructed in Section \ref{s:zero_BPS}  can be called an algebra of the Higgs branch (since we performed the localization with respect to the Higgs branch of our quantum field theory).
A generic approach to localization dictates that both ways of localization are expected to give isomorphic Hilbert spaces of BPS states, and this observation is referred to as the \emph{Higgs-Coulomb duality} in the literature.
However, the Higgs-Coulomb duality should be approached with some care (see e.g. \cite{Beaujard:2021fsk}).
The reason is that both Higgs and Coulomb branches may have singularities, so that not all the BPS states are well defined in both frames.
We will treat the following homomorphism:
\be\label{shuffle_homo}
\begin{split}
	\xi:\quad\hat\myE_{\tau}^+(Q)\longrightarrow {\bf Sh}(Q)\;,
\end{split}
\ee
as being reminiscent of the Higgs-Coulomb duality.

In the literature \cite{1995q.alg.....9021F,1998math......9036E,Negut:2020npc,Rapcak:2018nsl,2013arXiv1302.6202N} on the shuffle algebras related to examples of quantum Yangians, toroidal and elliptic algebras, it is proven that the homomorphism \eqref{shuffle_homo} may go both ways, promoting $\xi$ to an algebra isomorphism.

Analogously to \cite{Kontsevich:2010px} we consider a subspace ${\bf Sh}(Q,\vec d)$, where $\vec d$ is a dimension vector of the quiver, being generated by Laurent series in the variables $\{x_{i,a}\}$.
The index $a$ run over the quiver nodes, and $i=1,\ldots, d_a$.
The functions generating ${\bf Sh}(Q,\vec d)$ are symmetric in each group of variables $\{x_{i,a}\}_{i=1}^{d_a}$ labeled by $a\in Q_0$.
This symmetry is reminiscent of the Weyl subgroup of the broken gauge symmetry of the quantum filed theory in question.
The shuffle product is a binary operation acting in the following way:
\be
\shuffle:\quad {\bf Sh}(Q,\vec d)\times {\bf Sh}(Q,\vec d')\longrightarrow {\bf Sh}(Q,\vec d+\vec d')\;.
\ee
Explicitly the shuffle product acts on elements of ${\bf Sh}(Q)$ in the following way (compare to \cite[Theorem 2]{Kontsevich:2010px} and \cite[Proposition 2.3]{MR3805051} for generalized cohomology theories):
\be\label{shuffle_alg}
\begin{split}
&(f\,\shuffle\, g)\left(\{x_{i,a}\}\cup \{x_{i,a}'\}\right)=\\
&=\mathop{\rm Sym}\lm_{\rm shuffles}\left\{f\left(\{x_{i,a}\}\right)g\left(\{x_{i,a}'\}\right)\frac{\prod\lm_{a\in Q_0}\prod\lm_{b\in Q_0}\prod\lm_{I\in\{a\to b\}}\prod\lm_{i=1}^{d_a}\prod\lm_{j=1}^{d_b'}\zeta\left(x_{j,b}'-x_{i,a}-h_I\right)}{\prod\lm_{c\in Q_0}\prod\lm_{i=1}^{d_c}\prod\lm_{j=1}^{d_c'}\zeta\left(x_{j,c}'-x_{i,c}\right)}\right\},
\end{split}
\ee
where the symmetrization is re-shuffling the elements of the groups $\{x_{i,a}\}$ and $\{x_{i,a}'\}$ in such a way that the result is symmetric with respect to each subgroup $\{x_{i,a}\}_{i=1}^{d_a}\cup\{x_{i,a}'\}_{i=1}^{d_a'}$ for each $a\in Q_0$, and the total number of shuffles is
\begin{equation}
    \prod\lm_{a\in Q_0}\left(\begin{array}{c}
     d_a+d_a'  \\
     d_a
\end{array}\right)\,.
\end{equation}

Let us rewrite this definition in a more ``field-theoretic way." Let us introduce a coupling between two types of variables:
\be
\Lambda\left(x_{i,a},x_{j,b}'\right)\myequiv\lambda_{b,a}(x_{j,b},x_{i,a})\;,
\ee
and combine double indices $(i,a)$ in single indices $\alpha$.
Thus one could rewrite \eqref{shuffle_alg} in the following way:
\be\label{shuffle_alg_1}
	(f\,\shuffle\, g)\left(\{x\}\cup \{x'\}\right)=\mathop{\rm Sym}\lm_{\rm shuffles}\left\{f\left(\{x\}\right)g\left(\{x'\}\right)\prod\lm_{\alpha,\beta}\Lambda(x_{\alpha},x_{\beta}')\right\}\,.
\ee
In this form it becomes clear that the shuffle product is \emph{associative}.

The homomorphism $\xi$ is then given by a simple identification of generator modes $\hat \mye_n^{(a)}$ with a multiplication in the shuffle algebra by a monomial $x_{i,a}^n$, where each generator of  $\hat\myE_{\tau}^+(Q)$ corresponds to a new index $i$. 

To justify this definition it is enough to show that simple generators satisfy \eqref{E_plus} and extend it to the whole algebra by the shuffle product associativity.
Indeed notice that for simple generators we have the following relations:
\be
\begin{split}
	&\xi\left[\lambda_{a,b}(z,w)\hat\mye^{(a)}(z)\hat\mye^{(b)}(w)\right]=\lambda_{a,b}(z,w)\,\xi\left[\hat\mye^{(a)}(z)\right]\shuffle\; \xi\left[\hat\mye^{(b)}(w)\right]=\\
	&=\lambda_{a,b}(z,w)\,\left(\sum\lm_{n\in \IZ}\frac{x_{a}^n}{z^n}\right)\shuffle \left(\sum\lm_{k\in \IZ}\frac{y_{b}^k}{w^k}\right)=\lambda_{a,b}(z,w)\,\mathop{\rm Sym}\lm_{\rm shuffles}\left\{\sum\lm_{n\in \IZ}\frac{x_{a}^n}{z^n}\sum\lm_{k\in \IZ}\frac{y_{b}^k}{w^k}\lambda_{b,a}(y_b,x_a)\right\}\\
	&=\lambda_{a,b}(z,w)\lambda_{b,a}(w,z)\,\mathop{\rm Sym}\lm_{\rm shuffles}\left\{\sum\lm_{n\in \IZ}\frac{x_{a}^n}{z^n}\sum\lm_{k\in \IZ}\frac{y_{b}^k}{w^k}\right\}=\xi\left[\lambda_{b,a}(w,z)\hat\mye^{(b)}(w)\hat\mye^{(a)}(z)\right].
\end{split}
\ee

\section{Beyond elliptic BPS algebras: hyperelliptic curves and generalized cohomologies}\label{sec:hyperelliptic}

In our discussion of elliptic and toroidal quiver algebras, 
it has become clear that one of the essential ingredients
is the choice of an odd function $\zeta$: once we fix this we can 
define the generalized Euler class by \eqref{Eul}, and 
compute all the matrix elements of the algebra generators,
thus recovering the representations of the algebra.

In fact, one can be even bolder and try to consider compactifications of 
quiver gauge theories on more general Riemann surfaces $\Sigma$
with genus $g>1$, and consider an even larger class of functions $\zeta$ there. 
Notice that it is indeed possible to consider compactification on such a manifold, by
a partial topological twist of the theory along the surface (cf.\ \cite{Closset:2012ru,Closset:2016arn}).

The fact that surfaces with genus $g>1$ do not have Abelian additive laws defined on them 
is not a problem for us. As discussed around \eqref{Wilson_line},
the equivariant parameters, and hence the spectral parameter, of the algebra
take values in the space of Wilson lines  for the complex gauge fields, and since we are interested in the zero-vorticity sector
the space is the Jacobian $J(\Sigma)\simeq H^1(\Sigma, \mathbb{R})/H^1(\Sigma, \mathbb{Z})$ of the curve $\Sigma$. 
We can then choose an odd function $\zeta$ on the Jacobian $J(\Sigma)$ (which is an Abelian variety and has the additive structure),
and this is enough to define the quiver algebra --- we can define the bond factor by \eqref{eq-charge-atob},
and then define the algebra by the relations \eqref{eq-summary}.

In practice, which function should we choose as $\zeta$? Since we have the theta function for the elliptic case,
a natural generalization is to consider analogues of theta functions for the Jacobian $J(\Sigma)$ as sections of corresponding determinant bundles (see e.g. \cite{Alvarez-Gaume:1986rcs}).
The theta functions depend on the choice of the spin structures or equivalently the choice of the quadratic refinement of the intersection form.
While there are $2^{2g}$ such spin structures, there are $2^{g-1} (2^g-1)$ spin structures (odd spin structures) 
whose associated theta functions are odd. This means that we have a set of $2^{g-1} (2^g-1)$ possible functions $\zeta$, for a genus $g$ surface. 
 Since different odd spin structures are related by the large coordinate transformations,
it follows that we need to consider all the $2^{g-1} (2^g-1)$ choices of $\zeta$ simultaneously.
This is one crucial difference from the genus $g=1$ case.
Moreover, for each choice of the theta function we in practice need to choose the set of $A$ and $B$-cycles, and 
the functions $\zeta$ transform non-trivially under the large coordinate transformations of the Jacobian --- in this sense, the algebras themselves
behave more as a ``wavefunction", rather than a ``partition function". Due to these subtleties it would be of great interest to further study the algebras associated with higher genus surfaces.

In addition to the algebras, we can discuss the geometry, in the language of the associated generalized cohomology theories
for the moduli spaces of quiver gauge theories. To explain this,
let us recall some relevant facts and definitions from algebraic topology.
(Our discussion below will be far from complete, and readers are referred to e.g. \cite{Lurie} for a modern review.)  

Suppose that one has an oriented generalized cohomology theory (GCT) $E^*(-)$. From the 
axioms one finds that $E^*(\mathbb{CP}^{\infty})=E^*(\textrm{pt})[[x]]$ (for an extensive review of generalized cohomology and homology theories see also \cite{may1999concise}), where the generator $x$ represents the first Chern class $c_1(\mathcal{O}(1))$ of the 
universal line bundle $\mathcal{O}(1)$ on $\mathbb{CP}^{\infty}$.
Now, by considering the product of two line bundles we have a map $\mathbb{CP}^{\infty} \times \mathbb{CP}^{\infty} \to \mathbb{CP}^{\infty}$.
When translated into the cohomology theory this defines a map $E^*(\mathbb{CP}^{\infty})\to E^*(\mathbb{CP}^{\infty} \otimes \mathbb{CP}^{\infty}) \simeq E^*(\mathbb{CP}^{\infty}) \otimes E^*(\mathbb{CP}^{\infty})$,
where the last equality follows from the Atiyah-Hirzebruch spectral sequence.
This map is represented by the image $F_E(x\otimes 1, 1\otimes x)$ of the generator $x$, with $F_E(x,y) \in E^{\rm even}[[x,y]]$. It turns out that this image satisfies some
consistency conditions, and is a formal group law.

A  formal group law\footnote{
More precisely, this defines a one-dimensional commutative formal group law. In the following by a formal group law we always mean a one-dimensional commutative formal group law.} over a commutative ring $R$ is a power series $F(x,y)\in R[[x,y]]$,
such that the following three conditions are satisfied:
\begin{align}
(1):\quad &F(x,0) = x  \;, \quad F(0,y)=y \;, \\
(2):\quad &F(x,y) = F(y,x) \;, \\
(3):\quad &F(x, F(y,z) )  = F(F(x,y), z) \;.
\end{align}
The most obvious example of a formal group law is
$F(x, y)= x+y$: this is called the additive formal group law, and represents the ordinary cohomology theory.
While a general formal group law does not represent generalized cohomology theories,
certain necessary and sufficient conditions for representing generalized cohomology theories are known in the literature \cite{MR423332}.

The logarithm $\ell_F$ of a formal group $F$ is an isomorphism from the additive formal group law to the formal group law $F$,
so that 
\begin{align}\label{FGL}
\ell_F'(0)=1 \;, \quad
\ell_F(F(x,y)) =\ell_F(x) + \ell_F(y) \;.
\end{align}
For a given formal group law $F$ we can construct its logarithm by a formal integral
\begin{align}
\ell_F(x) = \int_0^x \left(\frac{\partial F(t,y)}{\partial y}\Bigg|_{y=0}\right)^{-1} dt \;.
\end{align}
For the additive formal group law we have a trivial logarithm
$\ell_F(x)=x$. 

By comparing the logarithm $\ell_F$ with the weight factor $\zeta$ \eqref{eq-def-zeta} playing the central role in our BPS algebra construction, we note that 
the two functions are inverse to each other:
$$
\zeta(u)=\ell_F^{-1}(u)\;.
$$

For simple line bundles the formal group law \eqref{FGL} boils down to a relation between their Chern characters:
\begin{equation}
    \zeta^{-1}\left(c_1\left(L_1\otimes L_2\right)\right)=\zeta^{-1}\left(c_1\left(L_1\right)\right)+\zeta^{-1}\left(c_1\left(L_2\right)\right).
\end{equation}
The function $\zeta$ specifies the relation between the first Chern class of some line bundle and its characteristic that behaves linearly under tensor multiplication.
The physical nature of such a quantity is transparent: it is a charge --- in our particular case when we consider simple flavor bundles this charge is the flavour charge, or the complex mass of the corresponding chiral field, or its equivariant weight.
This physical intuition allows us to restore the relation between an elementary contribution of a weighted subspace in the meson space to the Euler class coincident in this case with the Chern class and its equivariant weight:
\begin{equation}
    c_1(L_{\rm flavor})=\zeta(h_a)\;.
\end{equation}

We can now come back to our discussion of the cohomology theories underlying our algebra.
As the notation suggests, our $\zeta$ for each of the rational/trigonometric/elliptic versions of quiver BPS algebras can be regarded as a particular example of the inverse logarithm of a formal group law
underlying one of the ordinary cohomology/K-theory/elliptic cohomology theories.
We can however consider more general cohomology theories, or more general formal group laws, and we expect that such choices will 
lead to general quiver BPS algebras, including those associated with hyperelliptic curves (see \cite{MR3805051} for a related discussion on a generalized cohomology theory applied to Nakajima quiver varieties and \cite{Aganagic:2016jmx} for an elliptic chomology example).
In other words, we expect to have  a ``quiver BPS algebra of type $F$'' (which we can denote by $\mathsf{F}(Q,W)$) for each formal group law $F$.

Summarizing, we have a chain of concepts 
\begin{align}
\left(\begin{array}{c} \textrm{generalized} \\ \textrm{cohomology} \\ E^*(-)\end{array} \right)
\rightleftarrows  \left(\begin{array}{c} \textrm{formal} \\ \textrm{group law} \\ F_E(x,y) \end{array} \right)
\rightleftarrows  \left(\begin{array}{c} \textrm{logarithm} \\ \ell_F(x)   \\  =\zeta^{-1}(x) \end{array} \right)
\rightleftarrows  \left(\begin{array}{c} \textrm{quiver} \\ \textrm{BPS algebra} \\ \mathsf{F}(Q,W) \end{array} \right)
\;.
\end{align}

It is also worth pointing out that the cobordism can be regarded as a universal cohomology theory \cite{MR73925,MR253350}, and this suggests that 
we can define a ``universal BPS algebra'' directly from oriented cobordism. Related to this, our function $\zeta(u)$ can be expanded as 
\begin{align}
\zeta^{-1}(u) = u + \frac{\varphi(\mathbb{CP}^2)}{3} u^3 +  \frac{\varphi(\mathbb{CP}^4)}{5} u^5  + \dots \in  \mathbb{Q}[[u]] \;.
\end{align}
where $\varphi: \Omega_*^{\rm SO}\to \mathbb{Q}$ is a multiplicative genus from the oriented cobordism $\Omega_*^{\rm SO}$,
and the oriented cobordism classes of complex projective spaces $[\mathbb{CP}^{2n}]$ represent
the polynomial generators of the oriented cobordism with rational coefficients, i.e.\ $\Omega_*^{\rm SO}\otimes \mathbb{Q}$.

If we define a theory on a generic Riemann surface as a spatial slice,
why is the result related to a generalized cohomology theory of the target space?
Although we have already given a somewhat mathematical answer to this question,
another more physical argument
 is that we have already started with a cohomological theory by construction \eqref{Hilbert}.
Ideally, we wish to verify the axioms of GCT directly from this definition.
However, unfortunately, we can not claim that our theory is some GCT $E^*(-)$ since \eqref{Hilbert} is defined for a specific class of target spaces specified by a family of quivers, therefore we are unable to check if defined in this way cohomology $H^*(-,\CQ)$ satisfies the set of GCT axioms for arbitrary CW complexes.
We do not think that this issue is insurmountable.
To put forward our arguments let us make a brief step back to K-theory.
We may claim that the theory of maps from a circle represents K-theory since this situation mimics another well-known scenario in string theory (see e.g.\cite{Minasian:1997mm,Witten:2000cn,Freed:2002qp}).
If we Wick-rotate the temporal direction of the space-time cylinder $S^1\times \IR_t$, the partition function of the resulting configuration represents the Witten index of a BPS string state stretched between two branes.
The brane boundary conditions are well-known to carry the structure of coherent sheaves under derived equivalences of the resulting BPS string states as Chan-Paton factors \cite{D-book_1,D_book_2}.
The Witten indices of these BPS Hilbert spaces correspond to the Grothendieck group $K_0$ of the category of coherent sheaves.
There exists a crucial difference between this setup and the cohomology \eqref{Hilbert}: in the K-theory case we can choose a generic algebraic variety as the support of a stalk produced by a Wilson line operator of a superconnection defined in \cite{Herbst:2008jq} in a theory with, say, a $\IC\IP^n$ target space.
If we were able to present for a theory on a generic Riemann surface a surface operator commuting with the supercharges and  associated with a generic algebraic variety, we might try to argue that a theory defined as a cohomology of maps from a generic Riemann surface to some target space always corresponds to some GCT of the target space by checking whether it satisfies the GCT axioms.
As an eventual outcome of this program we might be able to construct a physically motivated cocycle GCT model (see \cite{2019arXiv190802868B} for such a construction for the equivariant elliptic cohomology theory).

\section*{Acknowledgements}

We would like to thank Mikhail Kapranov and Hiraku Nakajima for interesting discussion on formal group laws, Yegor Zenkevich for interesting discussion on coproduct structures in quantum elliptic algebras.
WL is grateful for support from NSFC No.\ 11875064 and 11947302, CAS Grant No.\ XDPB15, the Max-Planck Partnergruppen fund, and the hospitality of ETH Zurich and Albert-Einstein-Institut (Potsdam). 
The work of MY and DG was supported in part by WPI Research Center Initiative, MEXT, Japan. MY was also supported by the JSPS Grant-in-Aid for Scientific Research (17KK0087, 19K03820, 19H00689, 20H05850, 20H05860).


\appendix

\section{Analysis on  cylinder}\label{app:Cyl}

Throughout this section we use a complex parameterization of the cylinder:
	$$
	z\in\IC^{\times}=\IC\setminus \{0,\infty\} \;.
	$$
	In terms of this, the delta function on the cylinder reads:
	\be
	\delta(z)\myequiv\sum\lm_{k\in\IZ}z^k\,,
	\ee
	which has the following defining property:
	\be
	\delta(z/w)f(z)=\delta(z/w)f(w) \;,
	\ee
	where $f(z)$ is a Laurent series.
	
	Consider a rational function $g(z)$ of the complex variable $z$.
	In general, it is not holomorphic on the cylinder $\IC^{\times}$, therefore there is no unique decomposition in Laurent series.
	We can define two extreme forms of decomposition:
	\be
	\begin{split}
		&	\left[g(z)\right]_+ \myequiv \sum\lm_{k=-\infty}^{\fs_g^+}a_k\, z^k,\quad a_k=-\frac{1}{2\pi\I}\ointctrclockwise\lm_{\CC_{\infty}}\frac{dz}{z^{k+1}}g(z) \;,\\
		&	\left[g(z)\right]_-  \myequiv\sum\lm_{k=\fs_g^-}^{+\infty}b_k\, z^k, \quad b_k=\frac{1}{2\pi\I}\ointctrclockwise\lm_{\CC_{0}}\frac{dz}{z^{k+1}}g(z) \;,
	\end{split}
	\ee
	where $\CC_0$ and $\CC_{\infty}$ are small clockwise contours encircling zero and infinity, respectively, and $\fs_g^\pm$ are finite integers that bound the two sums from above and below correspondingly.
	
	Suppose a function $g(z)$ has the following form:
	\be
	g(z)=\frac{\prod\lm_{i\in\CI}\left(\alpha_i z-\alpha_i^{-1}z^{-1}\right)}{\prod\lm_{j\in\CJ}\left(\beta_j z-\beta_j^{-1}z^{-1}\right)} \;,
	\ee
	where $\CI$ and $\CJ$ are some sets of indices.
	Then we have the following ``residue formula":
	\be
	\left[g(z)\right]_+-\left[g(z)\right]_-=\sum\lm_{k\in\CJ}\left(\lim\lm_{t\to 1}\left(t-t^{-1}\right)g\left(t\beta_k^{-1}\right)\right)\left(z\beta_k\right)^{p}\delta\left(\left(z\beta_k\right)^2\right) \;,
	\ee
	where 
	\be
	p=\left(|\CI|-|\CJ|\right)\;{\rm mod}\; 2 \;.
	\ee
	
	Apparently, this analysis goes through if the function $g(z)$ is multiplied by any monomial $z^k$.
	This observation allows us to extend this formula to the situation when $g(z)$ is multiplied by a Laurent series of $z$: we just apply it term by term.
	
	In a similar fashion we deal with the cases when $g(z)$ is a ratio of theta-functions --- we expand the theta-functions in $q$ and apply this formula to each $q$-term that is represented by a rational function multiplied by a finite Laurent polynomial.
	We apply this tactics whenever we deal with the ratio of theta-functions, hence, in particular, the two decompositions of the theta function read:\footnote{
	Here traditionally $[n]_q=(q^{\frac{n}{2}}-q^{-\frac{n}{2}})/(q^{\frac{1}{2}}-q^{-\frac{1}{2}})$.}
\be
\begin{split}
	\left[\Theta_q(z)\right]_{\pm}=\pm Z^{\pm \frac{1}{2}}\exp\left(\left(q^{\frac{1}{2}}-q^{-\frac{1}{2}}\right)\left[\sum\lm_{k=1}^{\infty}\frac{q^{\mp\frac{k}{2}}Z^{-k}}{k\,[k]_q}+\sum\lm_{k=1}^{\infty}\frac{q^{\pm\frac{k}{2}}Z^{k}}{k\,[k]_q} \right]\right) \;.
\end{split}
\ee

\section{Consistency check of the algebra}
\label{app:consistency}

In this appendix, we check the consistency of the algebra defined in \eqref{eq-summary}. 
\subsection{Mutual consistency of algebraic relations}

The relations in \eqref{eq-summary} can be classified into three groups:
    \begin{enumerate}
        \item $\psi_{\pm}-\psi_{\pm}$ relations:
        \begin{equation}
          	\begin{aligned}\label{eq-pp}
			\myk^{(a)}_{\epsilon}(z)\, \myk^{(b)}_{\epsilon}(w) &\simeq C^{\epsilon\,\extra\,\chi_{ab}}\,\myk^{(b)}_{\epsilon}(w)\, \myk^{(a)}_{\epsilon}(z) \qquad \epsilon = \pm \,, \\
			\myk^{(a)}_{+}(z)\, \myk^{(b)}_{-}(w) &\simeq \frac{ \myphi^{a\Leftarrow b}\left(z+\frac{c}{2},w-\frac{c}{2}\right) }{\myphi^{a\Leftarrow b}\left(z-\frac{c}{2},w+\frac{c}{2}\right)}\, \myk^{(b)}_{-}(w)\, \myk^{(a)}_{+}(z)\,.
		\end{aligned}  
        \end{equation}
               \item $\psi_{\pm}-e$ and $\psi_{\pm}-f$ relations:
        \begin{equation}
          	\begin{aligned}\label{eq-pepf}
			\myk^{(a)}_{\pm}(z)\mye^{(b)}(w) &\simeq \myphi^{a\Leftarrow b}\left(z\pm \frac{c}{2},w\right) \mye^{(b)}(w) \myk^{(a)}_{\pm}(z)\,,\\
			\myk^{(a)}_{\pm}(z)\myf^{(b)}(w) &\simeq \myphi^{a\Leftarrow b}\left(z\mp \frac{c}{2},w\right)^{-1} \myf^{(b)}(w) \myk^{(a)}_{\pm}(z)\,.
		\end{aligned}  
        \end{equation}
               \item $e-e$, $f-f$, and $e-f$ relations:
        \begin{equation}
          	\begin{aligned}\label{eq-ef}
			\mye^{(a)}(z)\mye^{(b)}(w) &\simeq (-1)^{|a||b|}\,\myphi^{a\Leftarrow b}(z,w) \mye^{(b)}(w) \mye^{(a)}(z)\,,\\
			\myf^{(a)}(z)\myf^{(b)}(w) &\simeq (-1)^{|a||b|}\, \myphi^{a\Leftarrow b}(z,w)^{-1} \myf^{(b)}(w) \myf^{(a)}(z)\,,\\
			\left[\mye^{(a)}(z)\,,\myf^{(b)}(w)\right\}&\simeq -\delta_{a,b} \left(\myp(\Delta-c) \myk^{(a)}_{+}\left(z-\frac{c}{2}\right)- \myp(\Delta+c)  \myk^{(a)}_{-}\left(w-\frac{c}{2}\right) \right)\,,
		\end{aligned}  
        \end{equation}
    \end{enumerate}
We will start with the equations in group-$3$, and first derive the equations in group-$2$, and finally using the equations in group-$3$ and group-$2$ to derive those in group-$1$.

First, write the last equation in \eqref{eq-ef} (with a slight change of variables) as $L^{a,b}(z_1,z_2)\simeq R^{a,b}(z_1,z_2)$, with
\begin{equation}\label{eq-LR}
\begin{aligned}
 L^{a,b}(z_1,z_2)&\myequiv\left[\mye^{(a)}(z_1)\,,\myf^{(b)}(z_2)\right\} \;,
\\
 R^{a,b}(z_1,z_2)&\myequiv-\delta_{a,b} \left(\myp(z_1-z_2-c) \myk^{(a)}_{+}\left(z_1-\frac{c}{2}\right)- \myp(z_1-z_2+c)  \myk^{(a)}_{-}\left(z_2-\frac{c}{2}\right) \right)\,.
\end{aligned}
\end{equation}

Let us consider the product $L^{a,b}(z_1,z_2)\, e^{(d)}(w)$: passing the $e^{(d)}(w)$ to the other side of $L^{a,b}(z_1,z_2)$ using the $e-e$ and $e-f$ relations, we have
\begin{equation}\label{eq-Le}
    \begin{aligned}
    & L^{a,b}(z_1,z_2)\, e^{(d)}(w)=(-1)^{|d|(|a|+|b|)}\big(\myphi^{a\Leftarrow d}(z_1,w)\,   e^{(d)}(w)\,L^{a,b}(z_1,z_2)\\
    & \qquad -(-1)^{|a||d|}  e^{(a)}(z_1)\,L^{d,b}(w,z_2)+(-1)^{|a||b|}\myphi^{a\Leftarrow d}(z_1,w)\,L^{d,b}(w,z_2)\,e^{(a)}(z_1)\big) \;.
    \end{aligned}
\end{equation}
Using $L^{a,b}(z_1,z_2)\simeq R^{a,b}(z_1,z_2)$ to replace all the $L$'s in \eqref{eq-Le} by the $R$'s (with appropriate superscripts and variables), we have
\begin{equation}\label{eq-Re}
    \begin{aligned}
    & R^{a,b}(z_1,z_2)\, e^{(d)}(w)=(-1)^{|d|(|a|+|b|)}\big(\myphi^{a\Leftarrow d}(z_1,w)\,   e^{(d)}(w)\,L^{a,b}(z_1,z_2)\\
    & \qquad -(-1)^{|a||d|}  e^{(a)}(z_1)\,R^{d,b}(w,z_2)+(-1)^{|a||b|}\myphi^{a\Leftarrow d}(z_1,w)\,R^{d,b}(w,z_2)\,e^{(a)}(z_1)\big) \;.
    \end{aligned}
\end{equation}
Plugging in the definition of $R$'s (the 2nd equation of \eqref{eq-LR}) into \eqref{eq-Re}, we obtain a relation that involves $\psi_{\pm}$ and $e$.
Focusing on each delta function, one thus obtain the $\psi_{\pm}-e$ relations in \eqref{eq-pepf}.
In particular, the last line in \eqref{eq-Re} vanishes due to the reciprocity condition \eqref{eq-reciprocity-final} of the bond factor.

Now, repeating the procedure above, but replacing $e^{(b)}(w)$ by $f^{(b)}(w)$, we obtain the $\psi_{\pm}-f$ relations in \eqref{eq-pepf}.
We have thus obtained all the equations in group-2.

Finally, let's derive the equations in group-1. 
Again, we use the relation $L^{a,b}(z_1,z_2)\simeq R^{a,b}(z_1,z_2)$ with \eqref{eq-LR}.
First, consider the product $L^{a,b}(z_1,z_2)\, \psi_{\pm}^{(d)}(w)$: passing the $\psi_{\pm}^{(d)}(w)$ to the other side of $L^{a,b}(z_1,z_2)$ using the $\psi_{\pm}-e$ and $\psi_{\pm}-f$ relations that we just derived, we have
\begin{equation}\label{eq-Lp}
    \begin{aligned}
    & L^{a,b}(z_1,z_2)\, \psi_{\pm}^{(d)}(w)=\frac{ \myphi^{d\Leftarrow b}\left(w\mp\frac{c}{2},z_2\right) }{\myphi^{d\Leftarrow a}\left(w\pm\frac{c}{2},z_1\right)}\,   \psi_{\pm}^{(d)}(w) \,L^{a,b}(z_1,z_2) \;.
    \end{aligned}
\end{equation}
Using $L^{a,b}(z_1,z_2)\simeq R^{a,b}(z_1,z_2)$ to replace both $L^{a,b}(z_1,z_2)$ in \eqref{eq-Lp} by  $R^{a,b}(z_1,z_2)$, we have
\begin{equation}\label{eq-Rp}
   \begin{aligned}
    & R^{a,b}(z_1,z_2)\, \psi_{\pm}^{(d)}(w)=\frac{ \myphi^{d\Leftarrow b}\left(w\mp\frac{c}{2},z_2\right) }{\myphi^{d\Leftarrow a}\left(w\pm\frac{c}{2},z_1\right)}\,   \psi_{\pm}^{(d)}(w) \,R^{a,b}(z_1,z_2) \;.
    \end{aligned}
\end{equation}
Then plugging in the expression of  $R^{a,b}(z_1,z_2)$ from the second equation of \eqref{eq-LR} into \eqref{eq-Rp}, we obtain a relation among $\psi_{\pm}$.
Focusing on each delta function, one thus obtains the $\psi_{\pm}-\psi_{\pm}$ relations in \eqref{eq-pp}.
This concludes our check on the mutual consistency of the algebra \eqref{eq-summary}.

\subsection{Consistency of central extension}

Although we have just checked explicitly the mutual consistency of the algebraic relations in \eqref{eq-summary}, let us discuss further the potential subtlety of the central extension.
Ultimately, the check that the central extension is possible again boils down to checking the mutual consistency of the algebra \eqref{eq-summary} when $c\neq 0$.

	In \eqref{eq-summary} we have introduced a central extension of the toroidal and elliptic quiver BPS algebras by a central element $c$.
	These relations are a natural algebra deformation based on known extensions in quantum toroidal \cite{2019arXiv191208729B} and elliptic algebras \cite{Konno:2016fmh}.
	For the case $c=0$ we have constructed explicitly a non-trivial crystal representation.
	However this setup raises a natural question whether a generic elliptic quiver BPS algebra for $c\neq 0$ constructed as a quotient of a free algebra of generators $\left(\mye^{(a)}(z),\myf^{(a)}(z),\myk^{(a)}(z)\right)$ by the relations \eqref{eq-summary} is non-empty.
	
	To convince ourselves that the resulting algebra is non-trivial we propose the following construction mechanism.
	We start with a free associative algebra $A_0$ of only $\mye^{(a)}(z)$ and $\myf^{(a)}(z)$ elements.
	If one considers further the quotient $A_1$ of this algebra by the $\mye$-$\mye$ and $\myf$-$\myf$ relations in \eqref{eq-summary}, the resulting quotient is non-zero since the $\mye$-$\mye$ and $\myf$-$\myf$ relations are not affected by the central extension.
	Then for all commutators in the resulting algebra we introduce two novel generators $\myk_{\pm}^{(a)}(z)$ according to the rule dictated by the $\mye$-$\myf$ relation in \eqref{eq-summary}.
	This last operation is a standard introduction of a central extension.
	One can switch the order of taking the quotient and first take the quotient of $A_0$ by the $\mye$-$\myf$ relation to get $A_2$.
	The only information on the quiver entering $A_2$ is the number of generators and their statistics, and we can therefore easily mimic $A_2$ by the generators and relations of the suitable quantum toroidal algebra associated with $\widehat{\fg\fl}_{m|n}$ that is known to be non-empty.
		
	We have seen in the previous subsection that the remaining $\myk$-$\mye$, $\myk$-$\myf$ and $\myk$-$\myk$ relations can be derived from the $\mye$-$\mye$, $\myf$-$\myf$, $\mye$-$\myf$ relations.
	Since the $c\neq 0$ case is actually slightly simpler than the general case, we repeat the computation for $c\neq 0$ here again.
	
	First let us note that when $c\neq 0$, the $\mye$-$\myf$ relation allows us to define $\myk_{\pm}$ as the corresponding contact terms:
	\be\label{contact}
	\psi_{\pm}^{(a)}(z)\sim -\myp(0)^{-1}\left[\mye^{(a)}\left(z\pm c/2\right),\myf^{(a)}\left(z\mp c/2\right)\right\}.
	\ee
	\noindent It is a simple exercise to derive the remaining relations for $a\neq b$.
	
	For coincident indices $a=b$ let us multiply the last string in \eqref{eq-summary} by a generator $e^{(a)}(u)$ from the left.
	After massaging the resulting expression a little we arrive at the following relation:
	\be\label{long_eq}
	\begin{split}
		(-1)^{|a||a|}&\myp\left(u-w-c\right)\left(\psi_+^{(a)}\,\left(u-\frac{c}{2}\right)\,e^{(a)}(z)-\myphi^{a\Leftarrow a}(u,z)e^{(a)}(z)\,\psi_+^{(a)}\left(u-\frac{c}{2}\right)\right)\\
		-(-1)^{|a||a|}&\myp\left(u-w+c\right)\left(\psi_-^{(a)}\left(w-\frac{c}{2}\right)\,e^{(a)}(z)-\myphi^{a\Leftarrow a}(u,z)e^{(a)}(z)\,\psi_-^{(a)}\left(w-\frac{c}{2}\right)\right)\\
		+&\myp\left(z-w-c\right)\left(e^{(a)}(u)\,\psi_+^{(a)}\left(z-\frac{c}{2}\right)-\myphi^{a\Leftarrow a}(u,z)\,\psi_+^{(a)}\left(z-\frac{c}{2}\right)\,e^{(a)}(u)\right)\\
		-&\myp\left(z-w+c\right)\left(e^{(a)}(u)\,\psi_-^{(a)}\left(w-\frac{c}{2}\right)-\myphi^{a\Leftarrow a}(u,z)\,\psi_-^{(a)}\left(w-\frac{c}{2}\right)\,e^{(a)}(u)\right)=0 \;.
	\end{split}
	\ee
	Then, colliding different pairs of points out of a triplet $(z,w,u)$ such that the delta functions in the l.h.s. of \eqref{long_eq} diverge, we derive the $\myk$-$\mye$ relations for coincident indices $a=b$.
	The $\myk$-$\myf$  relations can be derived in the same fashion.
	Eventually, all the $\myk$-$\myk$ relations follow from the $\myk$-$\mye$ and $\myk$-$\myf$ relations and the identification \eqref{contact}.

\section{Free field realization for \texorpdfstring{$\myk_{\pm}$}{}}\label{app:free_field}

Before proceeding let us introduce a useful notation for the bond potential:
\be
U_k^{a\Leftarrow b}:=\sum\lm_{I\in\{a\to b\}}H_I^{-k}-\sum\lm_{J\in\{b\to a\}}H_J^{k},\quad |k|\geq 1\;.
\ee
We note that this bond potential function satisfies the following symmetry property:
\be
U_k^{a\Leftarrow b}=-U_{-k}^{a\Leftarrow b} \;. 
\ee
In addition we extend the notion of the bond potential to the value $k=0$:
\be
U_0^{a\Leftarrow b}:=\sum\lm_{I\in\{a\to b\}}\log\,H_I+\sum\lm_{J\in\{b\to a\}}\log \,H_J \;.
\ee

In terms of this notation, two Laurent expansions of the bare bond factors can be written in the following form:
\be
\left[\varphi^{a\Leftarrow b}(z)\right]_\pm=(-1)^{\frac{1\mp 1}{2}\chi_{ab}}Z^{\pm\frac{\chi_{ab}}{2}}e^{\pm\frac{1}{2}U_0^{a \Leftarrow b}}\exp\left(-\sum\lm_{k=1}^{\infty}\frac{Z^{\mp k}}{k}U_k^{a\Leftarrow b}\right)\;.
\ee

In this section we will construct a free field representation for a subalgebra generated by the generators $\myk_\pm^{(a)}(z)$ similarly to \cite{1996q.alg....11030S}.
Unfortunately, unlike the field OPE \eqref{eq-summary} this representation is not universal with respect to the hierarchy \eqref{red_diagram}.
We will restrict ourselves to $\myT(Q)$, assuming that the corresponding representation for $\myE(Q)$ can be derived along the lines of \cite{konno2009elliptic}.

In this representation, the expressions for the fields $\myk_\pm^{(a)}(z)$ can be rewritten in terms of the new generators:
\be
\begin{split}
\myk_+^{(a)}(z)&=\gamma_+^{(a)}\cdot\exp\left(-\sum\lm_{k=1}^{\infty}\frac{\alpha_k^{(a)}}{k}Z^{-k}\right) \;,\\
\myk_-^{(a)}(z)&=\gamma_-^{(a)}\cdot\exp\left(-\sum\lm_{k=1}^{\infty}\frac{\alpha_{-k}^{(a)}}{k}Z^{k}+\beta^{(a)}\log\, Z\right)\;.
\end{split}
\ee
Here the new generators form a Heisenberg algebra with relations:
\be
\begin{split}
&\left[\alpha_k^{(a)},\alpha_m^{(b)}\right]=\delta_{k+m} k\left(C^k-C^{-k}\right)U_k^{a\Leftarrow b}\;,\\
&\gamma_+^{(a)}\gamma_-^{(b)}=C^2\,\gamma_-^{(b)}\gamma_+^{(a)},\quad \gamma_+^{(a)}\gamma_+^{(b)}=C^{\chi_{ab}}\gamma_+^{(b)}\gamma_+^{(a)},\quad \gamma_-^{(a)}\gamma_-^{(b)}=C^{-\chi_{ab}}\gamma_-^{(b)}\gamma_-^{(a)}
\end{split}
\ee
and all the other commutators are trivial.

The commutators of the Heisenberg algebra generators with the operators $\mye^{(a)}(z)$ and $\myf^{(a)}(z)$ read:
\begin{equation}
	\begin{split}
		\left[\alpha_k^{(a)},\mye^{(b)}(z)\right]=Z^k C^{-\frac{k}{2}}\,U_k^{a\Leftarrow b}\cdot \mye^{(b)}(z) \;,\\
		\left[\alpha_k^{(a)},\myf^{(b)}(z)\right]=Z^k C^{\frac{k}{2}}\,U_k^{a\Leftarrow b}\cdot \myf^{(b)}(z) \;,\\
		\left[\beta^{(a)},\mye^{(b)}(z)\right]=\chi_{ab}\,\mye^{(b)}(z),\quad \left[\beta^{(a)},\myf^{(b)}(z)\right]=-\chi_{ab}\,\myf^{(b)}(z) \;,\\
		\gamma_+^{(a)}\mye^{(b)}(z)=C^{\frac{\chi_{ab}}{2}}e^{\frac{1}{2}U_0^{a\Leftarrow b}}\,\mye^{(b)}(z)\gamma_+^{(a)} \;,\\
		\gamma_+^{(a)}\myf^{(b)}(z)=C^{\frac{\chi_{ab}}{2}}e^{-\frac{1}{2}U_0^{a\Leftarrow b}}\,\myf^{(b)}(z)\gamma_+^{(a)} \;,\\
		\gamma_-^{(a)}\mye^{(b)}(z)=(-Z)^{\chi_{ab}}e^{-\frac{1}{2}U_0^{a\Leftarrow b}}\,\mye^{(b)}(z)\gamma_-^{(a)} \;,\\
		\gamma_-^{(a)}\myf^{(b)}(z)=(-Z)^{\chi_{ab}}e^{\frac{1}{2}U_0^{a\Leftarrow b}}\,\myf^{(b)}(z)\gamma_-^{(a)}\;.
	\end{split}
\end{equation}

For the case $c=0$ the Heisenberg algebra becomes commutative and we can rewrite the operators in terms of Wilson operators defined in \eqref{Wilson_line} with the help of additional operators $d_a$ that are diagonal in the crystal basis:
$$
d_a|\Kappa\rangle=\left|\Kappa^{(a)}\right|\cdot |\Kappa\rangle \;,
$$
thus we have:
\be
\begin{split}
	&\alpha_k^{(a)}=\sum\lm_{b\in Q_0}U_k^{a\Leftarrow b}\;\Tr\,{\bf W}_b^k\;,
	\quad \beta^{(a)}=e^{\sum\lm_{b\in Q_0}\chi_{ab}d_b}\;,\\
	&\gamma_+^{(a)}=e^{\frac{1}{2}\sum\lm_{b\in Q_0}U_0^{a\Leftarrow b}d_b} \;,
	\quad \gamma_-^{(a)}=e^{\sum\lm_{b\in Q_0}\left(\chi_{ab}\Tr\,\log\left(-{\bf W}_b\right)-\frac{1}{2}U_0^{a\Leftarrow b}d_b\right)}\;.
\end{split}
\ee


\section{On Serre relations for \texorpdfstring{$\myT_{\beta}\left(K_{\IP^2}\right)$}{TKP2} and \texorpdfstring{$\myT_{\beta}\left(K_{\IP^1\times \IP^1}\right)$}{TKP1xP1}}\label{app:Serre}
	
	Calabi-Yau threefolds in general may have chiral unframed quivers $Q$, corresponding quiver algebras $\myA(Q)$ can not be associated with a deformation of some affine Lie superalgebra $\widehat{\fg\fl}_{m|n}$.
	Here by $\myA$ we imply any of $\myY$, $\myT_{\beta}$ or $\myE_{\tau}$.
	This issue makes it difficult to assign some Serre relations to them.
	However it is possible to summarize some set of rules the Serre relations obey for quiver algebras associated with $\widehat{\fg\fl}_{m|n}$:
	\begin{enumerate}
		\item The Serre relations are automatically satisfied for an analytic continuation of \eqref{eq-summary} when $\mye$-$\mye$ relations are treated as an equality.
		\item The Serre relations for $\myA=\myY$ are independent of the $\Omega$-background parameters $\mathsf{h}_1$, $\mathsf{h}_2$.
		\item Consider the lowest weight module $\mathscr{M}_i$ of $\underline{\myA}(Q):=\myA(Q)/\{\mbox{Serre relations}\}$ satisfying 
		$$
		\mye^{(a)}(z)|\varnothing\rangle\sim \dfrac{\delta_{a,i}}{z}|\varnothing\rangle,\quad f^{(a)}(z)|\varnothing\rangle=0\,.
		$$
		Module $\mathscr{M}_i$ is isomorphic to a canonical crystal representation, where the canonical crystal starts its growth from an atom of color $i$.
		For details on this property of Serre relations see \cite[sections 5.3, 8.2.3.2, appendix A]{Li:2020rij}.
	\end{enumerate}
	We could take this set of rules as a guiding principle and suggest relations allowing one to reduce the corresponding algebra further consistently even for the cases of chiral quivers. 
	By analogy, we will call these relations Serre relations for our quiver algebras.
	Unfortunately, at present we do not have a generic algorithm to construct the Serre relations for an arbitrary quiver algebra $\myA(Q)$, however we can present corresponding Serre relations for few examples associated with chiral quivers to demonstrate that such a construction is valid in principle.
	
	We will consider examples of Calabi-Yau threefolds $K_{\IP^2}$ and $K_{\IP^1\times\IP^1}$ both having a compact four-cycle.
	Corresponding quivers and superpotentials read:
	\begingroup \renewcommand{\arraystretch}{1.4}
	\begin{equation}
		\begin{array}{c|c}
			K_{\IP^2} & K_{\IP^1\times\IP^1}\\
			\hline
			\begin{array}{c}
				\begin{tikzpicture}[scale=1]
					\begin{scope}[rotate=0]
						\begin{scope}[shift={(0,0.57735)}]
							\draw[postaction={decorate},decoration={markings, 
								mark= at position 0.65 with {\arrow{stealth}}}] (-1,0) -- (1,0);
							\draw[postaction={decorate},decoration={markings, 
								mark= at position 0.65 with {\arrow{stealth}}}] (-1,0) to[out=15,in=165] node[pos=0.5,above] {\tiny $X_{i=1,2,3}$} (1,0);
							\draw[postaction={decorate},decoration={markings, 
								mark= at position 0.65 with {\arrow{stealth}}}] (-1,0) to[out=345,in=195] (1,0);
						\end{scope}
					\end{scope}
					\begin{scope}[rotate=120]
						\begin{scope}[shift={(0,0.57735)}]
							\draw[postaction={decorate},decoration={markings, 
								mark= at position 0.65 with {\arrow{stealth}}}] (-1,0) -- (1,0);
							\draw[postaction={decorate},decoration={markings, 
								mark= at position 0.65 with {\arrow{stealth}}}] (-1,0) to[out=15,in=165] node[pos=0.5,below left] {\tiny $Z_{i=1,2,3}$} (1,0);
							\draw[postaction={decorate},decoration={markings, 
								mark= at position 0.65 with {\arrow{stealth}}}] (-1,0) to[out=345,in=195] (1,0);
						\end{scope}
					\end{scope}
					\begin{scope}[rotate=240]
						\begin{scope}[shift={(0,0.57735)}]
							\draw[postaction={decorate},decoration={markings, 
								mark= at position 0.65 with {\arrow{stealth}}}] (-1,0) -- (1,0);
							\draw[postaction={decorate},decoration={markings, 
								mark= at position 0.65 with {\arrow{stealth}}}] (-1,0) to[out=15,in=165] node[pos=0.5,below right] {\tiny $Y_{i=1,2,3}$} (1,0);
							\draw[postaction={decorate},decoration={markings, 
								mark= at position 0.65 with {\arrow{stealth}}}] (-1,0) to[out=345,in=195] (1,0);
						\end{scope}
					\end{scope}
					\foreach \i/\j in {0/1,1/3,2/2}
					{
						\begin{scope}[rotate = 120 * \i]
							\draw[fill=\myblue] (-1,0.57735) circle (0.15);
							\node at (-1.3, 0.750555) {\j};
						\end{scope}
					}
				\end{tikzpicture}
			\end{array} & \begin{array}{c}
			\begin{tikzpicture}[scale=1]
				\begin{scope}[rotate=0]
					\begin{scope}[shift={(0,1)}]
						\draw[postaction={decorate},decoration={markings, 
							mark= at position 0.65 with {\arrow{stealth}}}] (-1,0) to[out=10,in=170] node[pos=0.5,above] {\tiny $A_{i=1,2}$} (1,0);
						\draw[postaction={decorate},decoration={markings, 
							mark= at position 0.65 with {\arrow{stealth}}}] (-1,0) to[out=350,in=190] (1,0);
					\end{scope}
				\end{scope}
				\begin{scope}[rotate=90]
					\begin{scope}[shift={(0,1)}]
						\draw[postaction={decorate},decoration={markings, 
							mark= at position 0.65 with {\arrow{stealth}}}] (-1,0) to[out=10,in=170] node[pos=0.5,left] {\tiny $D_{i=1,2}$} (1,0);
						\draw[postaction={decorate},decoration={markings, 
							mark= at position 0.65 with {\arrow{stealth}}}] (-1,0) to[out=350,in=190] (1,0);
					\end{scope}
				\end{scope}
				\begin{scope}[rotate=180]
					\begin{scope}[shift={(0,1)}]
						\draw[postaction={decorate},decoration={markings, 
							mark= at position 0.65 with {\arrow{stealth}}}] (-1,0) to[out=10,in=170] node[pos=0.5,below] {\tiny $C_{i=1,2}$} (1,0);
						\draw[postaction={decorate},decoration={markings, 
							mark= at position 0.65 with {\arrow{stealth}}}] (-1,0) to[out=350,in=190] (1,0);
					\end{scope}
				\end{scope}
				\begin{scope}[rotate=270]
					\begin{scope}[shift={(0,1)}]
						\draw[postaction={decorate},decoration={markings, 
							mark= at position 0.65 with {\arrow{stealth}}}] (-1,0) to[out=10,in=170] node[pos=0.5,right] {\tiny $B_{i=1,2}$} (1,0);
						\draw[postaction={decorate},decoration={markings, 
							mark= at position 0.65 with {\arrow{stealth}}}] (-1,0) to[out=350,in=190] (1,0);
					\end{scope}
				\end{scope}
				\foreach \i/\j in {0/1,1/4,2/3,3/2}
				{
					\begin{scope}[rotate = 90 * \i]
						\draw[fill=\myblue] (-1,1) circle (0.15);
						\node at (-1.3,1.3) {\j};
					\end{scope}
				}
			\end{tikzpicture}
		\end{array}\\
		\hline
		W=\sum\lm_{i,j,k}\epsilon^{ijk}\Tr\left(Z_iY_jX_k\right) & W=\Tr\left(\sum\lm_i D_iC_iB_iA_i-\sum\lm_{i\neq j}D_iC_jB_iA_j\right)
		\end{array}
	\end{equation}
	\endgroup
	
	Correspondingly, we propose the following sets of Serre relations for $\myY\left(K_{\IP^2}\right)$ and $\myY\left(K_{\IP^1\times\IP^1}\right)$:
	\begin{equation}\label{Serre:rational}
		\begin{split}
			&\mathop{\rm CycSym}_{\{1,2,3\}}\;\left(z_1-2z_2+z_3\right)\;\mye^{(3)}(z_3)\, \mye^{(2)}(z_2)\, \mye^{(1)}(z_1)=0\,,\\
			&\mathop{\rm CycSym}_{\{1,2,3,4\}}\;\left(z_1-z_2-z_3+z_4\right)\;\mye^{(4)}(z_4)\, \mye^{(3)}(z_3)\, \mye^{(2)}(z_2)\, \mye^{(1)}(z_1)=0\,,
		\end{split}
	\end{equation}
	\noindent and for $\myT_{\beta}\left(K_{\IP^2}\right)$ and $\myT_{\beta}\left(K_{\IP^1\times \IP^1}\right)$ respectively:
	\begin{equation}\label{Serre:trigonometric}
		\begin{split}
			&\mathop{\rm CycSym}_{\{1,2,3\}}\;\left(1-\frac{Z_2^2}{Z_1Z_3}\right)\;\mye^{(3)}(z_3)\, \mye^{(2)}(z_2)\, \mye^{(1)}(z_1)=0\,,\\
			&\mathop{\rm CycSym}_{\{1,2,3,4\}}\;\left(1-\frac{Z_2Z_3}{Z_1Z_4}\right)\;\mye^{(4)}(z_4)\, \mye^{(3)}(z_3)\, \mye^{(2)}(z_2)\, \mye^{(1)}(z_1)=0\,.
		\end{split}
	\end{equation}
	\noindent Here by a symbol ${\rm CycSym}$ we imply a symmetrization with respect to all cyclic permutations of respective sets of indices and superscripts.
	
	These Serre relations \eqref{Serre:rational} and \eqref{Serre:trigonometric} are automatically satisfied for analytic continuation of the quiver algebras. Furthermore, they reduce the numbers of states at modules $\mathscr{M}_i$ to those of canonical molten crystals up to level 5 in both cases. For the number of states in modules $\mathscr{M}_i$ to match the number of molten crystals one needs to apply higher order Serre relations.


\section{Sign equations}\label{app:sign}

In this section we describe a method to solve the equations \eqref{epsilon_equation}.

The crystal atoms have internal statistics, therefore processes re-shuffling atoms may lead to additional $\pm$ signs multiplying the crystal wave functions.
Let us consider a process of adding/removing an atom $\sqbox{$a$}$ to/from the crystal $\Kappa$. This process produces a statistical multiplier that we denote as $P(\sqbox{$a$},\Kappa)$. 
Now suppose we add/remove first an atom $\sqbox{$a$}$ then another atom $\sqbox{$b$}$.
Then if we apply the same set of operations to the atoms $\sqbox{$a$}$ and $\sqbox{$b$}$ but in the reverse order, the result might be different:
\begin{equation}\label{sign_equation}
    \frac{P\left(\sqbox{$a$},\Kappa\right)P\left(\sqbox{$b$},\Kappa\pm\sqbox{$a$}\right)}{P\left(\sqbox{$b$},\Kappa\right)P\left(\sqbox{$a$},\Kappa\pm\sqbox{$b$}\right)}=(-1)^{\mathfrak{f}_{ab}}\,,
\end{equation}
where the function $\mathfrak{f}_{ab}$ only depends on the atom colors $a$ and $b$ and satisfies
\begin{equation}
    \mathfrak{f}_{ab}+\mathfrak{f}_{ba}=0\; ({\rm mod}\;2)\;.
\end{equation}

Following \cite[section 2.7]{Galakhov:2020vyb} we search for the solution of \eqref{sign_equation} using the following ansatz:
\begin{equation}\label{P_expression}
    P\left(\Box,\Kappa\right)=\prod\lm_{\Box'\in\Kappa}\upsilon\left(\Box,\Box'\right)\;,
\end{equation}
where the unknown function $\upsilon$ is a function of a pair of atoms, with the boundary condition $\upsilon\left(\Box,\Box\right)\equiv 1$.

Substituting \eqref{P_expression} into \eqref{sign_equation}, we derive the defining equation for $\upsilon$:
\begin{equation}\label{upsilon_eq}
    \upsilon\left(\sqbox{$a$}_1,\sqbox{$b$}_2\right)=(-1)^{\mathfrak{f}_{ab}}\upsilon\left(\sqbox{$b$}_2,\sqbox{$a$}_1\right) \;,
\end{equation}
where subscripts $1,2$ are used to stress that the atoms contributing to this equation never coincide, therefore the functions $\mathfrak{f}_{ab}$ are \emph{not} subject to the further constraint $\mathfrak{f}_{aa}=0\;({\rm mod}\;2)$.

To solve \eqref{upsilon_eq} further, let us assign a global ordering to  a generic set of atoms in the canonical crystal lattice, for example, by following a lexicographic ordering $<$ of monomials in the quiver path algebra associated to crystal atoms. We then introduce a function:
\be\label{sign_function}
{\rm sgn}(\Box_1,\Box_2)=\left\{\begin{array}{ll}
+1,& \Box_1>\Box_2\;,\\
-1,& \Box_1<\Box_2\;.
\end{array} \right.
\ee
Then a solution to \eqref{upsilon_eq} reads:
\begin{equation}
    \upsilon\left(\sqbox{$a$},\sqbox{$b$}\right)=\exp\left(\pi\I\,\mathfrak{f}_{ab}\frac{1+{\rm sgn}\left(\sqbox{$a$},\sqbox{$b$}\right)}{2}\right).
\end{equation}
We note that this is not the unique solution to \eqref{upsilon_eq}: for example, different choices of the ordering function for \eqref{sign_function} will lead to unequal solutions to \eqref{sign_equation}.

Let us introduce a notation for a solution to \eqref{sign_equation} that we will use throughout the paper:
\begin{equation}
    P(\Box,\Kappa)=\mathscr{S}_{\Box}\left[\mathfrak{f}_{ab}|\Kappa\right].
\end{equation}

\section{3d \texorpdfstring{$\CN=2$}{N=2} SUSY in curved space-time}\label{app:3d}

In this appendix we place a 4d $\CN=1$ theory on $\IR_t\times S^1\times \Sigma_g$ with appropriate topological twist, where $\IR_t$ is the temporal direction, and $\Sigma_g$ is a Riemann surface of genus $g$. 
Our aim is to study the 3d $\CN=2$ theory on $\IR_t\times \Sigma_g$ that can be derived from the 4d $\CN=1$ theory on $\IR_t\times S^1\times \Sigma_g$ by dimensional reduction along the $S^1$.
Nevertheless notations for the former theory are more compact and symmetric than those for the latter, therefore we construct supersymmetric Lagrangians and transformations
in 4d --- to reduce to the latter theory we simply assume that all the fields are constant along $S^1$, and the gauge field component $A_3$ gives rise to a real scalar $X$.

In this section and throughout the paper we adopt the notations of \cite{Wess:1992cp}.

On $\IR_t\times S^1\times \Sigma_g$ we choose a product metric and a metric along $\Sigma_g$ component in a conformally flat form:
\be
ds^2=g_{\mu\nu}dx^{\mu}dx^{\nu}=-\left(dx^0\right)^2+\left(dx^3\right)^2+e^{\varphi}dx_idx^i \;.
\ee
In these terms the spin connection reads ($\epsilon_{12}=-\epsilon_{21}=1$):
\be
\omega^{12}=\frac{1}{2}\epsilon_{ij}dx^i\p_j\varphi \;.
\ee
We gauge the R-symmetry with the following connection:
\be
A^{({\rm R})}=-\frac{1}{2}\omega^{12},\quad F^{(R)}=dA^{({\rm R})}=-\frac{R}{4}\,{\rm det}\,g\,d^2x,\quad \frac{1}{2\pi}\int_{\Sigma_g}F^{(R)}=g-1 \;,
\ee
where $R$ is a 2d scalar curvature of $\Sigma_g$:
\be
R=-e^{-\varphi}\;\p^i\p_i \varphi \;.
\ee

Recall that the canonical R-charges of the various components in the Lagrangian are:
\be
\begin{array}{c|c|c|c|c|c|c|c}
\xi_{\alpha} & A_{\mu} &  \lambda_{\alpha} &{\bf D} & \phi &\psi_{\alpha} & {\bf F} & W\\
\hline
1& 0&  1 & 0& r_{\phi}& r_{\phi}-1 & r_{\phi}-2 & 2\\
\end{array}
\ee

The 3d action consists of five components --- gauge, chiral, superpotential, Fayet-Iliopolous and Chern-Simons terms:
\be
S=S_{\rm gauge}+S_{\chi}+S_{W}+S_{\rm FI}+S_{\rm top} \;.
\ee

The gauge extension of the covariant derivatives reads:
\be
\begin{split}
D_{\mu}\phi&=\p_{\mu}\phi+\I A_{\mu}^{(a)}T^a\phi \;,\\
F_{\mu\nu}^{(a)}&=\p_{\mu}A_{\nu}^{(a)}-\p_{\nu}A_{\mu}^{(a)}-f^{abc}A_{\mu}^{(b)}A_{\nu}^{(c)} \;,\\
D_{\mu}\lambda^{(a)}&=\p_{\mu}\lambda^{(a)}-f^{abc}A_{\mu}^{(b)}\lambda^{(c)}\;.
\end{split}
\ee

The supersymmetry (SUSY) is generated by a covariantly-constant Killing spinor $\xi$:
\be
\CD_{\mu}\xi_{\alpha}=\p_{\mu}\xi_{\alpha}-\frac{1}{2}\omega_{\mu ab}\left(\sigma^{ab}\xi\right)_{\alpha}+\I A^{(R)}_{\mu}\xi_{\alpha}=0\;.
\ee
This equation has a constant solution:
\be
\xi_{\alpha}=(c,0)\;.
\ee

The action of the gauge fields reads:
\be
S_{\rm gauge}=\int\lm_{\IR_t\times S^1\times \Sigma_g} \hspace{-3mm} d^4x\,\sqrt{-g}\;\left(-\frac{1}{4}F_{\mu\nu}^{(a)}F^{\mu\nu(a)}-\I\bar\lambda^{(a)}\bar\sigma^{\mu}D_{\mu}\lambda^{(a)}+\frac{1}{2}{\bf D}^{(a)}{\bf D}^{(a)}\right),
\ee
with the SUSY transformations:
\be
\begin{split}
\delta A_{\mu}^{(a)}&=-\I\bar\lambda^{(a)}\bar\sigma_{\mu}\xi+\I\bar\xi\bar\sigma_{\mu}\lambda^{(a)} \;,\\
\delta\lambda^{(a)}_{\alpha}&=\left(\sigma^{\mu\nu}\xi\right)_{\alpha} F^{(a)}_{\mu\nu}+\I\xi_{\alpha} {\bf D}^{(a)} \;,\\
\delta\bar\lambda^{(a)}_{\dot\alpha}&=-\left(\bar\xi\bar\sigma^{\mu\nu}\right)_{\dot\alpha} F^{(a)}_{\mu\nu}-\I\bar\xi_{\dot\alpha} {\bf D}^{(a)} \;,\\
\delta{\bf D}^{(a)}&=D_{\mu}\bar\lambda^{(a)}\bar\sigma^{\mu}\xi+\bar\xi\bar\sigma^{\mu}D_{\mu}\lambda^{(a)}\;.
\end{split}
\ee

The chiral action is:
\be
\begin{split}
S_{\chi}=\int\lm_{\IR_t\times S^1\times \Sigma_g}  \hspace{-3mm} d^4x\,\sqrt{-g}\;\Big(& -D_{\mu}\phi^{\dagger}D^{\mu}\phi-\I\bar\psi\bar\sigma^{\mu}D_{\mu}\psi+{\bf F}^{\dagger}{\bf F}+\\
&+\I\sqrt{2}\left(\phi^{\dagger}T^a\psi\lambda^{(a)}-\bar\lambda^{(a)}\bar\psi T^a\phi\right)+{\bf D}^{(a)}\phi^{\dagger}T^a\phi+\frac{\I}{4}r_{\phi}R\,\phi^{\dagger}\phi\Big)\;,
\end{split}
\ee
with the SUSY transformations:
\be
\begin{split}
\delta \phi&=\sqrt{2}\xi\psi \;,\\
\delta\psi_\alpha&=\I\sqrt{2}\left(\sigma^{\mu}\bar\xi\right)_{\alpha} D_{\mu}\phi+\sqrt{2}\xi_\alpha{\bf F} \;,\\
\delta\bar\psi_{\dot\alpha}&=-\I\sqrt{2}\left(\xi\sigma^{\mu}\right)_{\dot\alpha} D_{\mu}\phi^{\dagger}+\sqrt{2}\bar\xi_{\dot\alpha}{\bf F}^{\dagger} \;,\\
\delta {\bf F}&=\I\sqrt{2}\bar\xi\bar\sigma^{\mu}D_{\mu}\psi+2\I\, T^a\phi\bar\xi\bar\lambda^{(a)} \;.
\end{split}
\ee

The superpotential term reads:
\be
S_W=\int\lm_{\IR_t\times S^1\times \Sigma_g}  \hspace{-3mm} d^4x\,\sqrt{-g}\;\Tr\left({\bf F}\p_{\phi}W-\frac{1}{2}\p^2_{\phi}W\,\psi\psi\right)+{\rm (c.c.)}\;.
\ee
This is invariant under SUSY only if the superpotential $W$ is gauge invariant and has R-charge +2.
In the case of the curved surface $\Sigma_g$, the R-charge contributes manifestly to the coupling with $A^{(R)}$.

The FI term is defined by a real FI parameter $r$: 
\be
S_{\rm FI}=\int\lm_{\IR_t\times S^1\times \Sigma_g} \hspace{-3mm}  d^4x\,\sqrt{-g}\; (-r)\,{\bf D}^{(a)} \; \Tr\, T^a \;.
\ee

Traditionally one considers in 3d a Chern-Simons topological term as well (see e.g. \cite{Marino:2011nm}).
We will omit it in our consideration since we are aiming to consider a sequence of dimensional compactifications from 4d super Yang-Mills theory towards 1d quiver quantum mechanics, and a Chern-Simons topological term is not universal in view of this reduction.

The anti-commutators for the fermionic fields are:
\be
\left\{ \psi_{\alpha}(\vec x),\bar\psi_{\dot\beta}(\vec y) \right\}=-\frac{1}{\sqrt{-g}}\;\sigma_{\alpha\dot\beta}^0\;\delta^{(3)}(\vec x-\vec y)\;.
\ee

The Gauss law constraint says that the following operator annihilates physical states:
\be
\begin{split}
&\CJ^a=\frac{1}{\sqrt{-g}}\frac{\delta S}{\delta A_0}=\\
&=D_{\mu}F^{\mu 0(a)}-\I f^{abc}\bar\lambda^{(b)}\bar\sigma^{(0)}\lambda^{(c)}+\I \left(\phi^{\dagger}T^aD^0\phi-D^0\phi^{\dagger}T^a\phi\right)+\bar\psi\bar\sigma^0T^a\psi\;.
\end{split}
\ee

The supercharges generate SUSY transformations:
$$
\delta \CO=\I\left[\xi Q+\bar\xi \bar Q,\CO\right] \;.
$$

The only supercharge components generated by $\xi$ are $Q_2$ and $\bar Q_{\dot 2}$. 
Expressions for them can be easily derived from covariantized supercharge expressions in a flat background:
\be\label{su_cha}
\begin{split}
Q_{\alpha}=\int\lm_{S^1\times \Sigma_g} \hspace{-1mm}  d^3x\,\sqrt{-g}\; \Big[&\sqrt{2}D_{\mu}\phi^{\dagger}\left(\sigma^{\mu}\bar\sigma^0\psi\right)_{\alpha}+\I\sqrt{2}\left(\sigma^0\bar\psi\right)_{\alpha}{\bf F}-\\
&-\I\left(\sigma^{\mu\nu}\sigma^0\bar\lambda^{(a)}\right)_\alpha F_{\mu
      \nu}^{(a)}-\left(\sigma^0\bar\lambda^{(a)}\right)_{\alpha}{\bf D}^{(a)}\Big]\;,\\
\bar Q_{\dot \alpha}=\int\lm_{S^1\times \Sigma_g}\hspace{-1mm}  d^3x\,\sqrt{-g}\; \Big[&\sqrt{2}\left(\bar\psi\bar\sigma^0\sigma^{\mu}\right)_{\dot \alpha}D_{\mu}\phi-\I\sqrt{2}{\bf F}^{\dagger}\left(\psi\sigma^0 \right)_{\dot \alpha}-\\
&-\I\left(\lambda^{(a)}\sigma^0\bar\sigma^{\mu\nu}\right)_{\dot \alpha}F_{\mu\nu}^{(a)}-\left(\lambda^{(a)}\sigma^0\right)_{\dot \alpha}{\bf D}^{(a)}\Big]\;.
\end{split}
\ee

These supercharges satisfy the following superalgebra:
\be
\begin{split}
\left\{Q_{2},\bar Q_{\dot 2} \right\}&=2H+2\I\hat G\left[A_0^{(a)}-A_3^{(a)}\right]-2Z  \;, \\
\left(Q_{2}\right)^2&=\left(\bar Q_{\dot 2}\right)^2=0\;,
\end{split}
\ee
where $H$ is the Hamiltonian, the gauge-transformation operator reads:
\be
\hat G[\epsilon]=\int\lm_{S^1\times \Sigma_g} \hspace{-1mm}  d^3x\,\sqrt{-g}\;\epsilon\,\CJ^{(a)} \;.
\ee

The central charge term is specific to a particular form of the spatial manifold $S^1\times \Sigma_g$ admitting non-trivial fluxes of magnetic field through $\Sigma_g$:
\be
Z={\rm vol}_{S^1}\times r\int\lm_{\Sigma_g} F^{(a)}\; \Tr\, T^a \;.
\ee


\newpage

\bibliographystyle{utphys} 
\bibliography{biblio}


\end{document}

%% file: figs/fig_partition.tex
\draw[style] (0.707107,3.67423) -- (0.707107,4.49073) -- (0.,4.08248) -- (0.,3.26599) -- cycle;
\draw[style] (0.707107,4.49073) -- (0.707107,5.30723) -- (0.,4.89898) -- (0.,4.08248) -- cycle;
\draw[style] (1.41421,2.44949) -- (1.41421,3.26599) -- (0.707107,2.85774) -- (0.707107,2.04124) -- cycle;
\draw[style] (0.707107,2.04124) -- (0.707107,2.85774) -- (0.,2.44949) -- (0.,1.63299) -- cycle;
\draw[style] (-0.707107,-2.04124) -- (-0.707107,-1.22474) -- (-1.41421,-1.63299) -- (-1.41421,-2.44949) -- cycle;
\draw[style] (2.12132,1.22474) -- (2.12132,2.04124) -- (1.41421,1.63299) -- (1.41421,0.816497) -- cycle;
\draw[style] (1.41421,0.816497) -- (1.41421,1.63299) -- (0.707107,1.22474) -- (0.707107,0.408248) -- cycle;
\draw[style] (0.707107,-1.22474) -- (0.707107,-0.408248) -- (0.,-0.816497) -- (0.,-1.63299) -- cycle;
\draw[style] (0.707107,-0.408248) -- (0.707107,0.408248) -- (0.,0.) -- (0.,-0.816497) -- cycle;
\draw[style] (2.82843,0.) -- (2.82843,0.816497) -- (2.12132,0.408248) -- (2.12132,-0.408248) -- cycle;
\draw[style] (2.12132,-0.408248) -- (2.12132,0.408248) -- (1.41421,0.) -- (1.41421,-0.816497) -- cycle;
\draw[style] (1.41421,-2.44949) -- (1.41421,-1.63299) -- (0.707107,-2.04124) -- (0.707107,-2.85774) -- cycle;
\draw[style] (2.82843,-1.63299) -- (2.82843,-0.816497) -- (2.12132,-1.22474) -- (2.12132,-2.04124) -- cycle;
\draw[style] (4.24264,-1.63299) -- (4.24264,-0.816497) -- (3.53553,-1.22474) -- (3.53553,-2.04124) -- cycle;
\draw[style] (3.53553,-2.85774) -- (3.53553,-2.04124) -- (2.82843,-2.44949) -- (2.82843,-3.26599) -- cycle;
\draw[style] (4.94975,-2.85774) -- (4.94975,-2.04124) -- (4.24264,-2.44949) -- (4.24264,-3.26599) -- cycle;
\draw[style] (-0.707107,3.67423) -- (-0.707107,4.49073) -- (0.,4.08248) -- (0.,3.26599) -- cycle;
\draw[style] (-0.707107,4.49073) -- (-0.707107,5.30723) -- (0.,4.89898) -- (0.,4.08248) -- cycle;
\draw[style] (-1.41421,1.63299) -- (-1.41421,2.44949) -- (-0.707107,2.04124) -- (-0.707107,1.22474) -- cycle;
\draw[style] (-1.41421,2.44949) -- (-1.41421,3.26599) -- (-0.707107,2.85774) -- (-0.707107,2.04124) -- cycle;
\draw[style] (-2.12132,-0.408248) -- (-2.12132,0.408248) -- (-1.41421,0.) -- (-1.41421,-0.816497) -- cycle;
\draw[style] (-2.12132,0.408248) -- (-2.12132,1.22474) -- (-1.41421,0.816497) -- (-1.41421,0.) -- cycle;
\draw[style] (-2.82843,-1.63299) -- (-2.82843,-0.816497) -- (-2.12132,-1.22474) -- (-2.12132,-2.04124) -- cycle;
\draw[style] (-0.707107,1.22474) -- (-0.707107,2.04124) -- (0.,1.63299) -- (0.,0.816497) -- cycle;
\draw[style] (-0.707107,2.04124) -- (-0.707107,2.85774) -- (0.,2.44949) -- (0.,1.63299) -- cycle;
\draw[style] (-1.41421,-0.816497) -- (-1.41421,0.) -- (-0.707107,-0.408248) -- (-0.707107,-1.22474) -- cycle;
\draw[style] (-1.41421,0.) -- (-1.41421,0.816497) -- (-0.707107,0.408248) -- (-0.707107,-0.408248) -- cycle;
\draw[style] (-2.12132,-2.04124) -- (-2.12132,-1.22474) -- (-1.41421,-1.63299) -- (-1.41421,-2.44949) -- cycle;
\draw[style] (0.,0.816497) -- (0.,1.63299) -- (0.707107,1.22474) -- (0.707107,0.408248) -- cycle;
\draw[style] (-0.707107,-2.04124) -- (-0.707107,-1.22474) -- (0.,-1.63299) -- (0.,-2.44949) -- cycle;
\draw[style] (-0.707107,-1.22474) -- (-0.707107,-0.408248) -- (0.,-0.816497) -- (0.,-1.63299) -- cycle;
\draw[style] (-0.707107,-0.408248) -- (-0.707107,0.408248) -- (0.,0.) -- (0.,-0.816497) -- cycle;
\draw[style] (0.707107,-1.22474) -- (0.707107,-0.408248) -- (1.41421,-0.816497) -- (1.41421,-1.63299) -- cycle;
\draw[style] (0.707107,-0.408248) -- (0.707107,0.408248) -- (1.41421,0.) -- (1.41421,-0.816497) -- cycle;
\draw[style] (0.,-2.44949) -- (0.,-1.63299) -- (0.707107,-2.04124) -- (0.707107,-2.85774) -- cycle;
\draw[style] (1.41421,-2.44949) -- (1.41421,-1.63299) -- (2.12132,-2.04124) -- (2.12132,-2.85774) -- cycle;
\draw[style] (1.41421,-1.63299) -- (1.41421,-0.816497) -- (2.12132,-1.22474) -- (2.12132,-2.04124) -- cycle;
\draw[style] (2.82843,-1.63299) -- (2.82843,-0.816497) -- (3.53553,-1.22474) -- (3.53553,-2.04124) -- cycle;
\draw[style] (2.12132,-2.85774) -- (2.12132,-2.04124) -- (2.82843,-2.44949) -- (2.82843,-3.26599) -- cycle;
\draw[style] (3.53553,-2.85774) -- (3.53553,-2.04124) -- (4.24264,-2.44949) -- (4.24264,-3.26599) -- cycle;
\draw[style] (0.,5.71548) -- (-0.707107,5.30723) -- (0.,4.89898) -- (0.707107,5.30723) -- cycle;
\draw[style] (-0.707107,3.67423) -- (-1.41421,3.26599) -- (-0.707107,2.85774) -- (0.,3.26599) -- cycle;
\draw[style] (-1.41421,1.63299) -- (-2.12132,1.22474) -- (-1.41421,0.816497) -- (-0.707107,1.22474) -- cycle;
\draw[style] (-2.12132,-0.408248) -- (-2.82843,-0.816497) -- (-2.12132,-1.22474) -- (-1.41421,-0.816497) -- cycle;
\draw[style] (0.707107,3.67423) -- (0.,3.26599) -- (0.707107,2.85774) -- (1.41421,3.26599) -- cycle;
\draw[style] (0.,3.26599) -- (-0.707107,2.85774) -- (0.,2.44949) -- (0.707107,2.85774) -- cycle;
\draw[style] (-0.707107,1.22474) -- (-1.41421,0.816497) -- (-0.707107,0.408248) -- (0.,0.816497) -- cycle;
\draw[style] (-1.41421,-0.816497) -- (-2.12132,-1.22474) -- (-1.41421,-1.63299) -- (-0.707107,-1.22474) -- cycle;
\draw[style] (1.41421,2.44949) -- (0.707107,2.04124) -- (1.41421,1.63299) -- (2.12132,2.04124) -- cycle;
\draw[style] (0.707107,2.04124) -- (0.,1.63299) -- (0.707107,1.22474) -- (1.41421,1.63299) -- cycle;
\draw[style] (0.,0.816497) -- (-0.707107,0.408248) -- (0.,0.) -- (0.707107,0.408248) -- cycle;
\draw[style] (2.12132,1.22474) -- (1.41421,0.816497) -- (2.12132,0.408248) -- (2.82843,0.816497) -- cycle;
\draw[style] (1.41421,0.816497) -- (0.707107,0.408248) -- (1.41421,0.) -- (2.12132,0.408248) -- cycle;
\draw[style] (0.707107,-1.22474) -- (0.,-1.63299) -- (0.707107,-2.04124) -- (1.41421,-1.63299) -- cycle;
\draw[style] (2.82843,0.) -- (2.12132,-0.408248) -- (2.82843,-0.816497) -- (3.53553,-0.408248) -- cycle;
\draw[style] (2.12132,-0.408248) -- (1.41421,-0.816497) -- (2.12132,-1.22474) -- (2.82843,-0.816497) -- cycle;
\draw[style] (3.53553,-0.408248) -- (2.82843,-0.816497) -- (3.53553,-1.22474) -- (4.24264,-0.816497) -- cycle;
\draw[style] (2.82843,-1.63299) -- (2.12132,-2.04124) -- (2.82843,-2.44949) -- (3.53553,-2.04124) -- cycle;
\draw[style] (4.24264,-1.63299) -- (3.53553,-2.04124) -- (4.24264,-2.44949) -- (4.94975,-2.04124) -- cycle;

%% file: figs/fig_partition_1.tex
\draw[style1] (0.707107,3.67423) -- (0.707107,4.49073) -- (0.,4.08248) -- (0.,3.26599) -- cycle;
\draw[style1] (0.707107,4.49073) -- (0.707107,5.30723) -- (0.,4.89898) -- (0.,4.08248) -- cycle;
\draw[style1] (1.41421,2.44949) -- (1.41421,3.26599) -- (0.707107,2.85774) -- (0.707107,2.04124) -- cycle;
\draw[style1] (0.707107,2.04124) -- (0.707107,2.85774) -- (0.,2.44949) -- (0.,1.63299) -- cycle;
\draw[style1] (-0.707107,-2.04124) -- (-0.707107,-1.22474) -- (-1.41421,-1.63299) -- (-1.41421,-2.44949) -- cycle;
\draw[style1] (2.12132,1.22474) -- (2.12132,2.04124) -- (1.41421,1.63299) -- (1.41421,0.816497) -- cycle;
\draw[style1] (1.41421,0.816497) -- (1.41421,1.63299) -- (0.707107,1.22474) -- (0.707107,0.408248) -- cycle;
\draw[style1] (0.707107,-1.22474) -- (0.707107,-0.408248) -- (0.,-0.816497) -- (0.,-1.63299) -- cycle;
\draw[style1] (0.707107,-0.408248) -- (0.707107,0.408248) -- (0.,0.) -- (0.,-0.816497) -- cycle;
\draw[style1] (2.82843,0.) -- (2.82843,0.816497) -- (2.12132,0.408248) -- (2.12132,-0.408248) -- cycle;
\draw[style1] (2.12132,-0.408248) -- (2.12132,0.408248) -- (1.41421,0.) -- (1.41421,-0.816497) -- cycle;
\draw[style1] (1.41421,-2.44949) -- (1.41421,-1.63299) -- (0.707107,-2.04124) -- (0.707107,-2.85774) -- cycle;
\draw[style1] (2.82843,-1.63299) -- (2.82843,-0.816497) -- (2.12132,-1.22474) -- (2.12132,-2.04124) -- cycle;
\draw[style1] (4.24264,-1.63299) -- (4.24264,-0.816497) -- (3.53553,-1.22474) -- (3.53553,-2.04124) -- cycle;
\draw[style1] (3.53553,-2.85774) -- (3.53553,-2.04124) -- (2.82843,-2.44949) -- (2.82843,-3.26599) -- cycle;
\draw[style1] (4.94975,-2.85774) -- (4.94975,-2.04124) -- (4.24264,-2.44949) -- (4.24264,-3.26599) -- cycle;
\draw[style1] (-0.707107,3.67423) -- (-0.707107,4.49073) -- (0.,4.08248) -- (0.,3.26599) -- cycle;
\draw[style1] (-0.707107,4.49073) -- (-0.707107,5.30723) -- (0.,4.89898) -- (0.,4.08248) -- cycle;
\draw[style1] (-1.41421,1.63299) -- (-1.41421,2.44949) -- (-0.707107,2.04124) -- (-0.707107,1.22474) -- cycle;
\draw[style1] (-1.41421,2.44949) -- (-1.41421,3.26599) -- (-0.707107,2.85774) -- (-0.707107,2.04124) -- cycle;
\draw[style1] (-2.12132,-0.408248) -- (-2.12132,0.408248) -- (-1.41421,0.) -- (-1.41421,-0.816497) -- cycle;
\draw[style1] (-2.12132,0.408248) -- (-2.12132,1.22474) -- (-1.41421,0.816497) -- (-1.41421,0.) -- cycle;
\draw[style1] (-2.82843,-1.63299) -- (-2.82843,-0.816497) -- (-2.12132,-1.22474) -- (-2.12132,-2.04124) -- cycle;
\draw[style1] (-0.707107,1.22474) -- (-0.707107,2.04124) -- (0.,1.63299) -- (0.,0.816497) -- cycle;
\draw[style1] (-0.707107,2.04124) -- (-0.707107,2.85774) -- (0.,2.44949) -- (0.,1.63299) -- cycle;
\draw[style1] (-1.41421,-0.816497) -- (-1.41421,0.) -- (-0.707107,-0.408248) -- (-0.707107,-1.22474) -- cycle;
\draw[style1] (-1.41421,0.) -- (-1.41421,0.816497) -- (-0.707107,0.408248) -- (-0.707107,-0.408248) -- cycle;
\draw[style1] (-2.12132,-2.04124) -- (-2.12132,-1.22474) -- (-1.41421,-1.63299) -- (-1.41421,-2.44949) -- cycle;
\draw[style1] (0.,0.816497) -- (0.,1.63299) -- (0.707107,1.22474) -- (0.707107,0.408248) -- cycle;
\draw[style1] (-0.707107,-2.04124) -- (-0.707107,-1.22474) -- (0.,-1.63299) -- (0.,-2.44949) -- cycle;
\draw[style1] (-0.707107,-1.22474) -- (-0.707107,-0.408248) -- (0.,-0.816497) -- (0.,-1.63299) -- cycle;
\draw[style1] (-0.707107,-0.408248) -- (-0.707107,0.408248) -- (0.,0.) -- (0.,-0.816497) -- cycle;
\draw[style1] (0.707107,-1.22474) -- (0.707107,-0.408248) -- (1.41421,-0.816497) -- (1.41421,-1.63299) -- cycle;
\draw[style1] (0.707107,-0.408248) -- (0.707107,0.408248) -- (1.41421,0.) -- (1.41421,-0.816497) -- cycle;
\draw[style1] (0.,-2.44949) -- (0.,-1.63299) -- (0.707107,-2.04124) -- (0.707107,-2.85774) -- cycle;
\draw[style1] (1.41421,-2.44949) -- (1.41421,-1.63299) -- (2.12132,-2.04124) -- (2.12132,-2.85774) -- cycle;
\draw[style1] (1.41421,-1.63299) -- (1.41421,-0.816497) -- (2.12132,-1.22474) -- (2.12132,-2.04124) -- cycle;
\draw[style1] (2.82843,-1.63299) -- (2.82843,-0.816497) -- (3.53553,-1.22474) -- (3.53553,-2.04124) -- cycle;
\draw[style1] (2.12132,-2.85774) -- (2.12132,-2.04124) -- (2.82843,-2.44949) -- (2.82843,-3.26599) -- cycle;
\draw[style1] (3.53553,-2.85774) -- (3.53553,-2.04124) -- (4.24264,-2.44949) -- (4.24264,-3.26599) -- cycle;
\draw[style1] (0.,5.71548) -- (-0.707107,5.30723) -- (0.,4.89898) -- (0.707107,5.30723) -- cycle;
\draw[style1] (-0.707107,3.67423) -- (-1.41421,3.26599) -- (-0.707107,2.85774) -- (0.,3.26599) -- cycle;
\draw[style1] (-1.41421,1.63299) -- (-2.12132,1.22474) -- (-1.41421,0.816497) -- (-0.707107,1.22474) -- cycle;
\draw[style1] (-2.12132,-0.408248) -- (-2.82843,-0.816497) -- (-2.12132,-1.22474) -- (-1.41421,-0.816497) -- cycle;
\draw[style1] (0.707107,3.67423) -- (0.,3.26599) -- (0.707107,2.85774) -- (1.41421,3.26599) -- cycle;
\draw[style1] (0.,3.26599) -- (-0.707107,2.85774) -- (0.,2.44949) -- (0.707107,2.85774) -- cycle;
\draw[style1] (-0.707107,1.22474) -- (-1.41421,0.816497) -- (-0.707107,0.408248) -- (0.,0.816497) -- cycle;
\draw[style1] (-1.41421,-0.816497) -- (-2.12132,-1.22474) -- (-1.41421,-1.63299) -- (-0.707107,-1.22474) -- cycle;
\draw[style1] (1.41421,2.44949) -- (0.707107,2.04124) -- (1.41421,1.63299) -- (2.12132,2.04124) -- cycle;
\draw[style1] (0.707107,2.04124) -- (0.,1.63299) -- (0.707107,1.22474) -- (1.41421,1.63299) -- cycle;
\draw[style1] (0.,0.816497) -- (-0.707107,0.408248) -- (0.,0.) -- (0.707107,0.408248) -- cycle;
\draw[style1] (2.12132,1.22474) -- (1.41421,0.816497) -- (2.12132,0.408248) -- (2.82843,0.816497) -- cycle;
\draw[style1] (1.41421,0.816497) -- (0.707107,0.408248) -- (1.41421,0.) -- (2.12132,0.408248) -- cycle;
\draw[style1] (0.707107,-1.22474) -- (0.,-1.63299) -- (0.707107,-2.04124) -- (1.41421,-1.63299) -- cycle;
\draw[style1] (2.82843,0.) -- (2.12132,-0.408248) -- (2.82843,-0.816497) -- (3.53553,-0.408248) -- cycle;
\draw[style1] (2.12132,-0.408248) -- (1.41421,-0.816497) -- (2.12132,-1.22474) -- (2.82843,-0.816497) -- cycle;
\draw[style1] (3.53553,-0.408248) -- (2.82843,-0.816497) -- (3.53553,-1.22474) -- (4.24264,-0.816497) -- cycle;
\draw[style1] (2.82843,-1.63299) -- (2.12132,-2.04124) -- (2.82843,-2.44949) -- (3.53553,-2.04124) -- cycle;
\draw[style1] (4.24264,-1.63299) -- (3.53553,-2.04124) -- (4.24264,-2.44949) -- (4.94975,-2.04124) -- cycle;
\draw[style2] (0.707107,5.30723) -- (0.707107,6.12372) -- (0.,5.71548) -- (0.,4.89898) -- cycle;
\draw[style2] (0.,3.26599) -- (0.,4.08248) -- (-0.707107,3.67423) -- (-0.707107,2.85774) -- cycle;
\draw[style2] (-0.707107,1.22474) -- (-0.707107,2.04124) -- (-1.41421,1.63299) -- (-1.41421,0.816497) -- cycle;
\draw[style2] (-1.41421,-0.816497) -- (-1.41421,0.) -- (-2.12132,-0.408248) -- (-2.12132,-1.22474) -- cycle;
\draw[style2] (-2.12132,-2.04124) -- (-2.12132,-1.22474) -- (-2.82843,-1.63299) -- (-2.82843,-2.44949) -- cycle;
\draw[style2] (1.41421,3.26599) -- (1.41421,4.08248) -- (0.707107,3.67423) -- (0.707107,2.85774) -- cycle;
\draw[style2] (2.12132,2.04124) -- (2.12132,2.85774) -- (1.41421,2.44949) -- (1.41421,1.63299) -- cycle;
\draw[style2] (0.,-2.44949) -- (0.,-1.63299) -- (-0.707107,-2.04124) -- (-0.707107,-2.85774) -- cycle;
\draw[style2] (2.82843,0.816497) -- (2.82843,1.63299) -- (2.12132,1.22474) -- (2.12132,0.408248) -- cycle;
\draw[style2] (1.41421,-1.63299) -- (1.41421,-0.816497) -- (0.707107,-1.22474) -- (0.707107,-2.04124) -- cycle;
\draw[style2] (3.53553,-0.408248) -- (3.53553,0.408248) -- (2.82843,0.) -- (2.82843,-0.816497) -- cycle;
\draw[style2] (2.12132,-2.85774) -- (2.12132,-2.04124) -- (1.41421,-2.44949) -- (1.41421,-3.26599) -- cycle;
\draw[style2] (3.53553,-2.04124) -- (3.53553,-1.22474) -- (2.82843,-1.63299) -- (2.82843,-2.44949) -- cycle;
\draw[style2] (4.94975,-2.04124) -- (4.94975,-1.22474) -- (4.24264,-1.63299) -- (4.24264,-2.44949) -- cycle;
\draw[style2] (4.24264,-3.26599) -- (4.24264,-2.44949) -- (3.53553,-2.85774) -- (3.53553,-3.67423) -- cycle;
\draw[style2] (5.65685,-3.26599) -- (5.65685,-2.44949) -- (4.94975,-2.85774) -- (4.94975,-3.67423) -- cycle;
\draw[style2] (-0.707107,5.30723) -- (-0.707107,6.12372) -- (0.,5.71548) -- (0.,4.89898) -- cycle;
\draw[style2] (-1.41421,3.26599) -- (-1.41421,4.08248) -- (-0.707107,3.67423) -- (-0.707107,2.85774) -- cycle;
\draw[style2] (-2.12132,1.22474) -- (-2.12132,2.04124) -- (-1.41421,1.63299) -- (-1.41421,0.816497) -- cycle;
\draw[style2] (-2.82843,-0.816497) -- (-2.82843,0.) -- (-2.12132,-0.408248) -- (-2.12132,-1.22474) -- cycle;
\draw[style2] (-3.53553,-2.04124) -- (-3.53553,-1.22474) -- (-2.82843,-1.63299) -- (-2.82843,-2.44949) -- cycle;
\draw[style2] (0.,3.26599) -- (0.,4.08248) -- (0.707107,3.67423) -- (0.707107,2.85774) -- cycle;
\draw[style2] (0.707107,2.04124) -- (0.707107,2.85774) -- (1.41421,2.44949) -- (1.41421,1.63299) -- cycle;
\draw[style2] (-1.41421,-2.44949) -- (-1.41421,-1.63299) -- (-0.707107,-2.04124) -- (-0.707107,-2.85774) -- cycle;
\draw[style2] (1.41421,0.816497) -- (1.41421,1.63299) -- (2.12132,1.22474) -- (2.12132,0.408248) -- cycle;
\draw[style2] (0.,-1.63299) -- (0.,-0.816497) -- (0.707107,-1.22474) -- (0.707107,-2.04124) -- cycle;
\draw[style2] (2.12132,-0.408248) -- (2.12132,0.408248) -- (2.82843,0.) -- (2.82843,-0.816497) -- cycle;
\draw[style2] (0.707107,-2.85774) -- (0.707107,-2.04124) -- (1.41421,-2.44949) -- (1.41421,-3.26599) -- cycle;
\draw[style2] (2.12132,-2.04124) -- (2.12132,-1.22474) -- (2.82843,-1.63299) -- (2.82843,-2.44949) -- cycle;
\draw[style2] (3.53553,-2.04124) -- (3.53553,-1.22474) -- (4.24264,-1.63299) -- (4.24264,-2.44949) -- cycle;
\draw[style2] (2.82843,-3.26599) -- (2.82843,-2.44949) -- (3.53553,-2.85774) -- (3.53553,-3.67423) -- cycle;
\draw[style2] (4.24264,-3.26599) -- (4.24264,-2.44949) -- (4.94975,-2.85774) -- (4.94975,-3.67423) -- cycle;
\draw[style2] (0.,6.53197) -- (-0.707107,6.12372) -- (0.,5.71548) -- (0.707107,6.12372) -- cycle;
\draw[style2] (-0.707107,4.49073) -- (-1.41421,4.08248) -- (-0.707107,3.67423) -- (0.,4.08248) -- cycle;
\draw[style2] (-1.41421,2.44949) -- (-2.12132,2.04124) -- (-1.41421,1.63299) -- (-0.707107,2.04124) -- cycle;
\draw[style2] (-2.12132,0.408248) -- (-2.82843,0.) -- (-2.12132,-0.408248) -- (-1.41421,0.) -- cycle;
\draw[style2] (-2.82843,-0.816497) -- (-3.53553,-1.22474) -- (-2.82843,-1.63299) -- (-2.12132,-1.22474) -- cycle;
\draw[style2] (0.707107,4.49073) -- (0.,4.08248) -- (0.707107,3.67423) -- (1.41421,4.08248) -- cycle;
\draw[style2] (1.41421,3.26599) -- (0.707107,2.85774) -- (1.41421,2.44949) -- (2.12132,2.85774) -- cycle;
\draw[style2] (-0.707107,-1.22474) -- (-1.41421,-1.63299) -- (-0.707107,-2.04124) -- (0.,-1.63299) -- cycle;
\draw[style2] (2.12132,2.04124) -- (1.41421,1.63299) -- (2.12132,1.22474) -- (2.82843,1.63299) -- cycle;
\draw[style2] (0.707107,-0.408248) -- (0.,-0.816497) -- (0.707107,-1.22474) -- (1.41421,-0.816497) -- cycle;
\draw[style2] (2.82843,0.816497) -- (2.12132,0.408248) -- (2.82843,0.) -- (3.53553,0.408248) -- cycle;
\draw[style2] (1.41421,-1.63299) -- (0.707107,-2.04124) -- (1.41421,-2.44949) -- (2.12132,-2.04124) -- cycle;
\draw[style2] (2.82843,-0.816497) -- (2.12132,-1.22474) -- (2.82843,-1.63299) -- (3.53553,-1.22474) -- cycle;
\draw[style2] (4.24264,-0.816497) -- (3.53553,-1.22474) -- (4.24264,-1.63299) -- (4.94975,-1.22474) -- cycle;
\draw[style2] (3.53553,-2.04124) -- (2.82843,-2.44949) -- (3.53553,-2.85774) -- (4.24264,-2.44949) -- cycle;
\draw[style2] (4.94975,-2.04124) -- (4.24264,-2.44949) -- (4.94975,-2.85774) -- (5.65685,-2.44949) -- cycle;

%% file: figs/fig_partition_2.tex
\draw[style1] (0.707107,3.67423) -- (0.707107,4.49073) -- (0.,4.08248) -- (0.,3.26599) -- cycle;
\draw[style1] (0.707107,4.49073) -- (0.707107,5.30723) -- (0.,4.89898) -- (0.,4.08248) -- cycle;
\draw[style1] (1.41421,2.44949) -- (1.41421,3.26599) -- (0.707107,2.85774) -- (0.707107,2.04124) -- cycle;
\draw[style1] (0.707107,2.04124) -- (0.707107,2.85774) -- (0.,2.44949) -- (0.,1.63299) -- cycle;
\draw[style1] (-0.707107,-2.04124) -- (-0.707107,-1.22474) -- (-1.41421,-1.63299) -- (-1.41421,-2.44949) -- cycle;
\draw[style1] (2.12132,1.22474) -- (2.12132,2.04124) -- (1.41421,1.63299) -- (1.41421,0.816497) -- cycle;
\draw[style1] (1.41421,0.816497) -- (1.41421,1.63299) -- (0.707107,1.22474) -- (0.707107,0.408248) -- cycle;
\draw[style1] (0.707107,-1.22474) -- (0.707107,-0.408248) -- (0.,-0.816497) -- (0.,-1.63299) -- cycle;
\draw[style1] (0.707107,-0.408248) -- (0.707107,0.408248) -- (0.,0.) -- (0.,-0.816497) -- cycle;
\draw[style1] (2.82843,0.) -- (2.82843,0.816497) -- (2.12132,0.408248) -- (2.12132,-0.408248) -- cycle;
\draw[style1] (2.12132,-0.408248) -- (2.12132,0.408248) -- (1.41421,0.) -- (1.41421,-0.816497) -- cycle;
\draw[style1] (1.41421,-2.44949) -- (1.41421,-1.63299) -- (0.707107,-2.04124) -- (0.707107,-2.85774) -- cycle;
\draw[style1] (2.82843,-1.63299) -- (2.82843,-0.816497) -- (2.12132,-1.22474) -- (2.12132,-2.04124) -- cycle;
\draw[style1] (4.24264,-1.63299) -- (4.24264,-0.816497) -- (3.53553,-1.22474) -- (3.53553,-2.04124) -- cycle;
\draw[style1] (3.53553,-2.85774) -- (3.53553,-2.04124) -- (2.82843,-2.44949) -- (2.82843,-3.26599) -- cycle;
\draw[style1] (4.94975,-2.85774) -- (4.94975,-2.04124) -- (4.24264,-2.44949) -- (4.24264,-3.26599) -- cycle;
\draw[style1] (-0.707107,3.67423) -- (-0.707107,4.49073) -- (0.,4.08248) -- (0.,3.26599) -- cycle;
\draw[style1] (-0.707107,4.49073) -- (-0.707107,5.30723) -- (0.,4.89898) -- (0.,4.08248) -- cycle;
\draw[style1] (-1.41421,1.63299) -- (-1.41421,2.44949) -- (-0.707107,2.04124) -- (-0.707107,1.22474) -- cycle;
\draw[style1] (-1.41421,2.44949) -- (-1.41421,3.26599) -- (-0.707107,2.85774) -- (-0.707107,2.04124) -- cycle;
\draw[style1] (-2.12132,-0.408248) -- (-2.12132,0.408248) -- (-1.41421,0.) -- (-1.41421,-0.816497) -- cycle;
\draw[style1] (-2.12132,0.408248) -- (-2.12132,1.22474) -- (-1.41421,0.816497) -- (-1.41421,0.) -- cycle;
\draw[style1] (-2.82843,-1.63299) -- (-2.82843,-0.816497) -- (-2.12132,-1.22474) -- (-2.12132,-2.04124) -- cycle;
\draw[style1] (-0.707107,1.22474) -- (-0.707107,2.04124) -- (0.,1.63299) -- (0.,0.816497) -- cycle;
\draw[style1] (-0.707107,2.04124) -- (-0.707107,2.85774) -- (0.,2.44949) -- (0.,1.63299) -- cycle;
\draw[style1] (-1.41421,-0.816497) -- (-1.41421,0.) -- (-0.707107,-0.408248) -- (-0.707107,-1.22474) -- cycle;
\draw[style1] (-1.41421,0.) -- (-1.41421,0.816497) -- (-0.707107,0.408248) -- (-0.707107,-0.408248) -- cycle;
\draw[style1] (-2.12132,-2.04124) -- (-2.12132,-1.22474) -- (-1.41421,-1.63299) -- (-1.41421,-2.44949) -- cycle;
\draw[style1] (0.,0.816497) -- (0.,1.63299) -- (0.707107,1.22474) -- (0.707107,0.408248) -- cycle;
\draw[style1] (-0.707107,-2.04124) -- (-0.707107,-1.22474) -- (0.,-1.63299) -- (0.,-2.44949) -- cycle;
\draw[style1] (-0.707107,-1.22474) -- (-0.707107,-0.408248) -- (0.,-0.816497) -- (0.,-1.63299) -- cycle;
\draw[style1] (-0.707107,-0.408248) -- (-0.707107,0.408248) -- (0.,0.) -- (0.,-0.816497) -- cycle;
\draw[style1] (0.707107,-1.22474) -- (0.707107,-0.408248) -- (1.41421,-0.816497) -- (1.41421,-1.63299) -- cycle;
\draw[style1] (0.707107,-0.408248) -- (0.707107,0.408248) -- (1.41421,0.) -- (1.41421,-0.816497) -- cycle;
\draw[style1] (0.,-2.44949) -- (0.,-1.63299) -- (0.707107,-2.04124) -- (0.707107,-2.85774) -- cycle;
\draw[style1] (1.41421,-2.44949) -- (1.41421,-1.63299) -- (2.12132,-2.04124) -- (2.12132,-2.85774) -- cycle;
\draw[style1] (1.41421,-1.63299) -- (1.41421,-0.816497) -- (2.12132,-1.22474) -- (2.12132,-2.04124) -- cycle;
\draw[style1] (2.82843,-1.63299) -- (2.82843,-0.816497) -- (3.53553,-1.22474) -- (3.53553,-2.04124) -- cycle;
\draw[style1] (2.12132,-2.85774) -- (2.12132,-2.04124) -- (2.82843,-2.44949) -- (2.82843,-3.26599) -- cycle;
\draw[style1] (3.53553,-2.85774) -- (3.53553,-2.04124) -- (4.24264,-2.44949) -- (4.24264,-3.26599) -- cycle;
\draw[style1] (0.,5.71548) -- (-0.707107,5.30723) -- (0.,4.89898) -- (0.707107,5.30723) -- cycle;
\draw[style1] (-0.707107,3.67423) -- (-1.41421,3.26599) -- (-0.707107,2.85774) -- (0.,3.26599) -- cycle;
\draw[style1] (-1.41421,1.63299) -- (-2.12132,1.22474) -- (-1.41421,0.816497) -- (-0.707107,1.22474) -- cycle;
\draw[style1] (-2.12132,-0.408248) -- (-2.82843,-0.816497) -- (-2.12132,-1.22474) -- (-1.41421,-0.816497) -- cycle;
\draw[style1] (0.707107,3.67423) -- (0.,3.26599) -- (0.707107,2.85774) -- (1.41421,3.26599) -- cycle;
\draw[style1] (0.,3.26599) -- (-0.707107,2.85774) -- (0.,2.44949) -- (0.707107,2.85774) -- cycle;
\draw[style1] (-0.707107,1.22474) -- (-1.41421,0.816497) -- (-0.707107,0.408248) -- (0.,0.816497) -- cycle;
\draw[style1] (-1.41421,-0.816497) -- (-2.12132,-1.22474) -- (-1.41421,-1.63299) -- (-0.707107,-1.22474) -- cycle;
\draw[style1] (1.41421,2.44949) -- (0.707107,2.04124) -- (1.41421,1.63299) -- (2.12132,2.04124) -- cycle;
\draw[style1] (0.707107,2.04124) -- (0.,1.63299) -- (0.707107,1.22474) -- (1.41421,1.63299) -- cycle;
\draw[style1] (0.,0.816497) -- (-0.707107,0.408248) -- (0.,0.) -- (0.707107,0.408248) -- cycle;
\draw[style1] (2.12132,1.22474) -- (1.41421,0.816497) -- (2.12132,0.408248) -- (2.82843,0.816497) -- cycle;
\draw[style1] (1.41421,0.816497) -- (0.707107,0.408248) -- (1.41421,0.) -- (2.12132,0.408248) -- cycle;
\draw[style1] (0.707107,-1.22474) -- (0.,-1.63299) -- (0.707107,-2.04124) -- (1.41421,-1.63299) -- cycle;
\draw[style1] (2.82843,0.) -- (2.12132,-0.408248) -- (2.82843,-0.816497) -- (3.53553,-0.408248) -- cycle;
\draw[style1] (2.12132,-0.408248) -- (1.41421,-0.816497) -- (2.12132,-1.22474) -- (2.82843,-0.816497) -- cycle;
\draw[style1] (3.53553,-0.408248) -- (2.82843,-0.816497) -- (3.53553,-1.22474) -- (4.24264,-0.816497) -- cycle;
\draw[style1] (2.82843,-1.63299) -- (2.12132,-2.04124) -- (2.82843,-2.44949) -- (3.53553,-2.04124) -- cycle;
\draw[style1] (4.24264,-1.63299) -- (3.53553,-2.04124) -- (4.24264,-2.44949) -- (4.94975,-2.04124) -- cycle;
\draw[style2] (0.707107,4.49073) -- (0.707107,5.30723) -- (0.,4.89898) -- (0.,4.08248) -- cycle;
\draw[style2] (0.707107,2.04124) -- (0.707107,2.85774) -- (0.,2.44949) -- (0.,1.63299) -- cycle;
\draw[style2] (-0.707107,-2.04124) -- (-0.707107,-1.22474) -- (-1.41421,-1.63299) -- (-1.41421,-2.44949) -- cycle;
\draw[style2] (1.41421,0.816497) -- (1.41421,1.63299) -- (0.707107,1.22474) -- (0.707107,0.408248) -- cycle;
\draw[style2] (0.707107,-0.408248) -- (0.707107,0.408248) -- (0.,0.) -- (0.,-0.816497) -- cycle;
\draw[style2] (2.12132,-0.408248) -- (2.12132,0.408248) -- (1.41421,0.) -- (1.41421,-0.816497) -- cycle;
\draw[style2] (1.41421,-2.44949) -- (1.41421,-1.63299) -- (0.707107,-2.04124) -- (0.707107,-2.85774) -- cycle;
\draw[style2] (2.82843,-1.63299) -- (2.82843,-0.816497) -- (2.12132,-1.22474) -- (2.12132,-2.04124) -- cycle;
\draw[style2] (4.24264,-1.63299) -- (4.24264,-0.816497) -- (3.53553,-1.22474) -- (3.53553,-2.04124) -- cycle;
\draw[style2] (3.53553,-2.85774) -- (3.53553,-2.04124) -- (2.82843,-2.44949) -- (2.82843,-3.26599) -- cycle;
\draw[style2] (4.94975,-2.85774) -- (4.94975,-2.04124) -- (4.24264,-2.44949) -- (4.24264,-3.26599) -- cycle;
\draw[style2] (-0.707107,4.49073) -- (-0.707107,5.30723) -- (0.,4.89898) -- (0.,4.08248) -- cycle;
\draw[style2] (-0.707107,2.04124) -- (-0.707107,2.85774) -- (0.,2.44949) -- (0.,1.63299) -- cycle;
\draw[style2] (-2.12132,-2.04124) -- (-2.12132,-1.22474) -- (-1.41421,-1.63299) -- (-1.41421,-2.44949) -- cycle;
\draw[style2] (0.,0.816497) -- (0.,1.63299) -- (0.707107,1.22474) -- (0.707107,0.408248) -- cycle;
\draw[style2] (-0.707107,-0.408248) -- (-0.707107,0.408248) -- (0.,0.) -- (0.,-0.816497) -- cycle;
\draw[style2] (0.707107,-0.408248) -- (0.707107,0.408248) -- (1.41421,0.) -- (1.41421,-0.816497) -- cycle;
\draw[style2] (0.,-2.44949) -- (0.,-1.63299) -- (0.707107,-2.04124) -- (0.707107,-2.85774) -- cycle;
\draw[style2] (1.41421,-1.63299) -- (1.41421,-0.816497) -- (2.12132,-1.22474) -- (2.12132,-2.04124) -- cycle;
\draw[style2] (2.82843,-1.63299) -- (2.82843,-0.816497) -- (3.53553,-1.22474) -- (3.53553,-2.04124) -- cycle;
\draw[style2] (2.12132,-2.85774) -- (2.12132,-2.04124) -- (2.82843,-2.44949) -- (2.82843,-3.26599) -- cycle;
\draw[style2] (3.53553,-2.85774) -- (3.53553,-2.04124) -- (4.24264,-2.44949) -- (4.24264,-3.26599) -- cycle;
\draw[style2] (0.,5.71548) -- (-0.707107,5.30723) -- (0.,4.89898) -- (0.707107,5.30723) -- cycle;
\draw[style2] (0.,3.26599) -- (-0.707107,2.85774) -- (0.,2.44949) -- (0.707107,2.85774) -- cycle;
\draw[style2] (-1.41421,-0.816497) -- (-2.12132,-1.22474) -- (-1.41421,-1.63299) -- (-0.707107,-1.22474) -- cycle;
\draw[style2] (0.707107,2.04124) -- (0.,1.63299) -- (0.707107,1.22474) -- (1.41421,1.63299) -- cycle;
\draw[style2] (0.,0.816497) -- (-0.707107,0.408248) -- (0.,0.) -- (0.707107,0.408248) -- cycle;
\draw[style2] (1.41421,0.816497) -- (0.707107,0.408248) -- (1.41421,0.) -- (2.12132,0.408248) -- cycle;
\draw[style2] (0.707107,-1.22474) -- (0.,-1.63299) -- (0.707107,-2.04124) -- (1.41421,-1.63299) -- cycle;
\draw[style2] (2.12132,-0.408248) -- (1.41421,-0.816497) -- (2.12132,-1.22474) -- (2.82843,-0.816497) -- cycle;
\draw[style2] (3.53553,-0.408248) -- (2.82843,-0.816497) -- (3.53553,-1.22474) -- (4.24264,-0.816497) -- cycle;
\draw[style2] (2.82843,-1.63299) -- (2.12132,-2.04124) -- (2.82843,-2.44949) -- (3.53553,-2.04124) -- cycle;
\draw[style2] (4.24264,-1.63299) -- (3.53553,-2.04124) -- (4.24264,-2.44949) -- (4.94975,-2.04124) -- cycle;

%% file: elliptic_draft_brief.bbl
\providecommand{\href}[2]{#2}\begingroup\raggedright\begin{thebibliography}{10}

\bibitem{Li:2020rij}
W.~Li and M.~Yamazaki, ``{Quiver Yangian from Crystal Melting},''
  \href{http://dx.doi.org/10.1007/JHEP11(2020)035}{{\em JHEP} {\bfseries 11}
  (2020) 035}, \href{http://arxiv.org/abs/2003.08909}{{\ttfamily
  arXiv:2003.08909 [hep-th]}}.

\bibitem{Galakhov:2020vyb}
D.~Galakhov and M.~Yamazaki, ``{Quiver Yangian and Supersymmetric Quantum
  Mechanics},'' \href{http://arxiv.org/abs/2008.07006}{{\ttfamily
  arXiv:2008.07006 [hep-th]}}.

\bibitem{Galakhov:2021xum}
D.~Galakhov, W.~Li, and M.~Yamazaki, ``{Shifted Quiver Yangians and
  Representations from BPS Crystals},''
  \href{http://arxiv.org/abs/2106.01230}{{\ttfamily arXiv:2106.01230
  [hep-th]}}.

\bibitem{Belavin-Drinfeld}
A.~A. Belavin and V.~G. Drinfel'd, ``Solutions of the classical {Y}ang-{B}axter
  equation for simple {L}ie algebras,'' {\em Funktsional. Anal. i Prilozhen.}
  {\bfseries 16} no.~3, (1982) 1--29, 96.

\bibitem{MR1324698}
V.~Ginzburg, M.~Kapranov, and E.~Vasserot, ``Langlands reciprocity for
  algebraic surfaces,'' \href{http://dx.doi.org/10.4310/MRL.1995.v2.n2.a4}{{\em
  Math. Res. Lett.} {\bfseries 2} no.~2, (1995) 147--160}.

\bibitem{Ding:1996mq}
J.-t. Ding and K.~Iohara, ``{Generalization and deformation of Drinfeld quantum
  affine algebras},'' \href{http://dx.doi.org/10.1023/A:1007341410987}{{\em
  Lett. Math. Phys.} {\bfseries 41} (1997) 181--193}.

\bibitem{Miki2007}
K.~Miki, ``A $(q,\gamma)$ analog of the $\mathcal{W}_{1+\infty}$ algebra,''
  \href{http://dx.doi.org/10.1063/1.2823979}{{\em Journal of Mathematical
  Physics} {\bfseries 48} (12, 2007) 123520--123520}.

\bibitem{MR2793271}
B.~Feigin, E.~Feigin, M.~Jimbo, T.~Miwa, and E.~Mukhin, ``Quantum continuous
  {$\mathfrak{gl}_\infty$}: semiinfinite construction of representations,''
  \href{http://dx.doi.org/10.1215/21562261-1214375}{{\em Kyoto J. Math.}
  {\bfseries 51} no.~2, (2011) 337--364}.

\bibitem{Feigin:2013fga}
B.~Feigin, M.~Jimbo, T.~Miwa, and E.~Mukhin, ``{Branching rules for quantum
  toroidal gl$_n$},'' \href{http://dx.doi.org/10.1016/j.aim.2016.03.019}{{\em
  Adv. Math.} {\bfseries 300} (2016) 229--274},
\href{http://arxiv.org/abs/1309.2147}{{\ttfamily arXiv:1309.2147 [math.QA]}}.

\bibitem{MR2566895}
B.~Feigin, K.~Hashizume, A.~Hoshino, J.~Shiraishi, and S.~Yanagida, ``A
  commutative algebra on degenerate {$\mathbb{CP}^1$} and {M}acdonald
  polynomials,'' \href{http://dx.doi.org/10.1063/1.3192773}{{\em J. Math.
  Phys.} {\bfseries 50} no.~9, (2009) 095215, 42}.

\bibitem{Bezerra:2019dmp}
L.~Bezerra and E.~Mukhin, ``{Quantum toroidal algebra associated with
  $\mathfrak{gl}_{m|n}$},''
\href{http://arxiv.org/abs/1904.07297}{{\ttfamily arXiv:1904.07297 [math.QA]}}.

\bibitem{Kels:2018xnf}
A.~P. Kels and M.~Yamazaki, ``{Lens Generalisation of $\tau$-Functions for the
  Elliptic Discrete Painlev\'e Equation},''
  \href{http://dx.doi.org/10.1093/imrn/rnz063}{{\em Int. Math. Res. Not.}
  {\bfseries 2021} no.~1, (2021) 754--765},
  \href{http://arxiv.org/abs/1810.12103}{{\ttfamily arXiv:1810.12103
  [nlin.SI]}}.

\bibitem{Kels:2015bda}
A.~P. Kels, ``{New solutions of the star\textendash{}triangle relation with
  discrete and continuous spin variables},''
  \href{http://dx.doi.org/10.1088/1751-8113/48/43/435201}{{\em J. Phys. A}
  {\bfseries 48} no.~43, (2015) 435201},
  \href{http://arxiv.org/abs/1504.07074}{{\ttfamily arXiv:1504.07074
  [math-ph]}}.

\bibitem{Kels:2017toi}
A.~P. Kels and M.~Yamazaki, ``{Elliptic hypergeometric sum/integral
  transformations and supersymmetric lens index},''
  \href{http://dx.doi.org/10.3842/SIGMA.2018.013}{{\em SIGMA} {\bfseries 14}
  (2018) 013}, \href{http://arxiv.org/abs/1704.03159}{{\ttfamily
  arXiv:1704.03159 [math-ph]}}.

\bibitem{Noshita:2021ldl}
G.~Noshita and A.~Watanabe, ``{A Note on Quiver Quantum Toroidal Algebra},''
  \href{http://arxiv.org/abs/2108.07104}{{\ttfamily arXiv:2108.07104
  [hep-th]}}.

\bibitem{MR3951025}
D.~Maulik and A.~Okounkov, ``Quantum groups and quantum cohomology,''
  \href{http://dx.doi.org/10.24033/ast}{{\em Ast\'{e}risque} no.~408, (2019)
  ix+209}, \href{http://arxiv.org/abs/1211.1287}{{\ttfamily arXiv:1211.1287
  [math.AG]}}.

\bibitem{MR3150250}
O.~Schiffmann and E.~Vasserot, ``Cherednik algebras, {W}-algebras and the
  equivariant cohomology of the moduli space of instantons on
  {$\mathbb{A}^2$},'' \href{http://dx.doi.org/10.1007/s10240-013-0052-3}{{\em
  Publ. Math. Inst. Hautes \'{E}tudes Sci.} {\bfseries 118} (2013) 213--342}.

\bibitem{Tsymbaliuk}
A.~Tsymbaliuk, ``{The affine Yangian of $\mathfrak{gl}_{1}$, and the
  infinitesimal Cherednik algebras},''. Ph.D. thesis.

\bibitem{Tsymbaliuk:2014fvq}
A.~Tsymbaliuk, ``{The affine Yangian of $\mathfrak{gl}_1$ revisited},''
  \href{http://dx.doi.org/10.1016/j.aim.2016.08.041}{{\em Adv. Math.}
  {\bfseries 304} (2017) 583--645},
\href{http://arxiv.org/abs/1404.5240}{{\ttfamily arXiv:1404.5240 [math.RT]}}.

\bibitem{Prochazka:2015deb}
T.~Proch\'azka, ``{$ \mathcal{W} $ -symmetry, topological vertex and affine
  Yangian},'' \href{http://dx.doi.org/10.1007/JHEP10(2016)077}{{\em JHEP}
  {\bfseries 10} (2016) 077}, \href{http://arxiv.org/abs/1512.07178}{{\ttfamily
  arXiv:1512.07178 [hep-th]}}.

\bibitem{Gaberdiel:2017dbk}
M.~R. Gaberdiel, R.~Gopakumar, W.~Li, and C.~Peng, ``{Higher Spins and Yangian
  Symmetries},'' \href{http://dx.doi.org/10.1007/JHEP04(2017)152}{{\em JHEP}
  {\bfseries 04} (2017) 152}, \href{http://arxiv.org/abs/1702.05100}{{\ttfamily
  arXiv:1702.05100 [hep-th]}}.

\bibitem{Awata:2017lqa}
H.~Awata, H.~Kanno, A.~Mironov, A.~Morozov, K.~Suetake, and Y.~Zenkevich,
  ``{$(q,t)$-KZ equations for quantum toroidal algebra and Nekrasov partition
  functions on ALE spaces},''
  \href{http://dx.doi.org/10.1007/JHEP03(2018)192}{{\em JHEP} {\bfseries 03}
  (2018) 192}, \href{http://arxiv.org/abs/1712.08016}{{\ttfamily
  arXiv:1712.08016 [hep-th]}}.

\bibitem{Awata:2018svb}
H.~Awata, H.~Kanno, A.~Mironov, A.~Morozov, K.~Suetake, and Y.~Zenkevich,
  ``{The MacMahon $R$-matrix},''
  \href{http://dx.doi.org/10.1007/JHEP04(2019)097}{{\em JHEP} {\bfseries 04}
  (2019) 097}, \href{http://arxiv.org/abs/1810.07676}{{\ttfamily
  arXiv:1810.07676 [hep-th]}}.

\bibitem{MR3262444}
Y.~Saito, ``Elliptic {D}ing-{I}ohara algebra and the free field realization of
  the elliptic {M}acdonald operator,''
  \href{http://dx.doi.org/10.4171/PRIMS/139}{{\em Publ. Res. Inst. Math. Sci.}
  {\bfseries 50} no.~3, (2014) 411--455}.

\bibitem{Kennaway:2007tq}
K.~D. Kennaway, ``{Brane Tilings},''
  \href{http://dx.doi.org/10.1142/S0217751X07036877}{{\em Int. J. Mod. Phys.}
  {\bfseries A22} (2007) 2977--3038},
\href{http://arxiv.org/abs/0706.1660}{{\ttfamily arXiv:0706.1660 [hep-th]}}.

\bibitem{Yamazaki:2008bt}
M.~Yamazaki, ``{Brane Tilings and Their Applications},''
  \href{http://dx.doi.org/10.1002/prop.200810536}{{\em Fortsch. Phys.}
  {\bfseries 56} (2008) 555--686},
\href{http://arxiv.org/abs/0803.4474}{{\ttfamily arXiv:0803.4474 [hep-th]}}.

\bibitem{Nagao:2009rq}
K.~Nagao and M.~Yamazaki, ``{The Non-commutative Topological Vertex and Wall
  Crossing Phenomena},''
  \href{http://dx.doi.org/10.4310/ATMP.2010.v14.n4.a3}{{\em Adv. Theor. Math.
  Phys.} {\bfseries 14} no.~4, (2010) 1147--1181},
\href{http://arxiv.org/abs/0910.5479}{{\ttfamily arXiv:0910.5479 [hep-th]}}.

\bibitem{2019arXiv191208729B}
L.~{Bezerra} and E.~{Mukhin}, ``{Braid actions on quantum toroidal
  superalgebras},'' \href{http://arxiv.org/abs/1912.08729}{{\ttfamily
  arXiv:1912.08729 [math.QA]}}.

\bibitem{Konno:2016fmh}
H.~Konno, ``{Elliptic Quantum Groups $U_{q,p} (\hat{gl}_N)$ and $E_{q,p}
  (\hat{gl}_N)$},'' \href{http://arxiv.org/abs/1603.04129}{{\ttfamily
  arXiv:1603.04129 [math.QA]}}.

\bibitem{konno2009elliptic}
H.~Konno, ``The elliptic quantum group $u_{q, p}(\widehat{\mathfrak{sl}}_2)$,''
  {\em {RIMS Kokyuroku Bessatsu}} {\bfseries 11} (2009) 53--73.

\bibitem{Okounkov:2003sp}
A.~Okounkov, N.~Reshetikhin, and C.~Vafa, ``{Quantum Calabi-Yau and classical
  crystals},'' \href{http://dx.doi.org/10.1007/0-8176-4467-9_16}{{\em Prog.
  Math.} {\bfseries 244} (2006) 597},
  \href{http://arxiv.org/abs/hep-th/0309208}{{\ttfamily arXiv:hep-th/0309208}}.

\bibitem{Iqbal:2003ds}
A.~Iqbal, N.~Nekrasov, A.~Okounkov, and C.~Vafa, ``{Quantum foam and
  topological strings},''
  \href{http://dx.doi.org/10.1088/1126-6708/2008/04/011}{{\em JHEP} {\bfseries
  04} (2008) 011}, \href{http://arxiv.org/abs/hep-th/0312022}{{\ttfamily
  arXiv:hep-th/0312022}}.

\bibitem{Szendroi}
B.~Szendr{\H{o}}i, ``Non-commutative {D}onaldson-{T}homas invariants and the
  conifold,'' {\em Geom. Topol.} {\bfseries 12} no.~2, (2008) 1171--1202,
  \href{http://arxiv.org/abs/0705.3419}{{\ttfamily arXiv:0705.3419 [math.AG]}}.

\bibitem{MR2836398}
K.~Nagao and H.~Nakajima, ``Counting invariant of perverse coherent sheaves and
  its wall-crossing,'' \href{http://dx.doi.org/10.1093/imrn/rnq195}{{\em Int.
  Math. Res. Not. IMRN} {\bfseries 2011} no.~17, (2011) 3885--3938},
  \href{http://arxiv.org/abs/0809.2992}{{\ttfamily arXiv:0809.2992 [math.AG]}}.

\bibitem{MR2592501}
S.~Mozgovoy and M.~Reineke, ``On the noncommutative {D}onaldson-{T}homas
  invariants arising from brane tilings,''
  \href{http://dx.doi.org/10.1016/j.aim.2009.10.001}{{\em Adv. Math.}
  {\bfseries 223} no.~5, (2010) 1521--1544}.

\bibitem{Ooguri:2009ijd}
H.~Ooguri and M.~Yamazaki, ``{Crystal Melting and Toric Calabi-Yau
  Manifolds},'' \href{http://dx.doi.org/10.1007/s00220-009-0836-y}{{\em Commun.
  Math. Phys.} {\bfseries 292} (2009) 179--199},
  \href{http://arxiv.org/abs/0811.2801}{{\ttfamily arXiv:0811.2801 [hep-th]}}.

\bibitem{Yamazaki:2010fz}
M.~Yamazaki, ``{Crystal Melting and Wall Crossing Phenomena},''
  \href{http://dx.doi.org/10.1142/S0217751X11051482}{{\em Int. J. Mod. Phys.}
  {\bfseries A26} (2011) 1097--1228},
\href{http://arxiv.org/abs/1002.1709}{{\ttfamily arXiv:1002.1709 [hep-th]}}.

\bibitem{Maldonado:2015gfa}
R.~Maldonado and N.~S. Manton, ``{Analytic vortex solutions on compact
  hyperbolic surfaces},''
  \href{http://dx.doi.org/10.1088/1751-8113/48/24/245403}{{\em J. Phys. A}
  {\bfseries 48} no.~24, (2015) 245403},
  \href{http://arxiv.org/abs/1502.01990}{{\ttfamily arXiv:1502.01990
  [hep-th]}}.

\bibitem{Miyake:2011yr}
A.~Miyake, K.~Ohta, and N.~Sakai, ``{Volume of Moduli Space of Vortex Equations
  and Localization},'' \href{http://dx.doi.org/10.1143/PTP.126.637}{{\em Prog.
  Theor. Phys.} {\bfseries 126} (2011) 637--680},
  \href{http://arxiv.org/abs/1105.2087}{{\ttfamily arXiv:1105.2087 [hep-th]}}.

\bibitem{Cordes:1994fc}
S.~Cordes, G.~W. Moore, and S.~Ramgoolam, ``{Lectures on 2-d Yang-Mills theory,
  equivariant cohomology and topological field theories},''
  \href{http://dx.doi.org/10.1016/0920-5632(95)00434-B}{{\em Nucl. Phys. B
  Proc. Suppl.} {\bfseries 41} (1995) 184--244},
  \href{http://arxiv.org/abs/hep-th/9411210}{{\ttfamily arXiv:hep-th/9411210}}.

\bibitem{Alvarez-Gaume:1986rcs}
L.~Alvarez-Gaume, G.~W. Moore, and C.~Vafa, ``{Theta Functions, Modular
  Invariance and Strings},'' \href{http://dx.doi.org/10.1007/BF01210925}{{\em
  Commun. Math. Phys.} {\bfseries 106} (1986) 1--40}.

\bibitem{Nakajima_book}
H.~Nakajima, \href{http://dx.doi.org/10.1090/ulect/018}{{\em Lectures on
  {H}ilbert schemes of points on surfaces}}, vol.~18 of {\em University Lecture
  Series}.
\newblock American Mathematical Society, Providence, RI, 1999.

\bibitem{Braverman:2016wma}
A.~Braverman, M.~Finkelberg, and H.~Nakajima, ``{Towards a mathematical
  definition of Coulomb branches of $3$-dimensional $\mathcal{N} = 4$ gauge
  theories, II},'' \href{http://dx.doi.org/10.4310/ATMP.2018.v22.n5.a1}{{\em
  Adv. Theor. Math. Phys.} {\bfseries 22} (2018) 1071--1147},
  \href{http://arxiv.org/abs/1601.03586}{{\ttfamily arXiv:1601.03586
  [math.RT]}}.

\bibitem{Taubes:1979tm}
C.~H. Taubes, ``{Arbitrary N: Vortex Solutions to the First Order
  Landau-Ginzburg Equations},''
  \href{http://dx.doi.org/10.1007/BF01197552}{{\em Commun. Math. Phys.}
  {\bfseries 72} (1980) 277--292}.

\bibitem{Manton:2010mj}
N.~S. Manton and N.~A. Rink, ``{Geometry and Energy of Non-abelian Vortices},''
  \href{http://dx.doi.org/10.1063/1.3574357}{{\em J. Math. Phys.} {\bfseries
  52} (2011) 043511}, \href{http://arxiv.org/abs/1012.3014}{{\ttfamily
  arXiv:1012.3014 [hep-th]}}.

\bibitem{Eto:2005yh}
M.~Eto, Y.~Isozumi, M.~Nitta, K.~Ohashi, and N.~Sakai, ``{Moduli space of
  non-Abelian vortices},''
  \href{http://dx.doi.org/10.1103/PhysRevLett.96.161601}{{\em Phys. Rev. Lett.}
  {\bfseries 96} (2006) 161601},
  \href{http://arxiv.org/abs/hep-th/0511088}{{\ttfamily arXiv:hep-th/0511088}}.

\bibitem{Bullimore:2016hdc}
M.~Bullimore, T.~Dimofte, D.~Gaiotto, J.~Hilburn, and H.-C. Kim, ``{Vortices
  and Vermas},'' \href{http://dx.doi.org/10.4310/ATMP.2018.v22.n4.a1}{{\em Adv.
  Theor. Math. Phys.} {\bfseries 22} (2018) 803--917},
  \href{http://arxiv.org/abs/1609.04406}{{\ttfamily arXiv:1609.04406
  [hep-th]}}.

\bibitem{Hanany:2003hp}
A.~Hanany and D.~Tong, ``{Vortices, instantons and branes},''
  \href{http://dx.doi.org/10.1088/1126-6708/2003/07/037}{{\em JHEP} {\bfseries
  07} (2003) 037}, \href{http://arxiv.org/abs/hep-th/0306150}{{\ttfamily
  arXiv:hep-th/0306150}}.

\bibitem{Bullimore:2018jlp}
M.~Bullimore, A.~Ferrari, and H.~Kim, ``{Twisted indices of 3d $ \mathcal{N} $
  = 4 gauge theories and enumerative geometry of quasi-maps},''
  \href{http://dx.doi.org/10.1007/JHEP07(2019)014}{{\em JHEP} {\bfseries 07}
  (2019) 014}, \href{http://arxiv.org/abs/1812.05567}{{\ttfamily
  arXiv:1812.05567 [hep-th]}}.

\bibitem{Bullimore:2021auw}
M.~Bullimore, A.~Ferrari, and H.~Kim, ``{Supersymmetric Ground States of 3d
  $\mathcal{N}=4$ Gauge Theories on a Riemann Surface},''
  \href{http://arxiv.org/abs/2105.08783}{{\ttfamily arXiv:2105.08783
  [hep-th]}}.

\bibitem{Borokhov:2002cg}
V.~Borokhov, A.~Kapustin, and X.-k. Wu, ``{Monopole operators and mirror
  symmetry in three-dimensions},''
  \href{http://dx.doi.org/10.1088/1126-6708/2002/12/044}{{\em JHEP} {\bfseries
  12} (2002) 044}, \href{http://arxiv.org/abs/hep-th/0207074}{{\ttfamily
  arXiv:hep-th/0207074}}.

\bibitem{Borokhov:2002ib}
V.~Borokhov, A.~Kapustin, and X.-k. Wu, ``{Topological disorder operators in
  three-dimensional conformal field theory},''
  \href{http://dx.doi.org/10.1088/1126-6708/2002/11/049}{{\em JHEP} {\bfseries
  11} (2002) 049}, \href{http://arxiv.org/abs/hep-th/0206054}{{\ttfamily
  arXiv:hep-th/0206054}}.

\bibitem{Intriligator:2013lca}
K.~Intriligator and N.~Seiberg, ``{Aspects of 3d N=2 Chern-Simons-Matter
  Theories},'' \href{http://dx.doi.org/10.1007/JHEP07(2013)079}{{\em JHEP}
  {\bfseries 07} (2013) 079}, \href{http://arxiv.org/abs/1305.1633}{{\ttfamily
  arXiv:1305.1633 [hep-th]}}.

\bibitem{Franco:2015tya}
S.~Franco, S.~Lee, and R.-K. Seong, ``{Brane Brick Models, Toric Calabi-Yau
  4-Folds and 2d (0,2) Quivers},''
  \href{http://dx.doi.org/10.1007/JHEP02(2016)047}{{\em JHEP} {\bfseries 02}
  (2016) 047}, \href{http://arxiv.org/abs/1510.01744}{{\ttfamily
  arXiv:1510.01744 [hep-th]}}.

\bibitem{Franco:2019bmx}
S.~Franco and A.~Hasan, ``{Graded Quivers, Generalized Dimer Models and Toric
  Geometry},'' \href{http://dx.doi.org/10.1007/JHEP11(2019)104}{{\em JHEP}
  {\bfseries 11} (2019) 104}, \href{http://arxiv.org/abs/1904.07954}{{\ttfamily
  arXiv:1904.07954 [hep-th]}}.

\bibitem{cao2015donaldsonthomas}
Y.~Cao and N.~C. Leung, ``{Donaldson-Thomas theory for Calabi-Yau 4-folds},''
  \href{http://arxiv.org/abs/1407.7659}{{\ttfamily arXiv:1407.7659 [math.AG]}}.

\bibitem{MR3861713}
Y.~Cao and M.~Kool, ``Zero-dimensional {D}onaldson-{T}homas invariants of
  {C}alabi-{Y}au 4-folds,''
  \href{http://dx.doi.org/10.1016/j.aim.2018.09.011}{{\em Adv. Math.}
  {\bfseries 338} (2018) 601--648}.

\bibitem{cao2020ktheoretic}
Y.~Cao, M.~Kool, and S.~Monavari, ``{$K$-theoretic DT/PT correspondence for
  toric Calabi-Yau 4-folds},''
  \href{http://arxiv.org/abs/1906.07856}{{\ttfamily arXiv:1906.07856
  [math.AG]}}.

\bibitem{Nekrasov:2017cih}
N.~Nekrasov, ``{Magnificent four},''
  \href{http://dx.doi.org/10.4310/ATMP.2020.v24.n5.a4}{{\em Adv. Theor. Math.
  Phys.} {\bfseries 24} no.~5, (2020) 1171--1202},
  \href{http://arxiv.org/abs/1712.08128}{{\ttfamily arXiv:1712.08128
  [hep-th]}}.

\bibitem{Nekrasov:2018xsb}
N.~Nekrasov and N.~Piazzalunga, ``{Magnificent Four with Colors},''
  \href{http://dx.doi.org/10.1007/s00220-019-03426-3}{{\em Commun. Math. Phys.}
  {\bfseries 372} no.~2, (2019) 573--597},
  \href{http://arxiv.org/abs/1808.05206}{{\ttfamily arXiv:1808.05206
  [hep-th]}}.

\bibitem{Nekrasov:2003rj}
N.~Nekrasov and A.~Okounkov, ``{Seiberg-Witten theory and random partitions},''
  \href{http://dx.doi.org/10.1007/0-8176-4467-9_15}{{\em Prog. Math.}
  {\bfseries 244} (2006) 525--596},
  \href{http://arxiv.org/abs/hep-th/0306238}{{\ttfamily arXiv:hep-th/0306238}}.

\bibitem{Smirnov:2013hh}
A.~Smirnov, ``{On the Instanton R-matrix},''
  \href{http://dx.doi.org/10.1007/s00220-016-2686-8}{{\em Commun. Math. Phys.}
  {\bfseries 345} no.~3, (2016) 703--740},
  \href{http://arxiv.org/abs/1302.0799}{{\ttfamily arXiv:1302.0799 [math.AG]}}.

\bibitem{MR1492989}
A.~Klimyk and K.~Schm\"{u}dgen,
  \href{http://dx.doi.org/10.1007/978-3-642-60896-4}{{\em Quantum groups and
  their representations}}.
\newblock Texts and Monographs in Physics. Springer-Verlag, Berlin, 1997.

\bibitem{Jimbo:1998bi}
M.~Jimbo, H.~Konno, S.~Odake, and J.~Shiraishi, ``{Elliptic algebra
  $U_{q,p}(\widehat{\mathfrak{sl}}_2)$: Drinfeld currents and vertex
  operators},'' \href{http://dx.doi.org/10.1007/s002200050514}{{\em Commun.
  Math. Phys.} {\bfseries 199} (1999) 605--647},
  \href{http://arxiv.org/abs/math/9802002}{{\ttfamily arXiv:math/9802002}}.

\bibitem{Jimbo:1999zz}
M.~Jimbo, H.~Konno, S.~Odake, and J.~Shiraishi, ``{Quasi-Hopf twistors for
  elliptic quantum groups},'' \href{http://dx.doi.org/10.1007/BF01238562}{{\em
  Transform. Groups} {\bfseries 4} (1999) 303--327},
  \href{http://arxiv.org/abs/q-alg/9712029}{{\ttfamily arXiv:q-alg/9712029}}.

\bibitem{Mironov:2021sfo}
A.~Mironov, A.~Morozov, and Y.~Zenkevich, ``{Duality in elliptic Ruijsenaars
  system and elliptic symmetric functions},''
  \href{http://dx.doi.org/10.1140/epjc/s10052-021-09248-9}{{\em Eur. Phys. J.
  C} {\bfseries 81} no.~5, (2021) 461},
  \href{http://arxiv.org/abs/2103.02508}{{\ttfamily arXiv:2103.02508
  [hep-th]}}.

\bibitem{FT}
B.~L. Feigin and A.~I. Tsymbaliuk, ``Equivariant {$K$}-theory of {H}ilbert
  schemes via shuffle algebra,''
  \href{http://dx.doi.org/10.1215/21562261-1424875}{{\em Kyoto J. Math.}
  {\bfseries 51} no.~4, (2011) 831--854}.

\bibitem{1995q.alg.....9021F}
B.~L. {Feigin} and A.~V. {Odesskii}, ``{Vector bundles on elliptic curve and
  Sklyanin algebras},'' \href{http://arxiv.org/abs/q-alg/9509021}{{\ttfamily
  arXiv:q-alg/9509021 [math.QA]}}.

\bibitem{Kontsevich:2010px}
M.~Kontsevich and Y.~Soibelman, ``{Cohomological Hall algebra, exponential
  Hodge structures and motivic Donaldson-Thomas invariants},''
  \href{http://dx.doi.org/10.4310/CNTP.2011.v5.n2.a1}{{\em Commun. Num. Theor.
  Phys.} {\bfseries 5} (2011) 231--352},
  \href{http://arxiv.org/abs/1006.2706}{{\ttfamily arXiv:1006.2706 [math.AG]}}.

\bibitem{Galakhov:2018lta}
D.~Galakhov, ``{BPS Hall Algebra of Scattering Hall States},''
  \href{http://dx.doi.org/10.1016/j.nuclphysb.2019.114693}{{\em Nucl. Phys. B}
  {\bfseries 946} (2019) 114693},
  \href{http://arxiv.org/abs/1812.05801}{{\ttfamily arXiv:1812.05801
  [hep-th]}}.

\bibitem{Beaujard:2021fsk}
G.~Beaujard, S.~Mondal, and B.~Pioline, ``{Multi-centered black holes, scaling
  solutions and pure-Higgs indices from localization},''
  \href{http://arxiv.org/abs/2103.03205}{{\ttfamily arXiv:2103.03205
  [hep-th]}}.

\bibitem{1998math......9036E}
B.~{Enriquez}, ``{On correlation functions of Drinfeld currents and shuffle
  algebras},'' \href{http://arxiv.org/abs/math/9809036}{{\ttfamily
  arXiv:math/9809036 [math.QA]}}.

\bibitem{Negut:2020npc}
A.~Negu\c{t}, ``{The R-matrix of the quantum toroidal algebra},''
  \href{http://arxiv.org/abs/2005.14182}{{\ttfamily arXiv:2005.14182
  [math.QA]}}.

\bibitem{Rapcak:2018nsl}
M.~Rapcak, Y.~Soibelman, Y.~Yang, and G.~Zhao, ``{Cohomological Hall algebras,
  vertex algebras and instantons},''
  \href{http://dx.doi.org/10.1007/s00220-019-03575-5}{{\em Commun. Math. Phys.}
  {\bfseries 376} no.~3, (2019) 1803--1873},
  \href{http://arxiv.org/abs/1810.10402}{{\ttfamily arXiv:1810.10402
  [math.QA]}}.

\bibitem{2013arXiv1302.6202N}
A.~{Negu{\c{t}}}, ``{Quantum toroidal and shuffle algebras},''
  \href{http://arxiv.org/abs/1302.6202}{{\ttfamily arXiv:1302.6202 [math.RT]}}.

\bibitem{MR3805051}
Y.~Yang and G.~Zhao, ``The cohomological {H}all algebra of a preprojective
  algebra,'' \href{http://dx.doi.org/10.1112/plms.12111}{{\em Proc. Lond. Math.
  Soc. (3)} {\bfseries 116} no.~5, (2018) 1029--1074},
  \href{http://arxiv.org/abs/1407.7994}{{\ttfamily arXiv:1407.7994 [math.RT]}}.

\bibitem{Closset:2012ru}
C.~Closset, T.~T. Dumitrescu, G.~Festuccia, and Z.~Komargodski,
  ``{Supersymmetric Field Theories on Three-Manifolds},''
  \href{http://dx.doi.org/10.1007/JHEP05(2013)017}{{\em JHEP} {\bfseries 05}
  (2013) 017}, \href{http://arxiv.org/abs/1212.3388}{{\ttfamily arXiv:1212.3388
  [hep-th]}}.

\bibitem{Closset:2016arn}
C.~Closset and H.~Kim, ``{Comments on twisted indices in 3d supersymmetric
  gauge theories},'' \href{http://dx.doi.org/10.1007/JHEP08(2016)059}{{\em
  JHEP} {\bfseries 08} (2016) 059},
  \href{http://arxiv.org/abs/1605.06531}{{\ttfamily arXiv:1605.06531
  [hep-th]}}.

\bibitem{Lurie}
J.~Lurie, ``Chromatic homotopy theory.'' Lecture notes, 2010, available at
  \url{ http://people.math.harvard.edu/~lurie/252x.html}.

\bibitem{may1999concise}
J.~P. May, {\em A concise course in algebraic topology}.
\newblock University of Chicago press, Chicago and London, 1999.

\bibitem{MR423332}
P.~S. Landweber, ``Homological properties of comodules over {$M{\rm U}\sb\ast
  (M{\rm U})$} and {BP{$\sb\ast $}}({BP}),''
  \href{http://dx.doi.org/10.2307/2373808}{{\em Amer. J. Math.} {\bfseries 98}
  no.~3, (1976) 591--610}.

\bibitem{Aganagic:2016jmx}
M.~Aganagic and A.~Okounkov, ``{Elliptic stable envelopes},''
  \href{http://dx.doi.org/10.1090/jams/954}{{\em J. Am. Math. Soc.} {\bfseries
  34} no.~1, (2021) 79--133}, \href{http://arxiv.org/abs/1604.00423}{{\ttfamily
  arXiv:1604.00423 [math.AG]}}.

\bibitem{MR73925}
M.~Lazard, ``Sur les groupes de {L}ie formels \`a un param\`etre,'' {\em Bull.
  Soc. Math. France} {\bfseries 83} (1955) 251--274.
  \url{http://www.numdam.org/item?id=BSMF_1955__83__251_0}.

\bibitem{MR253350}
D.~Quillen, ``On the formal group laws of unoriented and complex cobordism
  theory,'' \href{http://dx.doi.org/10.1090/S0002-9904-1969-12401-8}{{\em Bull.
  Amer. Math. Soc.} {\bfseries 75} (1969) 1293--1298}.

\bibitem{Minasian:1997mm}
R.~Minasian and G.~W. Moore, ``{K theory and Ramond-Ramond charge},''
  \href{http://dx.doi.org/10.1088/1126-6708/1997/11/002}{{\em JHEP} {\bfseries
  11} (1997) 002}, \href{http://arxiv.org/abs/hep-th/9710230}{{\ttfamily
  arXiv:hep-th/9710230}}.

\bibitem{Witten:2000cn}
E.~Witten, ``{Overview of K theory applied to strings},''
  \href{http://dx.doi.org/10.1142/S0217751X01003822}{{\em Int. J. Mod. Phys. A}
  {\bfseries 16} (2001) 693--706},
  \href{http://arxiv.org/abs/hep-th/0007175}{{\ttfamily arXiv:hep-th/0007175}}.

\bibitem{Freed:2002qp}
D.~S. Freed, ``{K theory in quantum field theory},'' in {\em {Current
  Developments in Mathematics}}.
\newblock 6, 2002.
\newblock \href{http://arxiv.org/abs/math-ph/0206031}{{\ttfamily
  arXiv:math-ph/0206031}}.

\bibitem{D-book_1}
K.~Hori, S.~Katz, A.~Klemm, R.~Pandharipande, R.~Thomas, C.~Vafa, R.~Vakil, and
  E.~Zaslow, {\em {Mirror symmetry}}, vol.~1 of {\em Clay mathematics
  monographs}.
\newblock AMS, Providence, USA, 2003.

\bibitem{D_book_2}
P.~S. Aspinwall, T.~Bridgeland, A.~Craw, M.~R. Douglas, A.~Kapustin, G.~W.
  Moore, M.~Gross, G.~Segal, B.~Szendr\"oi, and P.~M.~H. Wilson, {\em
  {Dirichlet branes and mirror symmetry}}, vol.~4 of {\em Clay Mathematics
  Monographs}.
\newblock AMS, Providence, RI, 2009.

\bibitem{Herbst:2008jq}
M.~Herbst, K.~Hori, and D.~Page, ``{Phases Of N=2 Theories In 1+1 Dimensions
  With Boundary},'' \href{http://arxiv.org/abs/0803.2045}{{\ttfamily
  arXiv:0803.2045 [hep-th]}}.

\bibitem{2019arXiv190802868B}
D.~{Berwick-Evans} and A.~{Tripathy}, ``{A de Rham model for complex analytic
  equivariant elliptic cohomology},''
  \href{http://arxiv.org/abs/1908.02868}{{\ttfamily arXiv:1908.02868
  [math.AT]}}.

\bibitem{1996q.alg....11030S}
Y.~{Saito}, ``{Quantum toroidal algebras and their vertex representations},''
  \href{http://arxiv.org/abs/q-alg/9611030}{{\ttfamily arXiv:q-alg/9611030
  [math.QA]}}.

\bibitem{Wess:1992cp}
J.~Wess and J.~Bagger, {\em {Supersymmetry and supergravity}}.
\newblock Princeton University Press, Princeton, NJ, USA, 1992.

\bibitem{Marino:2011nm}
M.~Marino, ``{Lectures on localization and matrix models in supersymmetric
  Chern-Simons-matter theories},''
  \href{http://dx.doi.org/10.1088/1751-8113/44/46/463001}{{\em J. Phys. A}
  {\bfseries 44} (2011) 463001},
  \href{http://arxiv.org/abs/1104.0783}{{\ttfamily arXiv:1104.0783 [hep-th]}}.

\end{thebibliography}\endgroup
